\documentclass[structabstract]{aa}  
\usepackage{graphicx}
\usepackage{txfonts}
\usepackage{natbib}

\newcommand{\teff}  {T$_\mathrm{eff}$}
\newcommand{\logg}  {$\log g$}
\newcommand{\loggf}  {$\log gf$}

\begin{document}
   \title{Chemical abundances of 1111 FGK stars from the HARPS GTO
planet search program II}

\subtitle{Cu, Zn, Sr, Y, Zr, Ba, Ce, Nd and Eu\thanks{
Based on observations collected at the La Silla Observatory, ESO
(Chile), with the HARPS spectrograph at the 3.6 m ESO telescope (ESO
runs ID 72.C---0488, 082.C---0212, and 085.C---0063).}}

   \author{E. Delgado Mena\inst{1}
      \and M. Tsantaki\inst{2}
      \and V. Zh. Adibekyan\inst{1}
      \and S. G. Sousa\inst{1,3}
      \and N.~C.~Santos\inst{1,3}
      \and J.~I.~Gonz\'alez Hern\'andez\inst{4,5}
      \and G. Israelian\inst{4,5} 
      }
      
\institute{
Instituto de Astrof\'isica e Ci\^encias do Espa\c{c}o, Universidade do Porto, CAUP, Rua das
Estrelas, PT4150-762 Porto, Portugal
             \email{Elisa.Delgado@astro.up.pt}
\and
Instituto de Radioastronom\'ia y Astrof\'isica, IRyA, UNAM, Campus Morelia, A.P. 3-72, C.P. 58089, Michoac\'an, Mexico      
\and
Departamento de F\'isica e Astronom\'ia, Faculdade de Ci\^encias, Universidade do Porto, Portugal
\and 
Instituto de Astrof\'{\i}sica de Canarias,
C/ Via Lactea, s/n, 38205, La Laguna, 
Tenerife, Spain 
\and 
Departamento de Astrof\'isica, Universidad de La Laguna, 38206 La Laguna, Tenerife, Spain
}


   \date{Received ...; accepted ...}

 
  \abstract
{}
{To understand the formation and evolution of the different stellar populations within our Galaxy it is essential to combine detailed kinematical and chemical information for large samples of stars. The aim of this work is to explore the chemical abundances of neutron capture elements which are a product of different nucleosynthesis processes taking place at diverse objects in the Galaxy, such as massive stars, asymptotic giant branch (AGB) stars and supernovae (SNe) explosions.}
{We derive chemical abundances of Cu, Zn, Sr, Y, Zr, Ba, Ce, Nd and Eu for a large sample of more than 1000 FGK dwarf stars with high-resolution ($R \sim$\,115000) and high-quality spectra from the HARPS-GTO program. The abundances are derived by a standard Local Thermodinamyc Equilibrium (LTE) analysis using measured Equivalent Widths (EWs) injected to the code MOOG and a grid of Kurucz ATLAS9 atmospheres.}
{We find that thick disk stars are chemically disjunct for Zn and Eu and also show on average higher Zr but lower Ba and Y when compared to the thin disk stars. We also discovered that the previously identified high-$\alpha$ metal-rich population is also enhanced in Cu, Zn, Nd and Eu with respect to the thin disk but presents Ba and Y abundances lower on average, following the trend of thick disk stars towards higher metallities and further supporting the different chemical composition of this population. By making a qualitative comparison of O (pure $\alpha$), Mg, Eu (pure \textit{r}-process) and \textit{s}-process elements we can distinguish between the contribution of the more massive stars (SNe II for $\alpha$ and \textit{r}-process elements) and the lower mass stars (AGBs) whose contribution to the enrichment of the Galaxy is delayed due to their longer lifetimes. The ratio of heavy-\textit{s} to light-\textit{s} elements of thin disk stars presents the expected behaviour (increasing towards lower metallicities) and can be explained by a major contribution of low-mass AGB stars for \textit{s}-process production at disk metallicities. However, the opposite trend found for thick disk stars suggests that intermediate-mass AGB stars played an important role in the enrichment of the gas from where these stars formed. Previous works in the literature also point to a possible primary production of light-\textit{s} elements at low metallicities to explain this trend. Finally, we also find an enhancement of light-\textit{s} elements in the thin disk at super solar metallicities which could be caused by the contribution of metal-rich AGB stars.}
{}


\maketitle
%

\section{Introduction}
In the last years several large spectroscopic surveys have started such as  Gaia ESO Survey \citep{gilmore12}, SEGUE \citep{yanni09}, APOGEE \citep{wilson10}, RAVE \citep{steinmetz03} or GALAH \citep{heijmans12}, helping to improve our vision and understanding of the Galaxy as well as the Galactic Chemical Evolution (GCE). In an interesting work, \citet{lindregen13} showed that when the precision is low even very large samples do not allow to separate different stellar populations. However, very interesting results can be obtained with smaller samples of high resolution and high S/N spectra. That is the case of the high quality HARPS GTO sample which allowed us to discover a new population of high-$\alpha$ metal-rich stars (hereafter \textit{h$\alpha$mr}) never unveiled before and with different properties than thin disk stars of similar iron metallicity ([Fe/H]) \citep{adibekyan11}. The objective of this work is to derive chemical abundances of heavy elements ($Z$\,$\geq$\,29) as a continuation of the work started by \citet{adibekyan12} for lighter elements using the same sample. Moreover, our volume-limited sample contains a significant number of metal rich stars ([Fe/H]\,$>$\,0.2\,dex) which permits to study the GCE at high metallicities, not very often explored in the literature.\\  

The nucleosyntheis of elements heavier than Fe cannot be produced by stellar fusion since it would require energy. Instead, they are created by neutron capture processes. There are two main kind of neutron capture processes. First, the \textit{s}-process (slow), in which the density of neutrons is low and the timescale between consecutives captures is relatively very long. If a recently created nuclei is unstable it will suffer a $\beta$ decay transforming neutrons into protons and hence producing heavier elements along the so-called \textit{s}-process path. Second, the \textit{r}-process (rapid) where the captures take place in very short timescales (shorter than the timescale for $\beta$ decay) and the density of neutrons is high. Finally, a marginal contribution of heavy elements is provided by the \textit{p}-process \citep{burbidge57}. They are also called \textit{p}-nuclei since they are relatively proton-rich nuclei built by ($p,\gamma$) and/or ($\gamma,n$) reactions. Among the elements studied in this work we have Sr, Zr and Y which belong to the first peak of the \textit{s}-process path (also called light-\textit{s}) and Ba, Ce and Nd which belong to the second peak (called heavy-\textit{s}). Eu is a \textit{r}-process element and Cu and Zn can be considered \textit{special} \textit{s}-process elements since their production sites are different than heavier \textit{s}-process elements (see next paragraph). Further information can be obtained in the reviews by \cite[e.g.][]{busso99,sneden08,kappeler11}.\\

\begin{figure}[t]
  \centering
   \includegraphics[width=1.0\linewidth]{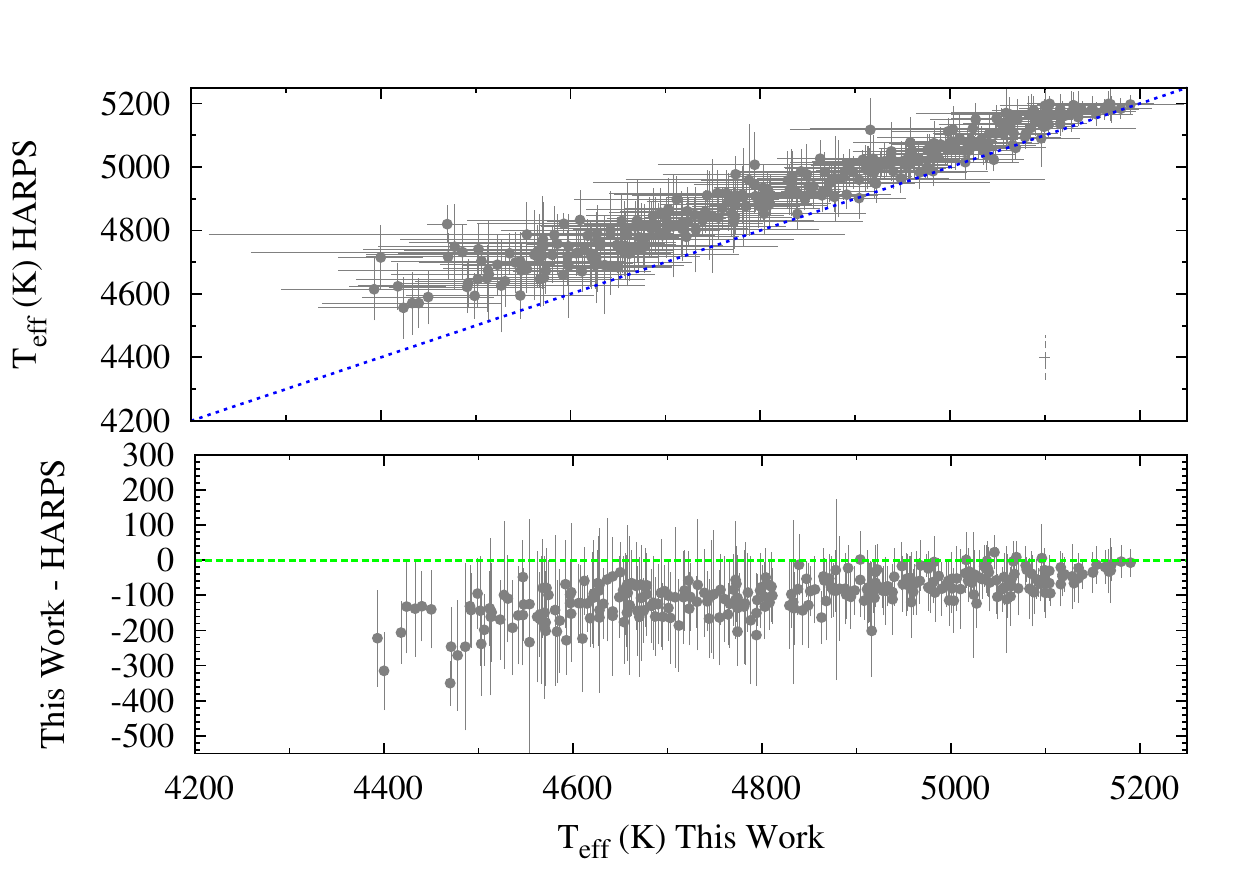} 
   \includegraphics[width=1.0\linewidth]{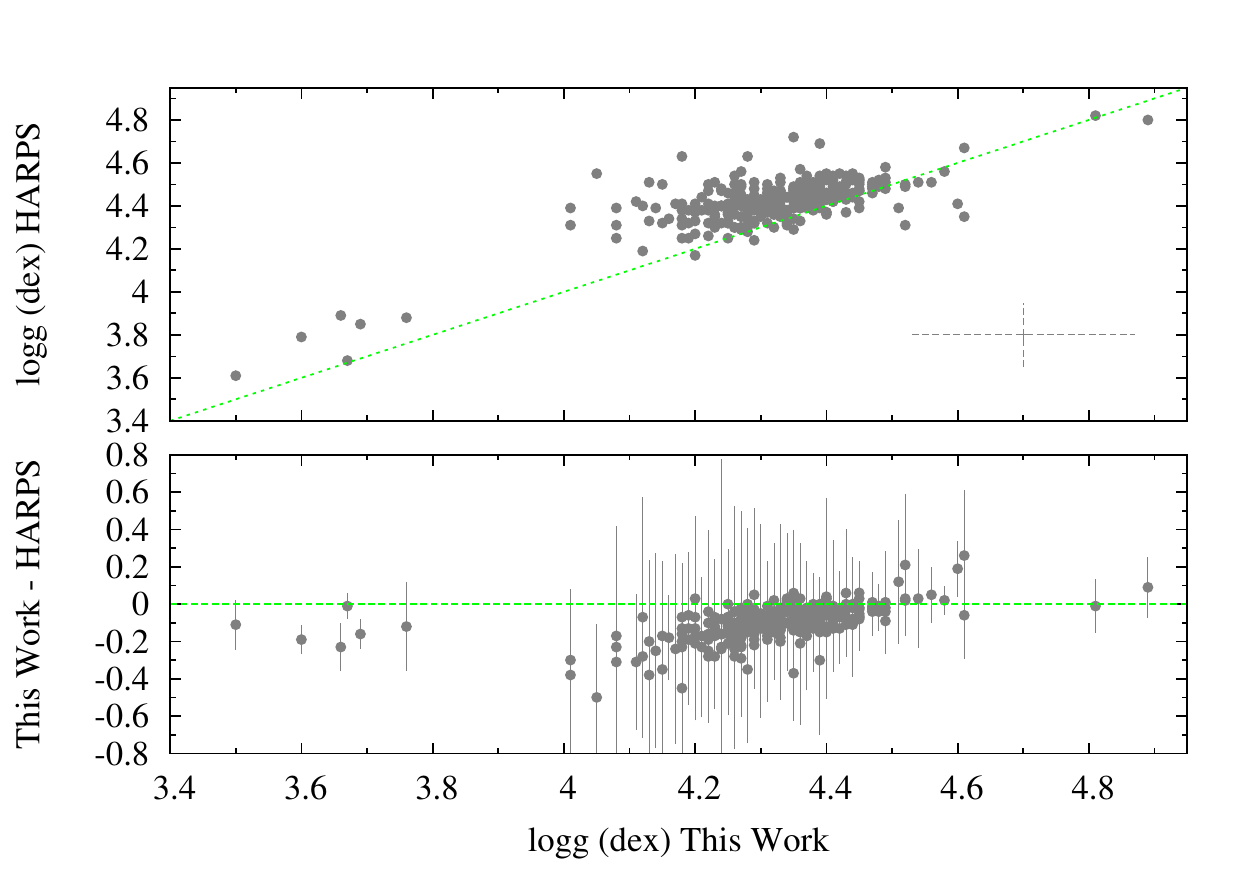} 
   \includegraphics[width=1.0\linewidth]{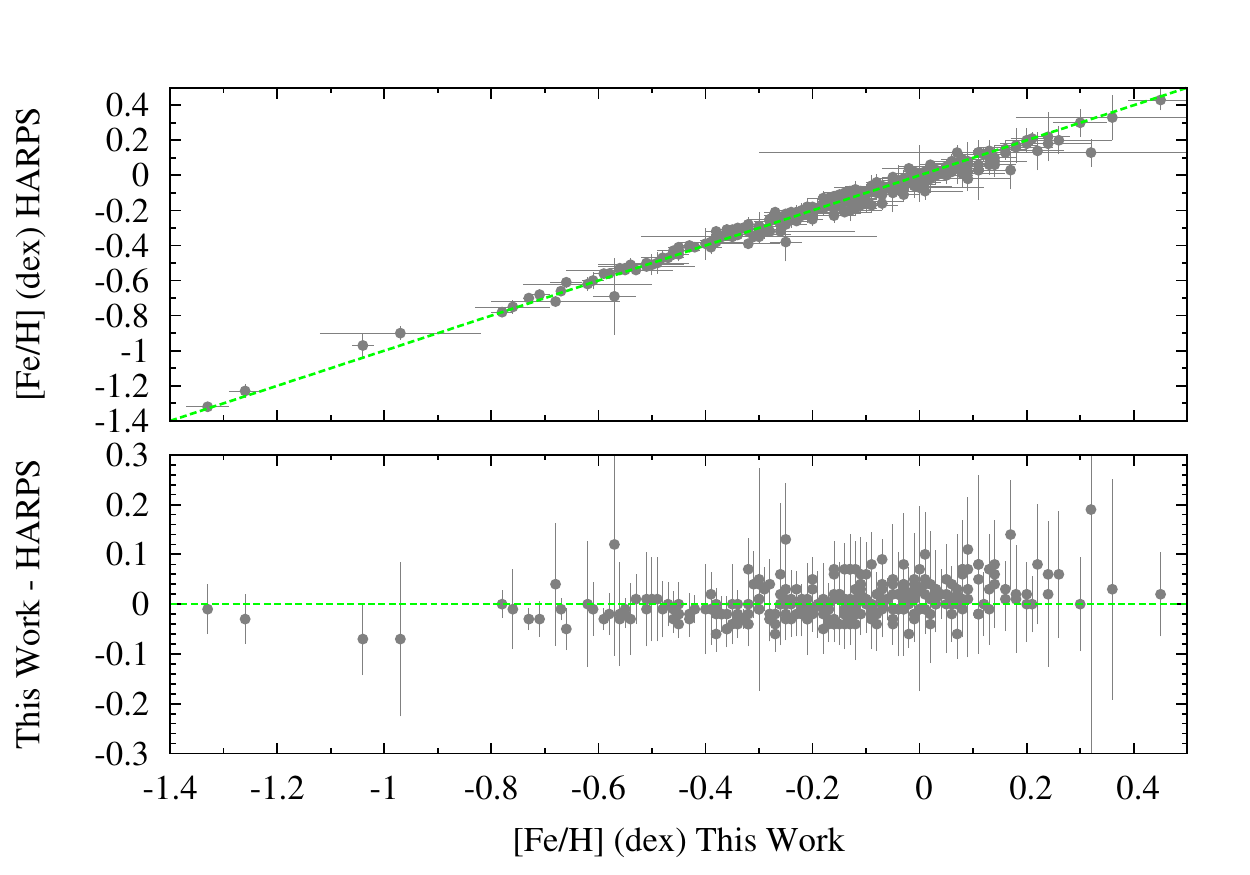} 
   \caption{\footnotesize{Comparison between the temperature derived with the cool line list of this work with the previous 
parameters (upper panel). Same for \logg\ (middle panel) and [Fe/H] (bottom panel).}}
\label{comparison}
\end{figure}
\begin{figure}[t]
  \centering
   \includegraphics[width=1.0\linewidth]{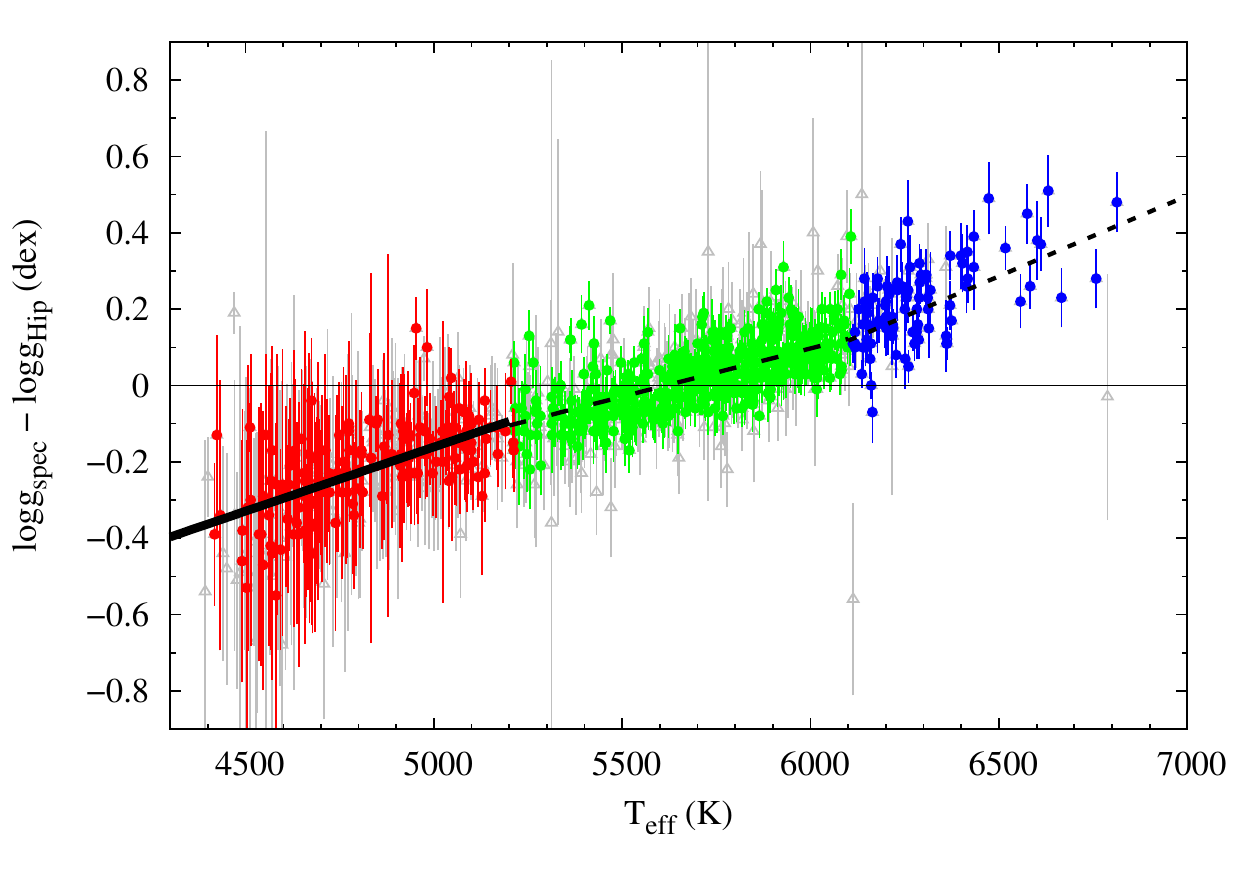} 
   \includegraphics[width=1.0\linewidth]{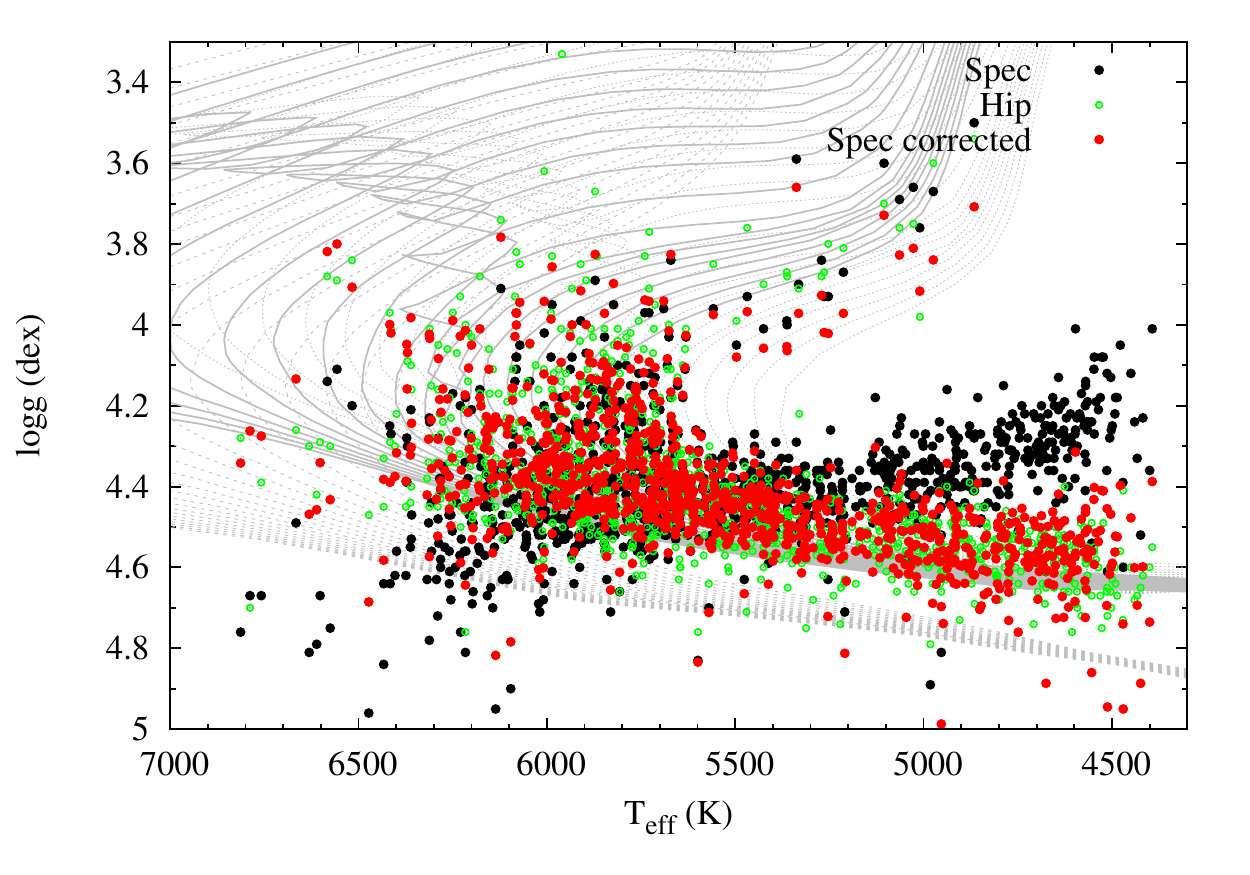} 
   \caption{Upper panel: Comparison between the \logg\ from spectroscopy and the one derived from \textit{Hipparcos} parallaxes as a function of \teff. The different lines represent linear fits in three different \teff\ bins. The stars with error in parallax higher than 5\% (not considered for the fit) are represented with grey triangles. Lower panel: HR diagram with our spectroscopic \logg, our corrected \logg\ values, and the \logg\ using the Hipparcos parallaxes.}
\label{gravities}
\end{figure}
\begin{center}
\begin{table*}
\caption{Sample table of the derived stellar parameters for each star including the spectroscopic, trigonometric and corrected \logg. The full table is available online.}
\centering
\label{table_param}
\begin{tabular}{lcccccccccc}
\hline
\noalign{\medskip} 
Star & T$_\mathrm{eff}$ & $\log g_{spec}$ & $\log g_{HIP}$ & [Fe/H] & $\xi_t$ & $\log g_{cor}$\\  
     & (K)  &  (cm\,s$^{-2}$) &(cm\,s$^{-2}$) & & (km\,s$^{-1}$) &  (cm\,s$^{-2}$)  \\
\noalign{\medskip} 
\hline
\hline
\noalign{\smallskip} 
HD144411   &  4839  $\pm$  72.0  &  4.45  $\pm$  0.14  &  4.60  $\pm$  0.08 &  -0.32  $\pm$  0.02  &  0.23 $\pm$   0.38  &  4.67  \\
HIP33392   &  4843  $\pm$  70.4  &  4.39  $\pm$  0.17  &  4.56  $\pm$  0.08 &  -0.05  $\pm$  0.04  &  0.49 $\pm$   0.19  &  4.60   \\
HD154577   &  4847  $\pm$  35.0  &  4.48  $\pm$  0.07  &  4.58  $\pm$  0.02 &  -0.73  $\pm$  0.01  &  0.31 $\pm$   0.19  &  4.69 \\
HIP32812   &  4849  $\pm$  72.4  &  4.27  $\pm$  0.18  &  4.55  $\pm$  0.09 &  -0.01  $\pm$  0.06  &  0.20 $\pm$   0.56  &  4.48 \\
HD2025     &  4851  $\pm$  49.0  &  4.49  $\pm$  0.13  &  4.58  $\pm$  0.03 &  -0.37  $\pm$  0.02  &  0.51 $\pm$   0.18  &  4.70  \\
... & ... & ... & ... & ... & ... & ... \\
\noalign{\smallskip}
\hline
\hline
\noalign{\medskip} 
\end{tabular}
\end{table*}
\end{center}

The bulk composition of \textit{s}-process elements is accounted for the main-\textit{s} component, which produces elements with 90\,$\lesssim$\,$A$\,$\lesssim$\,204\footnote{\textit{A} is the atomic mass number} in the He-intershell of asymptotic giant branch (AGB) stars (1.3\,M$_{\odot}$\,$\lesssim$\,M\,$\lesssim$\,8\,M$_{\odot}$) \citep{arlandini99,busso99}. The ejection of AGB envelopes, leaving the CO core as a white dwarf, produces the enrichment of the interstellar medium (ISM) with the elements previously processed. On the other hand, elements with 60\,$\lesssim$\,$A$\,$\lesssim$\,90 are thought to be produced by the weak-\textit{s} component, during He-core and C-shell burning in massive stars with M\,$\gtrsim$\,8\,M$_{\odot}$. Part of the material processed inside these massive stars is not altered by the explosive nucleosynthesis of SNe II and is ejected to enrich the ISM \cite[e.g.][]{kappeler89,prantzos90,raiteri93}. A third component, the strong-\textit{s}, provides the stable isotopes at the termination point of the \textit{s}-path, $^{208}$Pb and $^{209}$Bi \citep[e.g.][]{bisterzo16}. Finally, the production sites for \textit{r}-process elements are still debated but they are presumably associated to explosive conditions in supernovae due to the extreme neutron density needed \citep[e.g.][and refererences therein]{cowan04,sneden08}.\\

As we have seen, the \textit{s}-process produces the bulk composition of most of the elements analyzed in this work. However, the \textit{r}-process contribution is not negligible in several cases. For example, \textit{s}-process contribution at the time of the Solar System formation for the elements of the first peak is 85\% for Sr, 92\% for Y and 83\% for Zr. However, the heavy-\textit{s} have a smaller contribution from \textit{s}-process: 81\% for Ba, 77\% for Ce and 56\% for Nd. Finally, Eu is considered a pure-\textit{r}-process element since its \textit{s}-process contribution is only 7\% \citep{arlandini99}. A recent study by \cite{bisterzo16} gives updated percentages of \textit{s}-process contribution for Sr (67\%), Y (70\%), Zr (64\%), Ba (83\%) and Ce (81\%). \\

By studying different populations of stars at different metallicities we can understand which processes played a major role in the production of such elements at a given moment of the evolution of the Galaxy, providing constrains for the current models of GCE. This is the main objective of this paper which is organized as follows: Sect. 2 briefly describes the collected data together with the determination of stellar parameters. In Sect. 3 we detail the derivation of abundances and the error treatment. In Sect. 4 we discuss the behaviour of different abundance ratios for the population of thin disk, thick disk and \textit{h$\alpha$mr} stars. Finally, we present our conclusions in Sect. 5.\\

\begin{center}
\begin{table}
\caption{Atomic parameters for the lines used in this work together with EWs and absolute abundances in the Sun from our Vesta combined spectrum.}
\label{lineas}
\centering
\begin{tabular}{lccrcc}
\hline
\noalign{\medskip} 
Element & $\lambda$ ($\AA{}$) & $\chi_{l}$ (eV) & log \textit{gf} & EW(m$\AA{}$)  & log (A) \\
\noalign{\medskip} 
\hline
\hline
\noalign{\smallskip} 
\ion{Cu}{I}\tablefootmark{*} & 5105.55 &  1.39 &   --1.516 &  93.0 & 4.245  \\
\ion{Cu}{I}\tablefootmark{*} & 5218.21 &  3.82 &    0.476 &  53.5 & 4.088  \\
\ion{Cu}{I}\tablefootmark{*} & 5220.09 &  3.82 &   --0.448 &  16.2 & 4.088  \\
\ion{Cu}{I}\tablefootmark{*} & 5782.12 &  1.64 &   --1.720 &  79.9 & 4.093  \\
\noalign{\smallskip} 
\hline
\noalign{\smallskip} 
\ion{Zn}{I} & 4722.16 &  4.03 &   --0.338 &  67.1 & 4.553  \\
\ion{Zn}{I} & 4810.54 &  4.08 &   --0.137 &  74.0 & 4.530  \\
\ion{Zn}{I} & 6362.35 &  5.80 &    0.150 &  20.4 & 4.506  \\
\noalign{\smallskip} 
\hline
\noalign{\smallskip} 
\ion{Sr}{I} & 4607.34 &  0.00 &    0.283 &  47.2 &  2.780 \\
\noalign{\smallskip} 
\hline
\noalign{\smallskip} 
\ion{Y}{II}  & 4374.94 &  0.41 &    0.160 &  86.2 & 2.318  \\
\ion{Y}{II}  & 4398.01 &  0.13 &   --1.000 &  46.5 & 2.097  \\
\ion{Y}{II}\tablefootmark{a}  & 4854.87 &  0.99 &   --0.380 &  48.8 & 2.326  \\
\ion{Y}{II}  & 4900.12 &  1.03 &   --0.090 &  55.7 & 2.258  \\
\ion{Y}{II}  & 5087.43 &  1.08 &   --0.170 &  48.5 & 2.178  \\
\ion{Y}{II}  & 5200.42 &  0.99 &   --0.570 &  37.5 & 2.189  \\
\ion{Y}{II}  & 5402.78 &  1.84 &   --0.630 &  11.7 & 2.273  \\
\noalign{\smallskip} 
\hline
\noalign{\smallskip} 
\ion{Zr}{I}  & 4805.87 &  0.69 &  --0.420  &   1.7 & 2.596 \\
\ion{Zr}{I}  & 4815.63 &  0.60 &  --0.030  &   3.0 & 2.378 \\
\ion{Zr}{I}  & 6127.44 &  0.15 &  --1.060  &   2.7 & 2.818 \\
\ion{Zr}{I}  & 6134.55 &  0.00 &  --1.280  &   2.8 & 2.899 \\
\ion{Zr}{I}  & 6143.20 &  0.07 &  --1.100  &   3.0 & 2.820 \\
\noalign{\smallskip} 
\hline
\noalign{\smallskip} 
\ion{Zr}{II} & 4050.32 &  0.71 &   --1.060 &  23.3 & 2.604  \\
\ion{Zr}{II} & 4208.98 &  0.71 &   --0.510 &  43.7 & 2.622  \\
\ion{Zr}{II} & 4379.74 &  1.53 &   --0.356 &  28.7 & 2.809  \\
\ion{Zr}{II} & 5112.28 &  1.67 &   --0.850 &   9.3 & 2.697  \\
\noalign{\smallskip} 
\hline
\noalign{\smallskip} 
\ion{Ba}{II}\tablefootmark{*} & 5853.69 &  0.60 &   --1.010 &  64.0 & 2.298  \\
\ion{Ba}{II}\tablefootmark{*} & 6141.73 &  0.70 &   --0.070 & 112.7 & 2.256  \\
\ion{Ba}{II}\tablefootmark{*} & 6496.91 &  0.60 &   --0.377 &  97.4 & 2.210  \\
\noalign{\smallskip} 
\hline
\noalign{\smallskip} 
\ion{Ce}{II} & 4523.08 &  0.52 &    0.040 &  15.3 & 1.630  \\
\ion{Ce}{II} & 4628.16 &  0.52 &    0.230 &  20.4 & 1.609  \\
\ion{Ce}{II} & 4773.96 &  0.92 &    0.250 &  10.8 & 1.603  \\
\ion{Ce}{II} & 5274.23 &  1.04 &    0.130 &   8.7 & 1.696  \\
\noalign{\smallskip} 
\hline
\noalign{\smallskip} 
\ion{Nd}{II}\tablefootmark{*} & 4811.34 &  0.06 &  --1.140  &  10.3 & 1.928 \\
\ion{Nd}{II}\tablefootmark{*} & 4989.95 &  0.63 &  --0.500  &   7.9 & 1.700 \\
\ion{Nd}{II}\tablefootmark{*} & 5092.80 &  0.38 &  --0.610  &   8.0 & 1.566 \\
\ion{Nd}{II}\tablefootmark{*} & 5130.59 &  1.30 &   0.450  &  14.6 & 1.739 \\
\noalign{\smallskip} 
\hline
\noalign{\smallskip} 
\ion{Eu}{II}\tablefootmark{*} & 6645.13 &  1.38 &  --0.200  &   5.8 & 0.670 \\
\noalign{\smallskip} 
\hline
\hline
\noalign{\smallskip} 
\end{tabular}
\tablefoottext{*}{Lines for which HFS is considered.}\\
\tablefoottext{a}{Discarded line.}\\
\end{table}
\end{center}

\section{Observations and stellar parameters}

The baseline sample used in this work is formed by 1111 FGK stars observed within the context of the HARPS GTO programs. It is a combination of three HARPS sub-samples hereafter called HARPS-1 \citep{mayor03}, HARPS-2 \citep{locurto} and HARPS-4 \citep{santos_harps4}. The individual spectra of each star were reduced using the HARPS pipeline and then combined with IRAF\footnote{IRAF is distributed by National Optical Astronomy Observatories, operated by the Association of Universities for Research in Astronomy, Inc., under contract with the National Science Foundation, USA.} after correcting for its radial velocity shift. The final spectra have a resolution of R $\sim$115000 and high signal-to-noise ratio (55\% of the spectra have S/N higher than 200). The total sample is composed by 136 stars with planets and 975 stars without detected planets. Chemical abundances of these samples for refractory elements with A $<$ 29 can be found in \citet{adibekyan12} together with oxygen \citep{bertrandelis15}, carbon \citep{suarez-andres17}, lithium \citep{delgado14,delgado15} and nitrogen abundances \citep[][only for a small fraction of stars]{suarez-andres16}.\\

\subsection{Stellar parameters}

The stellar parameters, namely the effective temperature (\teff), surface gravity (\logg), metallicity ([Fe/H]), and 
microturbulence ($\xi_{t}$), were taken from \citet{sousa08,sousa_harps4,sousa_harps2}. All atmospheric parameters were determined in a homogeneous way based on the measurements of the equivalent widths (EW) of \ion{Fe}{I} and \ion{Fe}{II} lines, and on iron 
excitation and ionization equilibrium. The effective temperatures of cool stars derived with the linelist of \cite{sousa08} were overestimated compared to the infrared flux method. Therefore, in \citet{tsantaki13} we compile a reduced linelist specially selected to eliminate lines that suffer from blending effects which are strongly present for the cooler stars and have a significant effect mainly on the determination of temperature. We used this linelist in the complete sample to re-derive the parameters of stars cooler than 5200\,K since only the stars in the subsample HARPS--1 were corrected in \citet{tsantaki13}. In Fig.~\ref{comparison} we present the comparison between the previous parameters and the parameters of this work for the 297 stars with \teff\,$<$\,5200K. The new \teff\ are corrected to lower values, with an average difference of --95\,$\pm$\,56\,K, whereas \logg\ is not affected sigificantly (the mean difference is -0.09\,$\pm$\,0.09\,dex) and even less for [Fe/H] (the average difference is 0.005\,$\pm$\,0.03\,dex). The remaining stars (above 5200\,K) kept their initial parameters. For some of these cool stars the derivation of parameters with the new linelist did not converge and thus the final sample is composed of 1059 stars. The estimation of errors in parameters is done in the same way as in \citep{tsantaki13}.\\  

Additionally, we calculated the trigonometric \logg\ (see Eq.\,1 from \citealt{santos04}) using the new \textit{Hipparcos} 
parallaxes \citep{hip}, $V$ magnitudes from Simbad, bolometric corrections based on \cite{flower96} and \cite{torres10}, 
solar magnitudes from \citep{bessell98}, stellar masses and spectroscopic \teff. Stellar masses are derived from the PARAM v1.3 tool\footnote{http://stev.oapd.inaf.it/cgi-bin/param} using the PARSEC theoretical isochrones from \citep{bressan12} and a Bayesian estimation method \citep{dasilva06_param}. The stellar masses are obtained using the observational information (V mag teff, [Fe/H], and parallax) to compute the probability density functions of the main stellar properties (mass, luminosity, and age).
The Bayesian inference is applied taking into account priors for the initial mass function \cite{chabrier01} and a constant Star Formation Rate. No correction for interstellar reddening is needed since all stars are in close distance.\\

The correlation of $\log g_{spec}$ -- $\log g_{HIP}$ with $T_{\mathrm{eff}}$ has already been reported in some studies but no clear explanation has been proposed why ionisation balance is not properly handled \citep[e.g.,][]{tsantaki13, bensby14}. Interestingly, the differences between $\log g_{spec}$ and $\log g$ derived from other more model-independent methods, such as from the transit fit of planet hosts and from the asteroseismic analysis show a similar correlation with \teff\ \citep{mortier14}. Moreover, it has been shown that \logg\ derived from the ionisation balance does not follow isochrones in the Hertzsprung-Russell (HR) diagram, opposite to what happens to trigonometric \logg\ (see lower panel of Fig. \ref{gravities}). Therefore, we decided to derive a correction of our spectroscopic \logg\ to have more realistic values. The comparison between the trigonometric gravities and the spectroscopic ones is presented in the upper panel of Fig.~\ref{gravities}. We calculated the correction to the spectroscopic values using linear fits in three different \teff\ ranges corresponding to different spectral types and only for stars with parallax errors smaller than 5\% (67\% of our sample). A cut in the parallax error is necessary because biases in our distance estimations are introduced by simply inverting the parallax \citep[e.g.][]{Astraatmadja2016}. The average differences between the spectrocopic \logg\ and the trigonometric \logg\ are --0.22\,$\pm$\,0.13, 0.02\,$\pm$\,0.12 and 0.21\,$\pm$\,0.13\,dex for the cool, solar temperature and hot stars, respectively. \\

The corrections are presented in Eq.~\ref{logg_cor}--3 and are suggested to correct \logg values derived from this method. With the new parallax releases of Gaia mission, we will obtain very precise trigonometric gravities for millions of stars that will help us improve our spectroscopic gravities from calibrations such as in this work.
In the HR diagram of Fig.~\ref{gravities} we can see how the new \logg\ corrected values, follow better the isochrones.
Thus, for the derivation of abundances explained in next section we used the corrected \logg.The errors for the corrected \logg\ are the same (the difference is less than 0.01\,dex) as for the spectroscopic values. The complete table with the updated parameters is available in electronic format, a sample of our results is shown in Table \ref{table_param}. 

\begin{equation}
\label{logg_cor}
\log g_{cor} = \log g_{spec} - 3.364 \times10^{-4} \, T_{\mathrm{eff}} + 1.843 (T_{\mathrm{eff}}<5200\,K) \\
\end{equation}

\begin{equation}
\log g_{cor} = \log g_{spec} - 2.521 \times10^{-4} \, T_{\mathrm{eff}} + 1.416 (5200\,K \leq T_{\mathrm{eff}} \leq 6100\,K) \\
\end{equation}

\begin{equation}
\log g_{cor} = \log g_{spec} - 4.217 \times10^{-4} \, T_{\mathrm{eff}} + 2.455 (T_{\mathrm{eff}} > 6100\,K)\\
\end{equation}

\begin{figure}
\centering
\includegraphics[width=9.0cm]{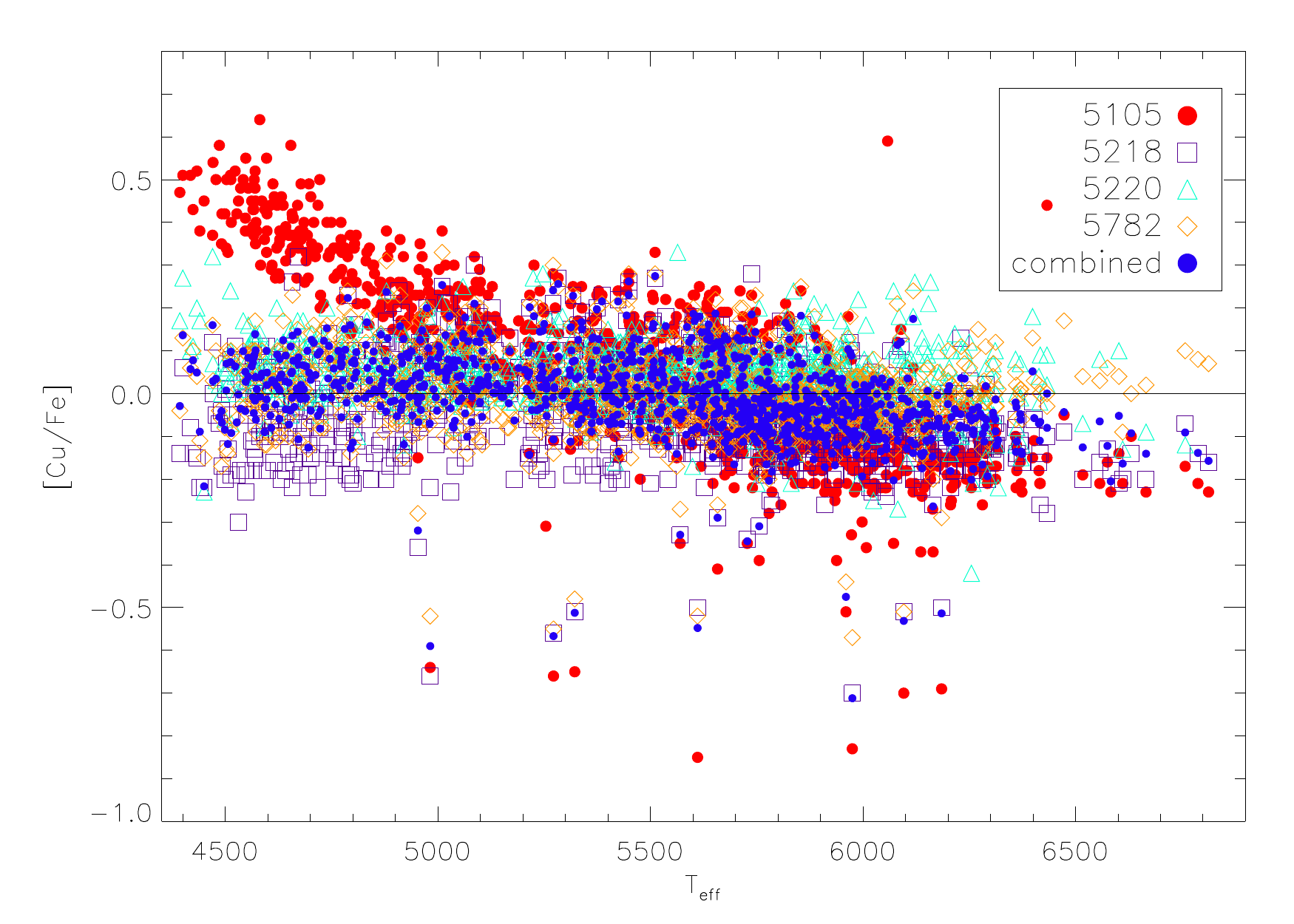}
\includegraphics[width=9.0cm]{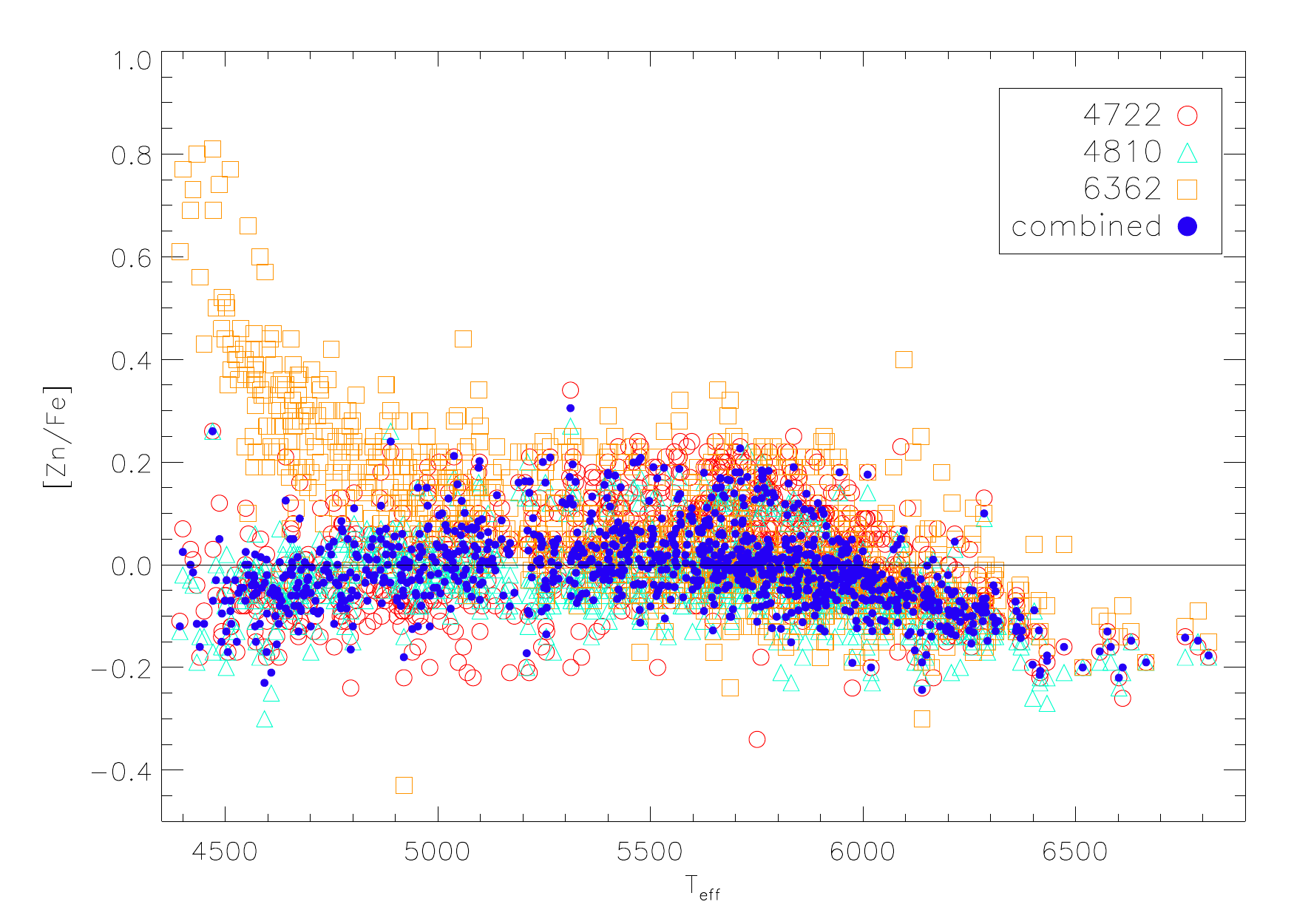}
\caption{[Cu/Fe] and [Zn/Fe] ratios for each line as a function of \teff\ together with the final adopted abundances for each star (blue filled circles).} 
\label{CuZnteff}
\end{figure}

\begin{figure}
\centering
\includegraphics[width=9.0cm]{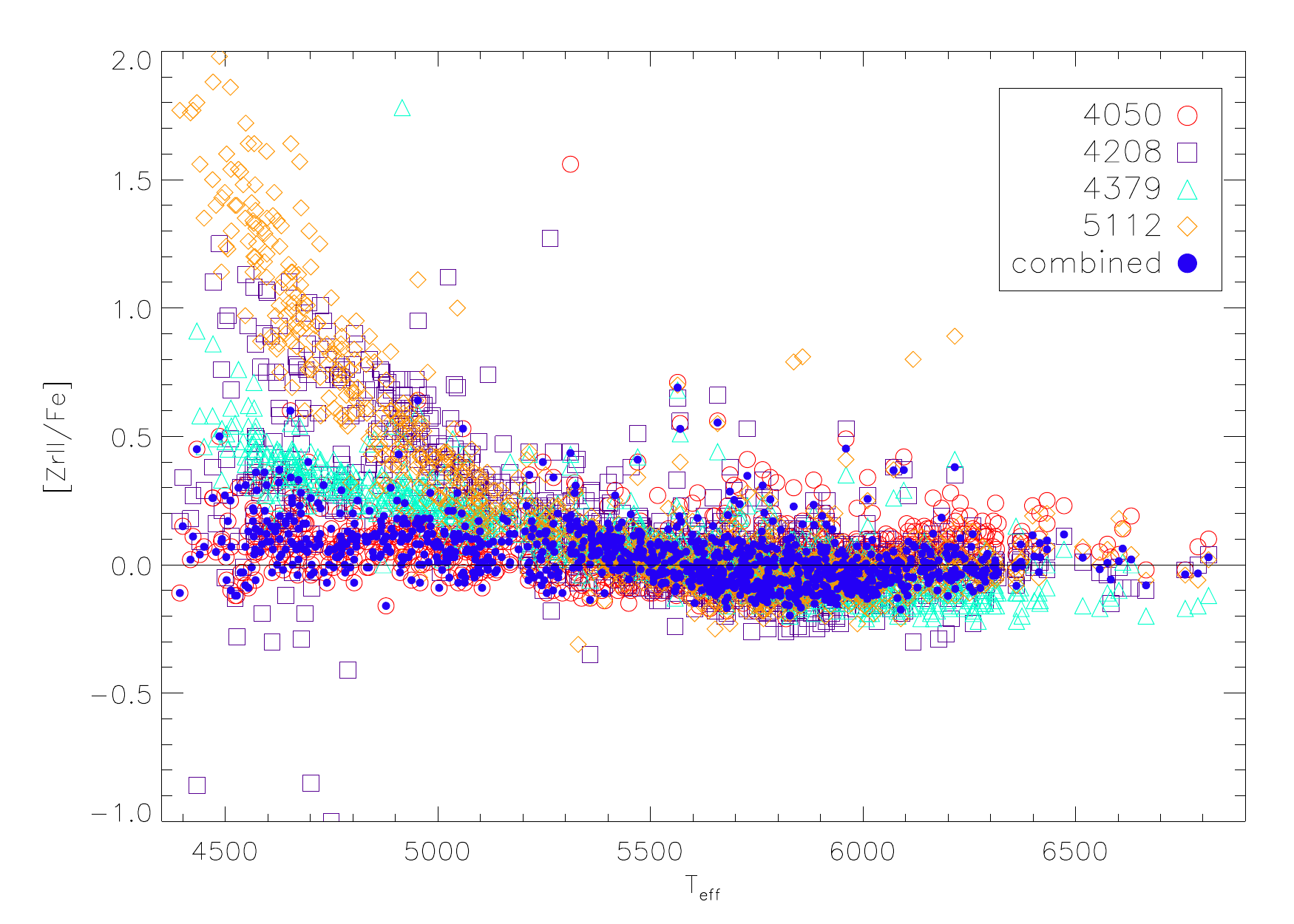}
\includegraphics[width=9.0cm]{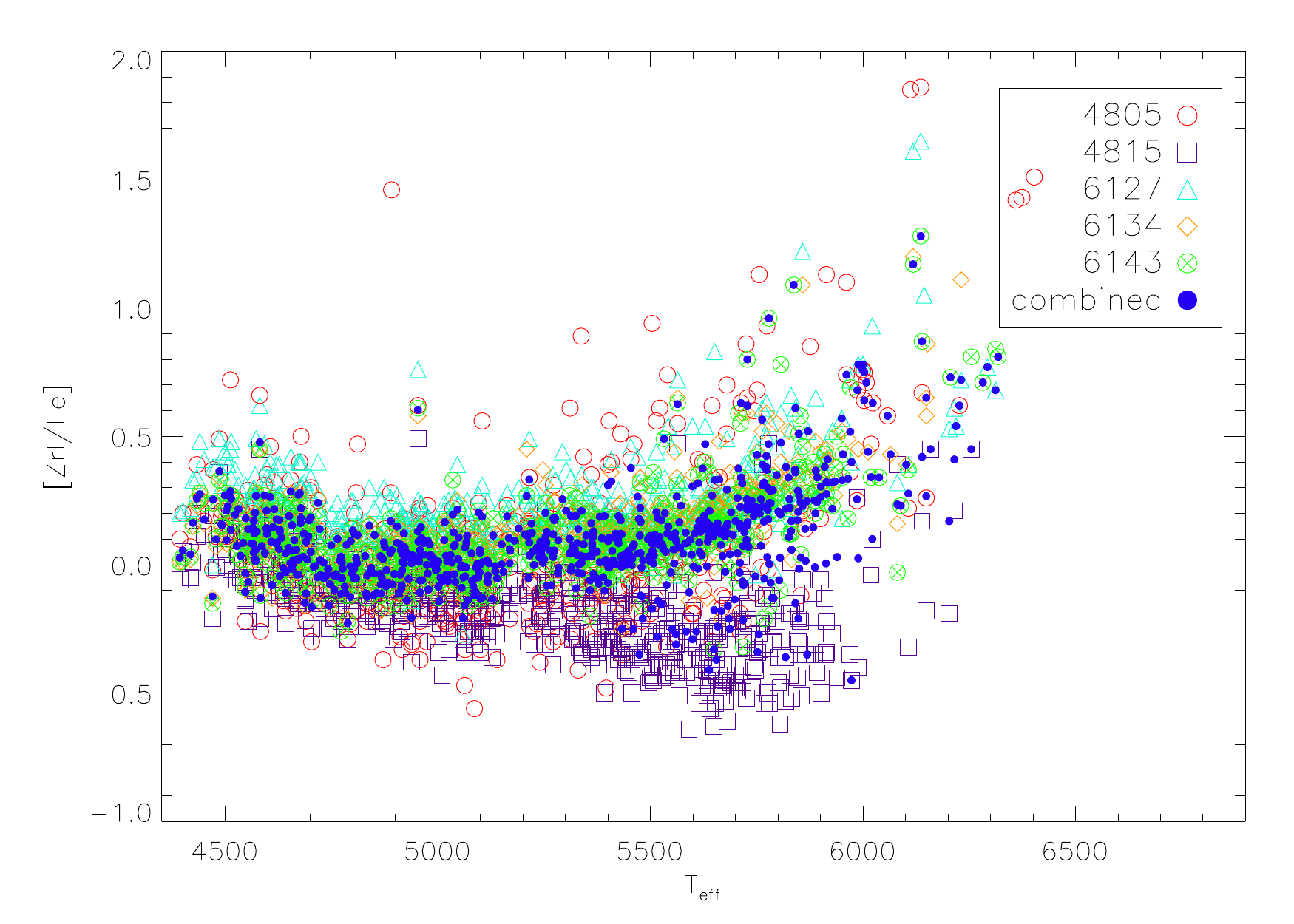}
\caption{[ZrII/Fe] and [ZrI/Fe] ratios for each line as a function of \teff\ together with the final abundance adopted for each star (blue filled circles).} 
\label{ZrIIteff}
\end{figure}

\section{Derivation of chemical abundances}
The chemical abundances for most of the elements were derived under a standard LTE analysis with 
the 2014 version of the code MOOG \citep{sneden} using the \textit{abfind} driver. For the lines affected by Hyperfine Splitting (HFS) we used the \textit{blends} driver. A grid of Kurucz ATLAS9 atmospheres \citep{kurucz} were used as input along with the equivalent widths (EWs) and the atomic parameters, wavelength ($\lambda$), excitation energy of the lower energy level ($\chi$), and oscillator strength (log \textit{gf}) of each line. The EWs of the different lines were measured automatically with the version 2 of the ARES program\footnote{The ARES code can be downloaded at http://www.astro.up.pt/~sousasag/ares/} \citep{sousa_ares2,sousa_ares}. The input parameters for ARES were adjusted for the different atomic lines used in this study. We measured the EWs of ten stars with different \teff\ and [Fe/H] manually, with the task \textit{splot} in IRAF, and compared them with the obtained values by ARES using different input parameters until we found the best approach for each line. We also measured the EWs in a solar reflected light spectrum of the Vesta asteroid (obtained by combining several high S/N spectra observed with HARPS) in order to derive our reference abundances. The atomic data, EWs and derived abundances for the Sun are shown in Table \ref{lineas}. If not specified in following subsections, the atomic parameters of the lines and the Van der Waals damping constants, log ($\gamma _{6}/N_{H}$), were retrieved from VALD3 database\footnote{http://vald.astro.univie.ac.at/$\sim$vald3/php/vald.php} \citep{vald15}. When no damping constants were available in VALD3, the Uns$\mathrm{\ddot{o}}$ld approximation with an enhancement factor was considered. This enhancement, E, was calculated as recommended by the Blackwell group (E\,=\,1\,+\,0.67\,$\chi_{l}$, damping option 2 within MOOG). We did not attempt to calibrate the \loggf\ values with standard solar abundances, therefore we derived the [X/Fe] ratios line by line with respect to the solar abundances shown in Table \ref{lineas} to later combine them as explained in next subsections. All the abundances are provided in electronic tables, a portion of our results is shown in Table \ref{table_abun}.\\

\subsection{Error determination}

In order to determine the uncertainties on the abundances we have to consider the errors on the EW measurements, the errors on the atomic parameters of the lines and the errors due to the uncertainties in the atmospheric parameters. We consider the line-to-line scatter as an approximation of the error on the measurement of EW (uncertainties in continuum position, blends, etc...) and the errors on atomic parameters. On the other hand the abundance uncertainties due to the errors on stellar parameters were estimated by calculating the abundance differences when one of each of the stellar parameters was modified by its individual error. The average abundance sensitivities for each element are shown in Table \ref{table_errors} for the same three groups of stars depending on \teff\ as done in \citet{adibekyan12}: ``low \teff'' stars -- stars with \teff\,$<$ 5277\,K, ``\textit{solar}'' -- stars with \teff\,=\,T$_{\odot}$\,$\pm$\,500\,K, and ``high Teff'' – stars with \teff\,$>$ 6277\,K. The average errors on \teff\ are 64, 24, and 46 \emph{K} for cool, Sun-like, and hot star groups, respectively. The average errors in $\log{g}$ are 0.17, 0.03, and 0.05\,dex, in \emph{$\xi{}_{\mathrm{t}}$} - 0.33, 0.04, and 0.08\,km\,s$^{-1}$, and in [Fe/H] - 0.04, 0.02, 0.03\,dex for the three groups, respectively. 
We can see that for solar \teff\ and hotter stars the dominant error is due to the line-to-line scatter. However, for cool stars the errors on the stellar parameters are larger and thus they translate into significant errors on abundances, especially those related with \logg\ and $\xi_{\mathrm{t}}$ uncertainties. The neutral species are more sensitive to errors in \teff\ meanwhile ionized species suffer a higher variation due to changes in \logg. Finally, the errors on microturbulence mainly affect the abundances of species with strong lines such as Sr, Y and Ba, which are more important for cool stars. The final abundance errors are given by the quadratic sum of all these individual errors.

\begin{center}
\begin{table*}
\caption{Sample table of the derived abundances of the elements, error, and number of measured lines for each star. The full table is available online.}
\label{table_abun}
\centering
\begin{tabular}{lcc|rcc|rcc|rcc|c}
\hline
\noalign{\medskip} 
Star & S/N  & Classif. & [Cu/Fe] & $\sigma_{Cu}$ & n$_{Cu}$  & [Zn/Fe] & $\sigma_{Zn}$ & n$_{Zn}$  & [Sr/Fe] & $\sigma_{Sr}$ & n$_{Sr}$ & ... \\  
\noalign{\medskip} 
\hline
\hline
\noalign{\smallskip} 
    HD181720  & 695 & thick &  -0.039  & 0.056  &  4 &   0.131  & 0.023  &   3  &  -0.071  &  0.023  &   1 & ...  \\
    HD131218  &  61 & thin  &  -0.131  & 0.028  &  4 &  -0.101  & 0.020  &   3  &   0.054  &  0.146  &   1 & ...  \\
    HD111031  & 987 & thin  &   0.101  & 0.059  &  4 &   0.009  & 0.059  &   3  &  -0.044  &  0.027  &   1 & ...  \\
    HD183658  & 370 & thin  &   0.005  & 0.017  &  4 &  -0.003  & 0.034  &   3  &  -0.030  &  0.029  &   1 & ...  \\
    HD107148  & 110 & thin  &   0.102  & 0.029  &  4 &  -0.007  & 0.027  &   3  &  -0.014  &  0.092  &   1 & ...  \\
... & ... & ... & ... & ... & ... & ... ... & ... & ... & ... & ... & ... & ... \\
\noalign{\smallskip} 
\hline
\hline
\noalign{\medskip} 
\end{tabular}
\end{table*}
\end{center}

\subsection{Cu abundances}
Copper abundances are based on four neutral lines for which we consider HFS splitting. The atomic parameters and isotopic ratios for those lines were taken from Kurucz database \footnote{http://kurucz.harvard.edu/linelists/gfhyperall/}. The lines at 5105\AA{} and 5782\AA{} are the most affected by HFS. The former line becomes very strong at low temperatures, probably due to an unknown blend, thus the abundances are also very high as can be seen in Fig \ref{CuZnteff}. In the Sun we also obtain the highest Cu abundance with this line (see Table \ref{lineas}). Therefore, we decided to discard this line for stars cooler than 5200K based on visual inspection of the abundances. The line at 5218\AA{} gets blended for cool stars and we had to measure it manually in several stars because ARES was not able to deblend it in a correct way. Furthermore, we also inspected visually the measurement of Cu lines in our most metal poor stars and in some cases we had to measure with IRAF some weak lines that ARES could not fit. We discard the line at 5220\AA{} for metal-poor stars because it becomes very weak (EW $<$ 5 mA). To obtain the final abundances we calculated the weighted mean (WM) whereby we consider the distance from the median abundance as a weight. As described in \citet{adibekyan15c} this method is a good approach that can be used without removing suspected outlier lines.

\subsection{Zn abundances}
Zinc abundances are determined with three neutral lines (see Fig \ref{CuZnteff}). The line at 6362\AA{} is located in a depressed continuum region due to a \ion{Ca}{I} auto-ionization line \citep[e.g.][and references therein]{barbuy15} and for some stars ARES cannot measure it correctly, thus we remeasure this line when the abundance is very different to the other lines. Moreover, for cool stars this line becomes weaker and blended with a nearby feature thus, giving higher than average abundances. Therefore, we decided to discard this line for stars cooler than 5000K so the final abundance for these stars is the average of the abundance given by the other two lines. For hotter stars we derived the WM as final abundance. In order to check that the line at 6362\AA{} is not affecting our final results we compared our final abundances with the abundance obtained using only the first two lines and the results are very similar except for a few of the most metal-rich stars where the abundances can increase up to $\sim$\,0.1\,dex. Since this difference affects to a very small percentage of our sample we keep using it for \teff\,$>$\,5000\,K.

\begin{table*}
\centering
\caption{Average abundance sensitivities of the studied elements to changes of each parameter by their individual $\sigma$.}
\label{table_errors}
\begin{tabular}{lcccccccccc}
\hline
\noalign{\vskip0.02\columnwidth}
 & \ion{Cu}{I} & \ion{Zn}{I} & \ion{Sr}{I} & \ion{Y}{II} & \ion{Zr}{I} & \ion{Zr}{II} & \ion{Ba}{II} & \ion{Ce}{II} & \ion{Nd}{II} & \ion{Eu}{II}\tabularnewline[0.01\columnwidth]
\hline 
\hline
\noalign{\smallskip} 
\multicolumn{11}{c}{\emph{line-to-line scatter/continuum error}}\tabularnewline
\noalign{\smallskip} 
 low \emph{T$_\mathrm{eff}$} & $\pm$0.04 & $\pm$0.04 & $\pm$0.04 & $\pm$0.08 & $\pm$0.07 & $\pm$0.07 & $\pm$0.04 & $\pm$0.06 & $\pm$0.06 & $\pm$0.10\tabularnewline
                  \emph{solar} & $\pm$0.04 & $\pm$0.03 & $\pm$0.05 & $\pm$0.04 &     --    & $\pm$0.04 & $\pm$0.02 & $\pm$0.05 & $\pm$0.04 & $\pm$0.09\tabularnewline
high \emph{T$_\mathrm{eff}$} & $\pm$0.07 & $\pm$0.03 & $\pm$0.06 & $\pm$0.06 &     --    & $\pm$0.07 & $\pm$0.03 & $\pm$0.05 & $\pm$0.06 & $\pm$0.12\tabularnewline
\noalign{\smallskip} 
\hline 
\noalign{\smallskip}
\multicolumn{11}{c}{\emph{$\Delta$T$_\mathrm{eff}$} $=\pm$ $\sigma_{T_\mathrm{eff}}$}\tabularnewline
\noalign{\smallskip} 
 low \emph{T$_\mathrm{eff}$} & $\pm$0.01 & $\mp$0.02 & $\pm$0.10 & $\pm$0.01 & $\pm$0.10 & $\pm$0.01 & $\pm$0.02 & $\pm$0.02 & $\pm$0.02 & $\mp0.00$\tabularnewline
                  \emph{solar} & $\pm$0.02 & $\pm$0.00 & $\pm$0.03 & $\pm$0.00 &      --   & $\pm$0.00 & $\pm$0.01 & $\pm$0.01 & $\pm$0.01 & $\pm0.00$\tabularnewline
high \emph{T$_\mathrm{eff}$} & $\pm$0.03 & $\pm$0.02 & $\pm$0.04 & $\pm$0.02 &      --   & $\pm$0.01 & $\pm$0.02 & $\pm$0.02 & $\pm$0.02 & $\pm0.01$\tabularnewline
\noalign{\smallskip} 
\hline
\noalign{\smallskip} 
\multicolumn{11}{c}{\emph{$\Delta$}{[}Fe/H{]}\emph{ }$=\pm$ $\sigma_{[Fe/H]}$}\tabularnewline
\noalign{\smallskip} 
 low \emph{T$_\mathrm{eff}$} & $\pm$0.01 & $\pm$0.02 & $\pm$0.01 & $\pm$0.02 & $\pm$0.00 & $\pm$0.02 & $\pm$0.02 & $\pm$0.02 & $\pm$0.01 & $\pm$0.01\tabularnewline
                  \emph{solar} & $\pm$0.00 & $\pm$0.00 & $\pm$0.00 & $\pm$0.01 &      --   & $\pm$0.00 & $\pm$0.01 & $\pm$0.01 & $\pm$0.01 & $\pm$0.00\tabularnewline
high \emph{T$_\mathrm{eff}$} & $\pm$0.00 & $\pm$0.00 & $\pm$0.00 & $\pm$0.01 &      --   & $\pm$0.00 & $\pm$0.01 & $\pm$0.01 & $\pm$0.01 & $\pm$0.00\tabularnewline
\noalign{\smallskip} 
\hline
\noalign{\smallskip} 
\multicolumn{11}{c}{\emph{$\Delta$}$\log{g}$\emph{ }$=\pm$ $\sigma_{\log{g}}$}\tabularnewline
\noalign{\smallskip} 
 low \emph{T$_\mathrm{eff}$} & $\pm$0.03 & $\pm$0.01 & $\mp$0.08 & $\pm$0.04 & $\pm$0.00 & $\pm$0.07 & $\pm$0.01 & $\pm$0.07 & $\pm$0.04 & $\pm$0.07\tabularnewline
                  \emph{solar} & $\pm$0.00 & $\pm$0.00 & $\pm$0.00 & $\pm$0.01 &      --   & $\pm$0.01 & $\pm$0.00 & $\pm$0.01 & $\pm$0.01 & $\pm$0.01\tabularnewline
high \emph{T$_\mathrm{eff}$} & $\pm$0.00 & $\pm$0.00 & $\pm$0.00 & $\pm$0.02 &      --   & $\pm$0.02 & $\pm$0.01 & $\pm$0.02 & $\pm$0.02 & $\pm$0.02\tabularnewline
\noalign{\smallskip} 
\hline
\noalign{\smallskip} 
\multicolumn{11}{c}{\emph{$\Delta$}$\xi_{\mathrm{t}}$ $=\pm$ $\sigma_{\xi_{\mathrm{t}}}$}\tabularnewline                   
\noalign{\smallskip} 
 low \emph{T$_\mathrm{eff}$} & $\mp$0.03 & $\mp$0.04 & $\mp$0.08 & $\mp$0.08 & $\mp$0.04 & $\mp$0.04 & $\mp$0.07 & $\mp$0.03 & $\mp$0.01 & $\mp$0.00\tabularnewline
                  \emph{solar} & $\mp$0.01 & $\mp$0.01 & $\mp$0.01 & $\mp$0.01 &      --   & $\mp$0.01 & $\mp$0.02 & $\mp$0.00 & $\mp$0.00 & $\mp$0.00\tabularnewline
high \emph{T$_\mathrm{eff}$} & $\mp$0.00 & $\mp$0.02 & $\mp$0.00 & $\mp$0.02 &      --   & $\mp$0.01 & $\mp$0.03 & $\mp$0.00 & $\mp$0.00 & $\mp$0.00\tabularnewline
\noalign{\smallskip} 
\hline
\noalign{\smallskip} 
\multicolumn{11}{c}{\emph{total error}}\tabularnewline                   
\noalign{\smallskip} 
 low \emph{T$_\mathrm{eff}$} & $\pm$0.07 & $\pm$0.07 & $\pm$0.16 & $\pm$0.12 & $\pm$0.14 & $\pm$0.12 & $\pm$0.08 & $\pm$0.11 & $\pm$0.08 & $\pm$0.12\tabularnewline
                  \emph{solar} & $\pm$0.05 & $\pm$0.04 & $\pm$0.06 & $\pm$0.05 &      --   & $\pm$0.04 & $\pm$0.03 & $\pm$0.09 & $\pm$0.05 & $\pm$0.09\tabularnewline
high \emph{T$_\mathrm{eff}$} & $\pm$0.08 & $\pm$0.04 & $\pm$0.08 & $\pm$0.07 &      --   & $\pm$0.07 & $\pm$0.06 & $\pm$0.06 & $\pm$0.07 & $\pm$0.12\tabularnewline
\noalign{\smallskip} 
\hline
\hline
\end{tabular}
\end{table*}

\begin{figure}
\centering
\includegraphics[width=9.0cm]{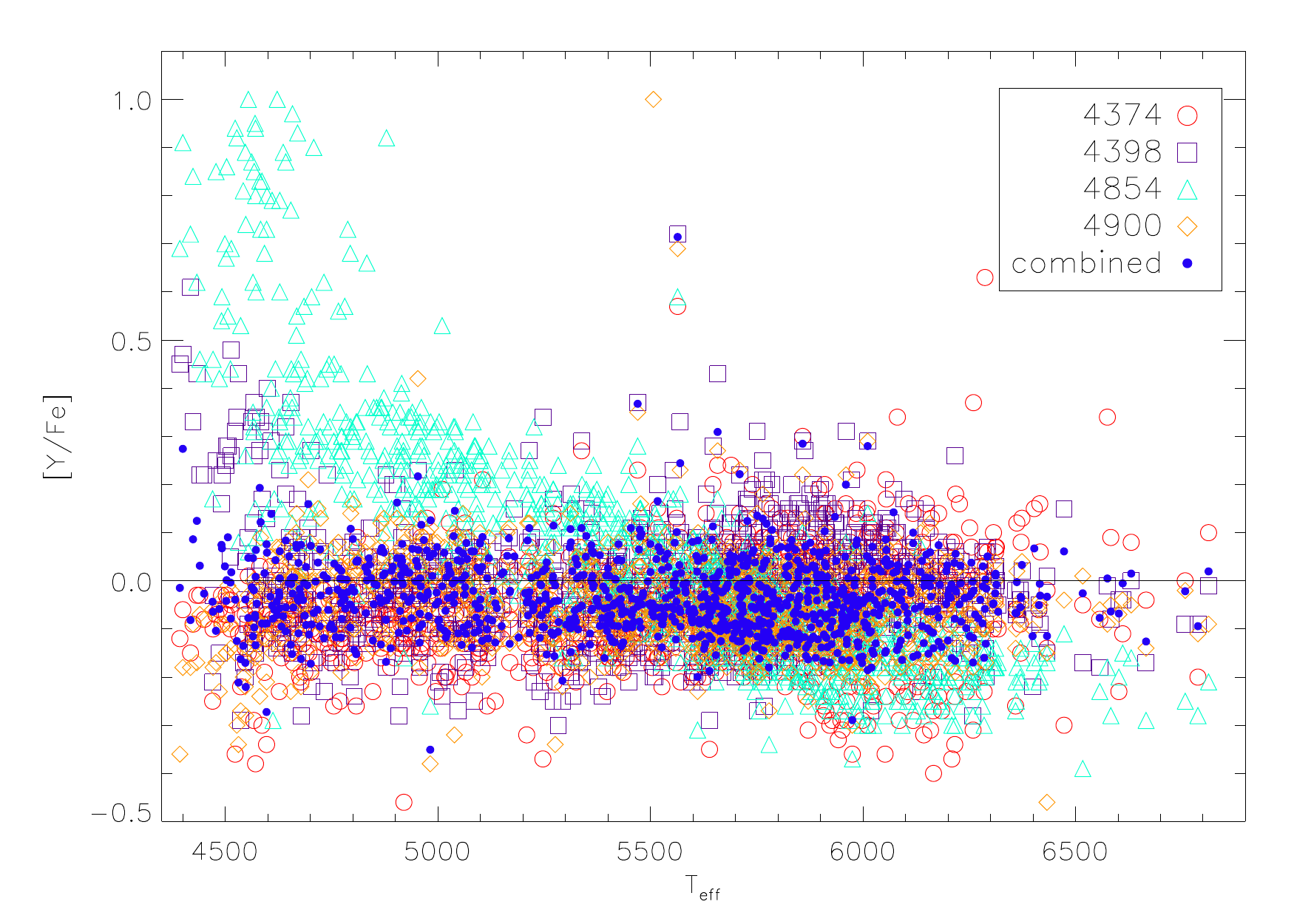}
\includegraphics[width=9.0cm]{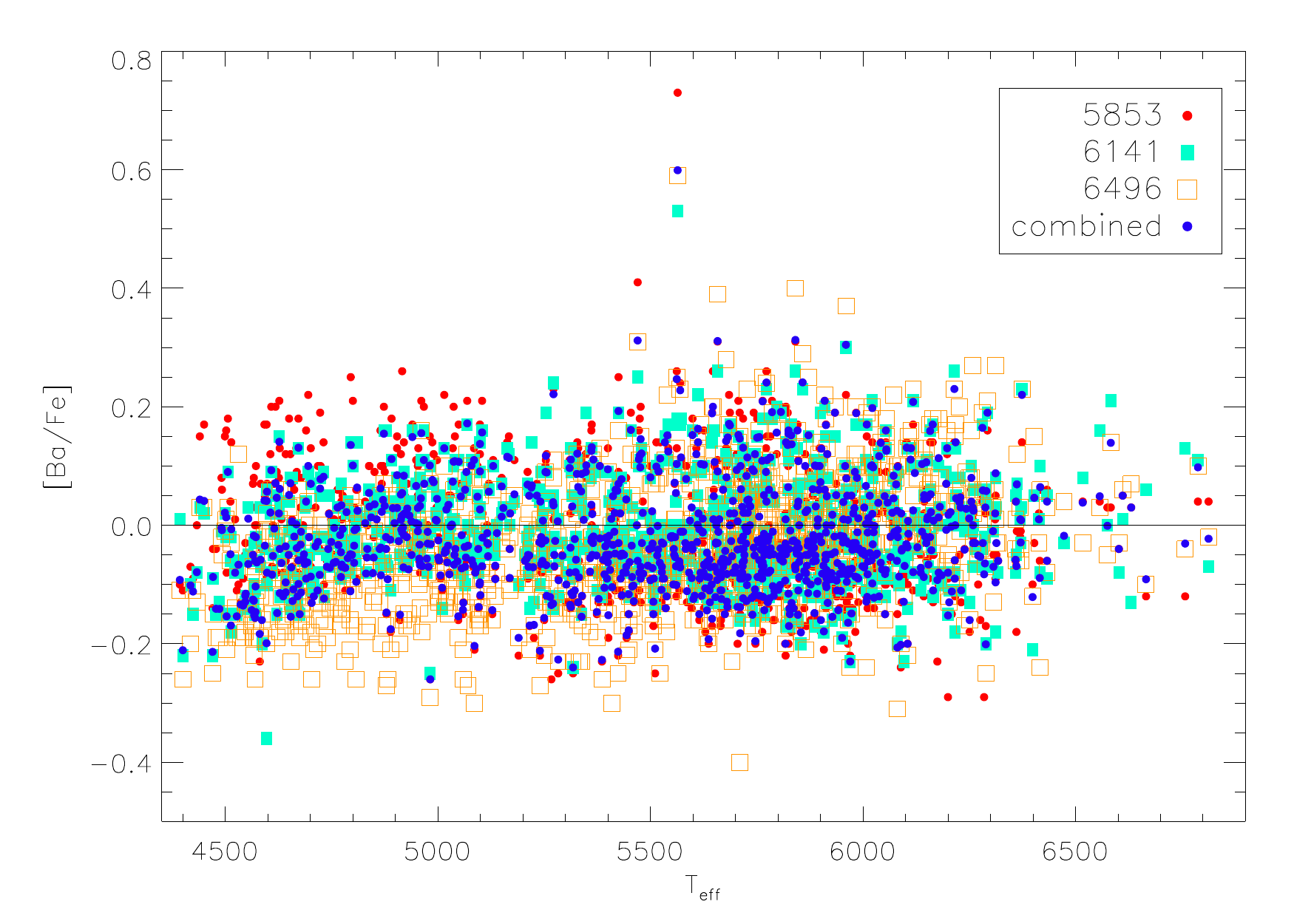}
\caption{Upper panel: [Y/Fe] ratios for three of the lines we use, together with the discarded line (blue triangles) and the final combined abundance from the six chosen lines. Lower panel: [Ba/Fe] ratios for each line as a function of \teff\ together with the final adopted values for each star (blue filled circles).} 
\label{YBateff}
\end{figure}

\subsection{Sr abundances}
To derive Sr abundances we only used one strong neutral line at 4607\AA{}. Thus, we cannot estimate the error due to the uncertainty in the continuum placement using the line-to-line scatter. For this case we calculated the errors on EWs following \citet{cayrel88} by using the FWHM of the lines provided by ARES. The calculated uncertainty takes into account the statistical photometric error due to the noise in each pixel and the error related to the continuum placement, which is the dominant contribution to the error \citep{cayrel88,bertrandelis15}. Then, these errors are propagated to derive the abundance uncertainties for each line. We show the [Sr/Fe] ratios as a function of \teff\ in Fig. \ref{SrEuteff}. 

\subsection{Y abundances}
We analyzed the Y abundances for seven \ion{Y}{II} lines. We found that the line at 4854\AA{} shows a strong trend of growing abundances for lower \teff\ and decreasing abundances for hotter stars, thus we discarded that line. The final abundances are derived as the WM of the six remaining lines. In Fig. \ref{YBateff} we show the discarded line together with only three out of the six final lines for clarity.  

\subsection{Zr abundances}
Zr abundances are based on four \ion{Zr}{II} lines. The lines at 4208\AA{} and 5112\AA{} become very blended for stars cooler than 5300\,K and ARES cannot separate them, therefore we used only these lines for hotter stars (see Fig. \ref{ZrIIteff}). The line at 4379\AA{} also seems to give overestimated abundances for cooler stars. Therefore, for stars cooler than 5300\,K we only considered the line at 4050\AA{} because it does not show a strong trend with \teff\ as happens for the other three lines. For hotter stars the final abundances are calculated using the WM of the four lines. In order to improve our abundances for cool stars we searched for reliable \ion{Zr}{I} lines since they are stronger for these stars. However, these lines have EWs smaller than 10 m\AA{} for stars above $\sim$5500K, thus \ion{Zr}{I} abundances for hotter stars must be considered with caution. Indeed, those lines are below 3 m\AA{} for the Sun and even with our high quality solar spectra the errors are not negligible. On the other hand, for cool stars the agreement between the absolute abundances among the \ion{Zr}{I} lines is much better than the line-by-line differential values with respect to the Sun because the solar absolute abundances of \ion{Zr}{I} show a great scatter (see Table \ref{lineas}). For cool stars, the absolute abundances of \ion{Zr}{I} lines are similar to the abundances of \ion{Zr}{II}---4050\AA{}. Therefore, to calculate [\ion{Zr}{I}/Fe] ratios we derived the WM of the absolute abundances and then subtracted the solar value for \ion{Zr}{II}---4050\AA{}, i.e. 2.60\,dex.

\begin{figure}
\centering
\includegraphics[width=9cm]{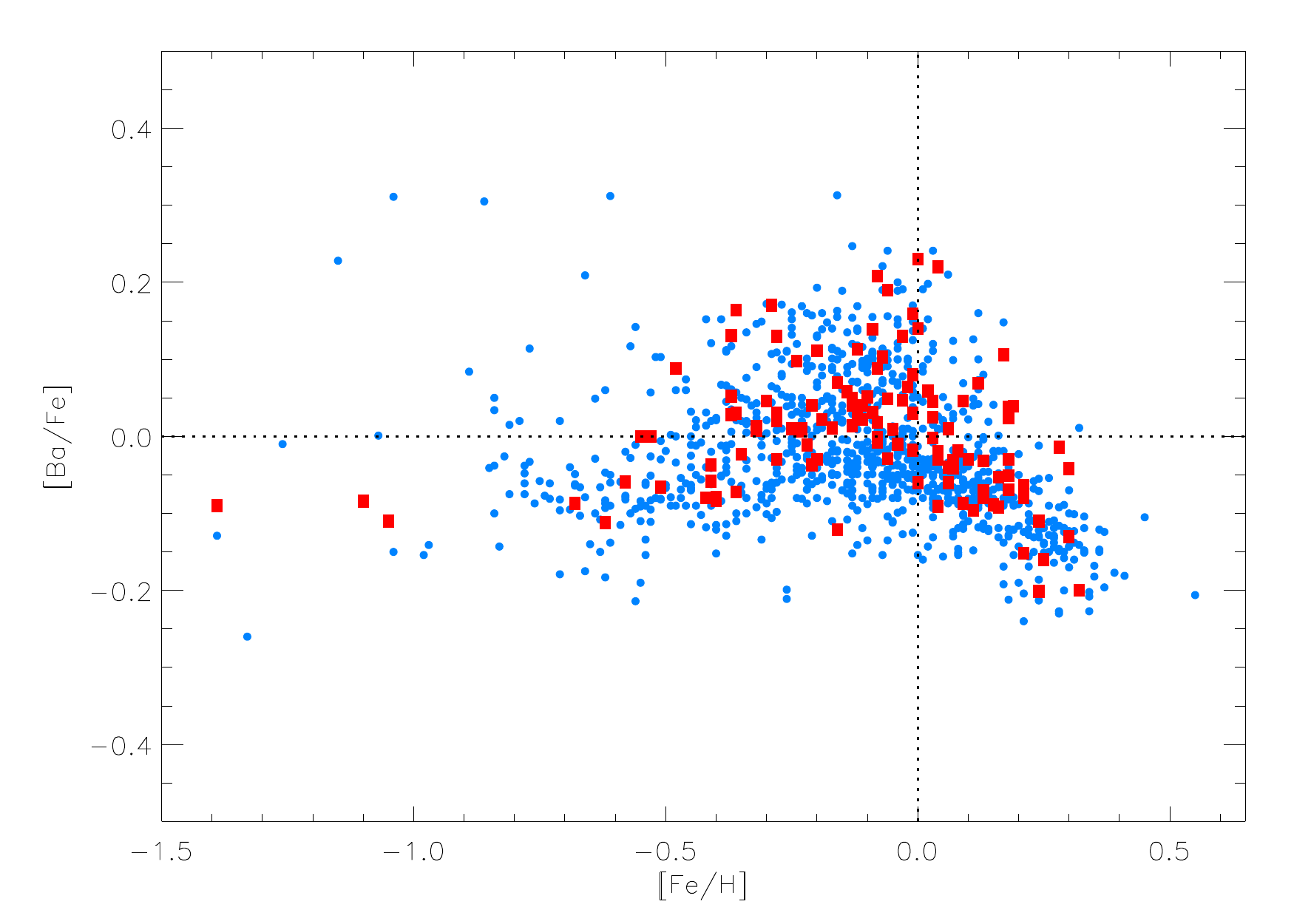}
\caption{Final [Ba/Fe] ratios as a function of [Fe/H] for the full sample. Stars hotter than 6100\,K are shown with red squares.} 
\label{Bateff}
\end{figure}

\subsection{Ba abundances}
Ba abundances are derived by measuring three strong \ion{Ba}{II} lines for which we consider HFS splitting. The atomic parameters and isotopic ratios for those lines were taken from \citet{prochaska00}. The three lines do not show any strong trend with \teff\ (see Fig. \ref{YBateff}) thus the final abundances are derived as the WM of them. Previous works have warned about the possible NLTE effects on Ba abundances of hot stars \citep[e.g.][]{bensby14}. In Fig. \ref{Bateff} we show that although Ba abundances for stars with \teff\,$>$\,6100\,K are higher on average around solar metallicity they are well mixed among the complete sample. Therefore, we decided not to remove any of these hot stars from our sample of Ba abundances.

\subsection{Ce abundances}
Ce abundances are based on four ionized lines. The atomic parameters for these lines were initially taken from VALD3 database, however the \loggf\ values provided there produced a high dispersion in abundances among the lines, thus we decided to take the calibrated \loggf\ values from \citet{reddy03}. The lines at 4523\AA{} and 4773\AA{} become very blended for stars cooler than 5300\,K, therefore we used only these lines for hotter stars (see Fig. \ref{CeNdteff}).

\subsection{Nd abundances}
\begin{figure}
\centering
\includegraphics[width=9.0cm]{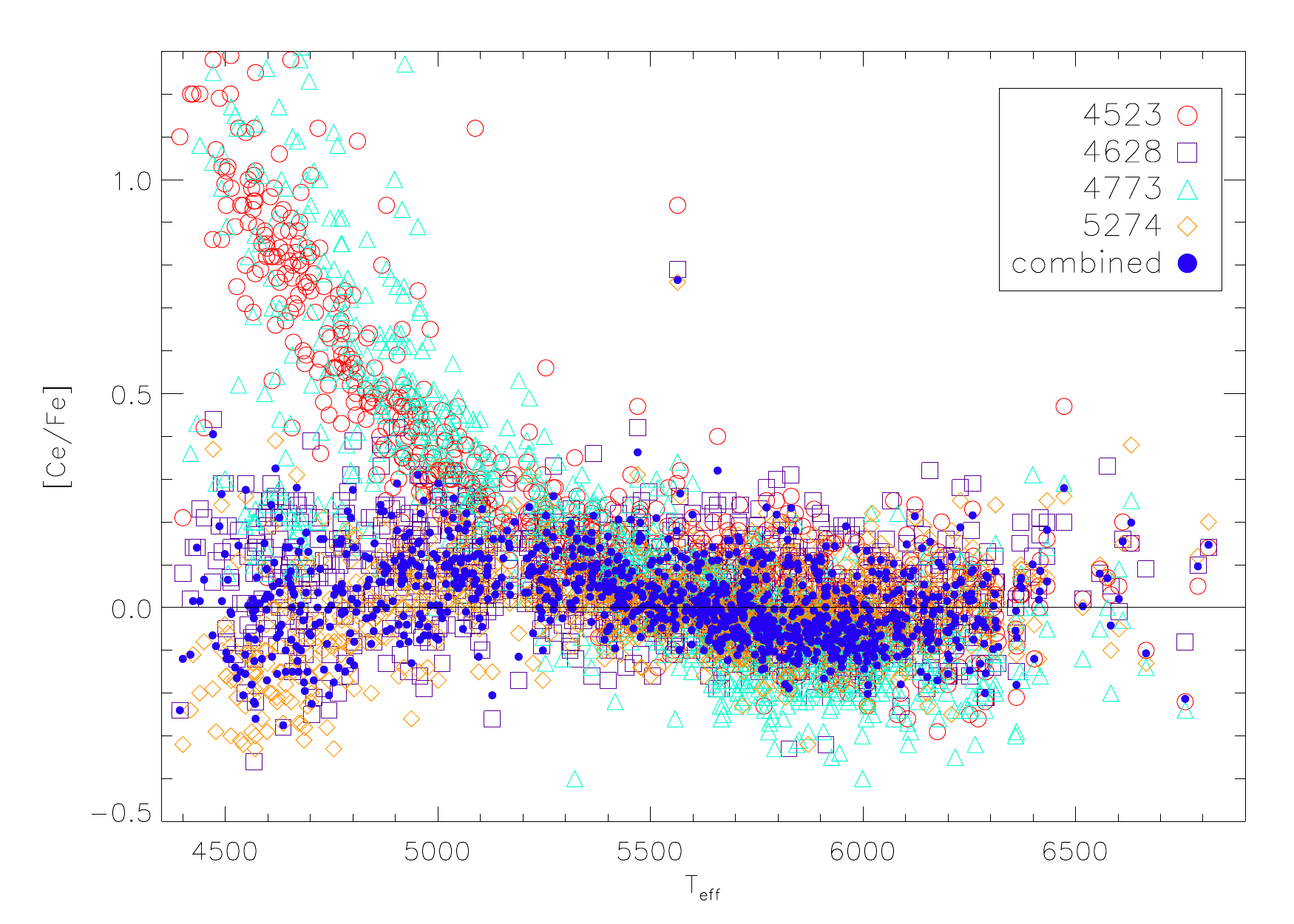}
\includegraphics[width=9.0cm]{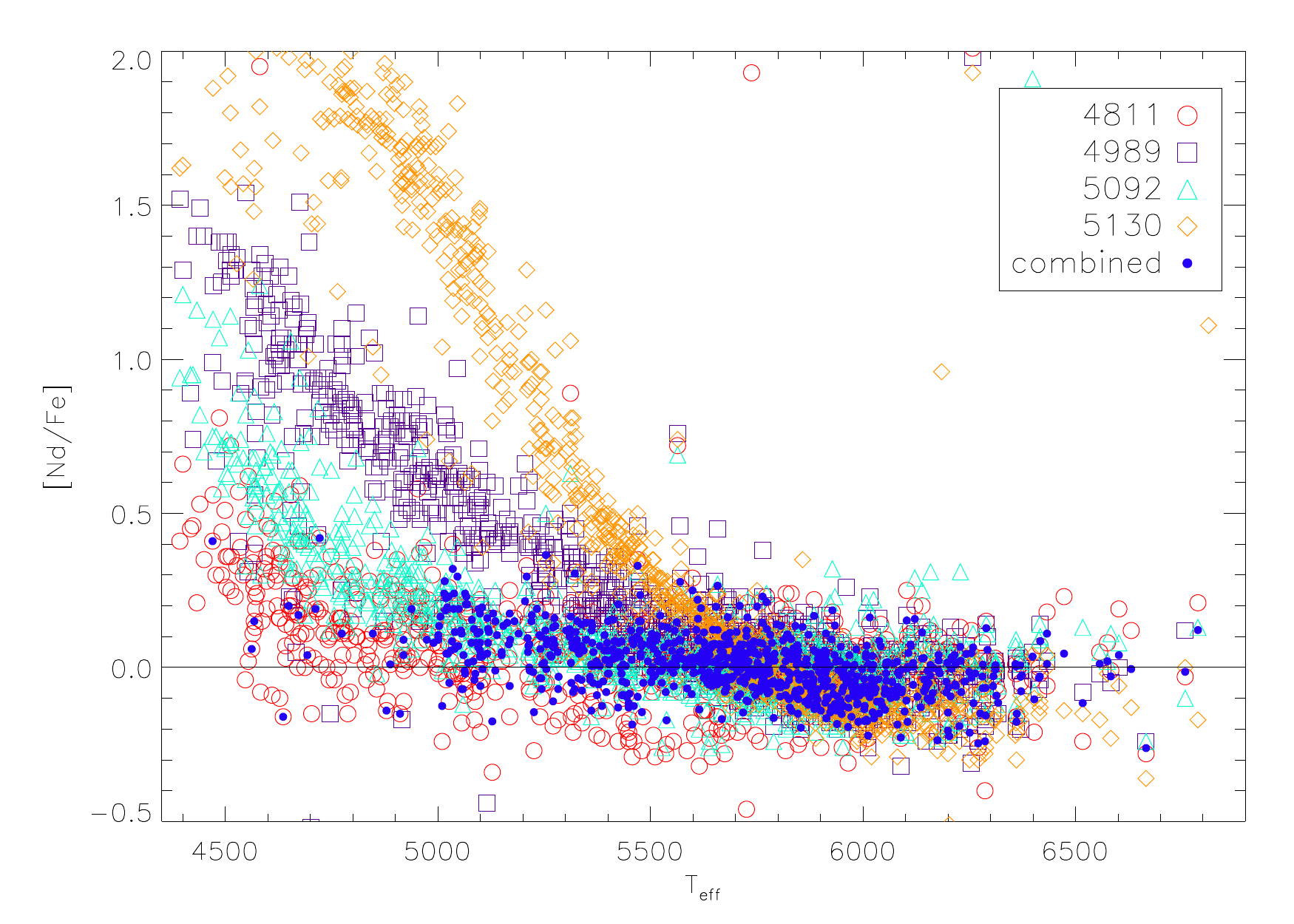}
\caption{[Ce/Fe] and [Nd/Fe] ratios for each line as a function of \teff\ together with the final adopted values for each star (blue filled circles).} 
\label{CeNdteff}
\end{figure}

\begin{figure}
\centering
\includegraphics[width=9.0cm]{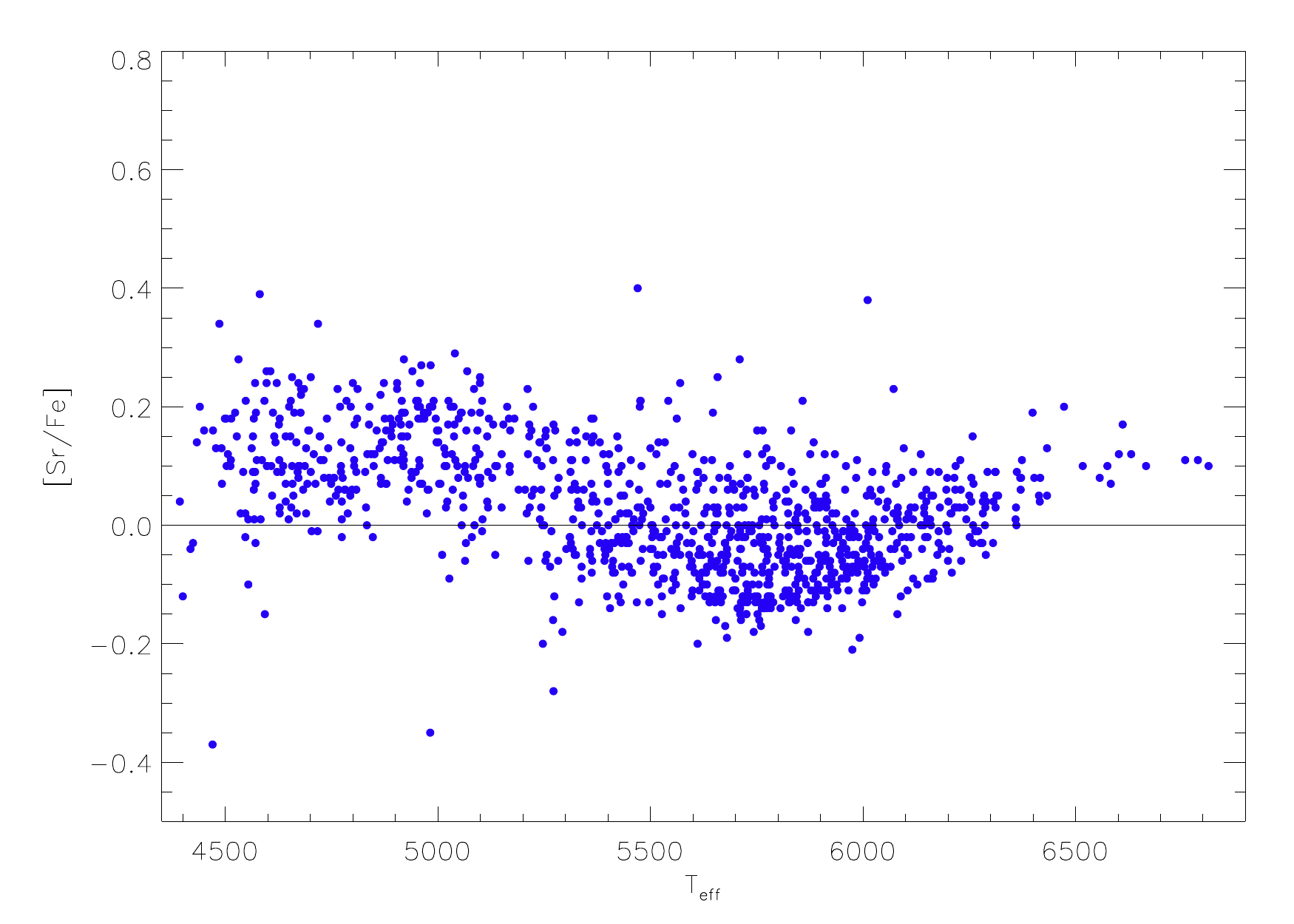}
\includegraphics[width=9.0cm]{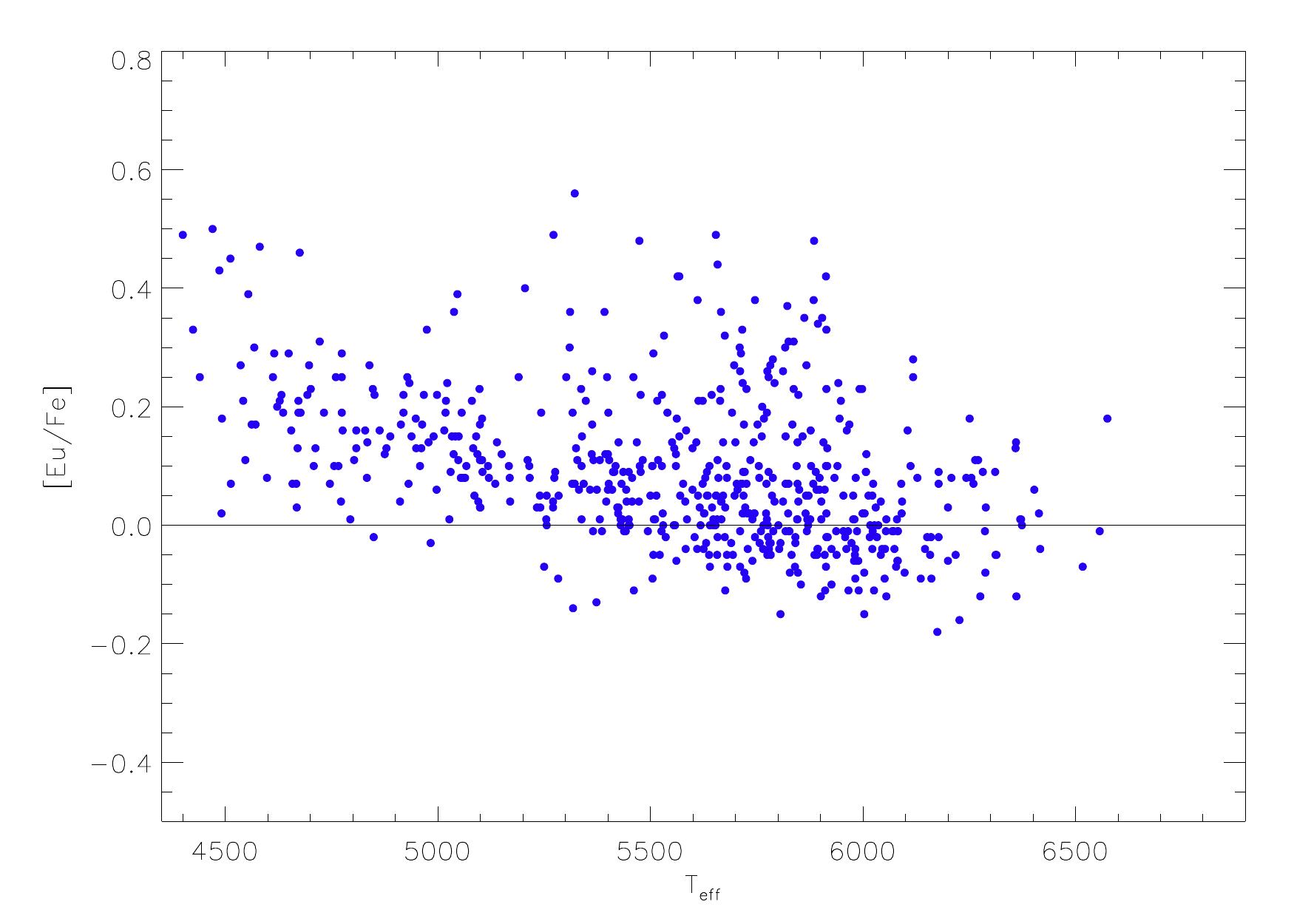}
\caption{[Sr/Fe] and [Eu/Fe] ratios as a function of \teff.} 
\label{SrEuteff}
\end{figure}

Nd abundances are calculated by using 4 \ion{Nd}{II} lines. We considered HFS for all the lines and the atomic parameters were taken from the Gaia-ESO Survey linelist \citep{heiter15}. All the lines become very blended for \teff\,$<$\,5000K and only in spectra with S/N\,$>$\,500 we can use the line at 4811\AA{}. The lines at 4989\AA{} and 5130\AA{} are only used at temperatures higher than 5500K and 5600K respectively since below those \teff\ the abundances show a strong upwards trend with decreasing \teff\ (see Fig. \ref{CeNdteff}). Finally, the line at 5092\AA{} is only used for \teff\,$>$\,5000K.
 
\subsection{Eu abundances}
Eu abundances are based on the weak ionized line at 6645.13\AA{}. This line is blended with another line at 6645.35\AA{}, thus the automatic measurement was not possible for many of the stars. We selected only the stars with S/N $>$ 200 and measured by hand the lines which ARES could not deblend. We considered HFS for this line and the atomic parameters were taken from the Gaia-ESO Survey linelist \citep{heiter15}. The error due to continuum placement was calculated in the same way as for Sr.

\begin{figure}
\centering
\includegraphics[width=9cm]{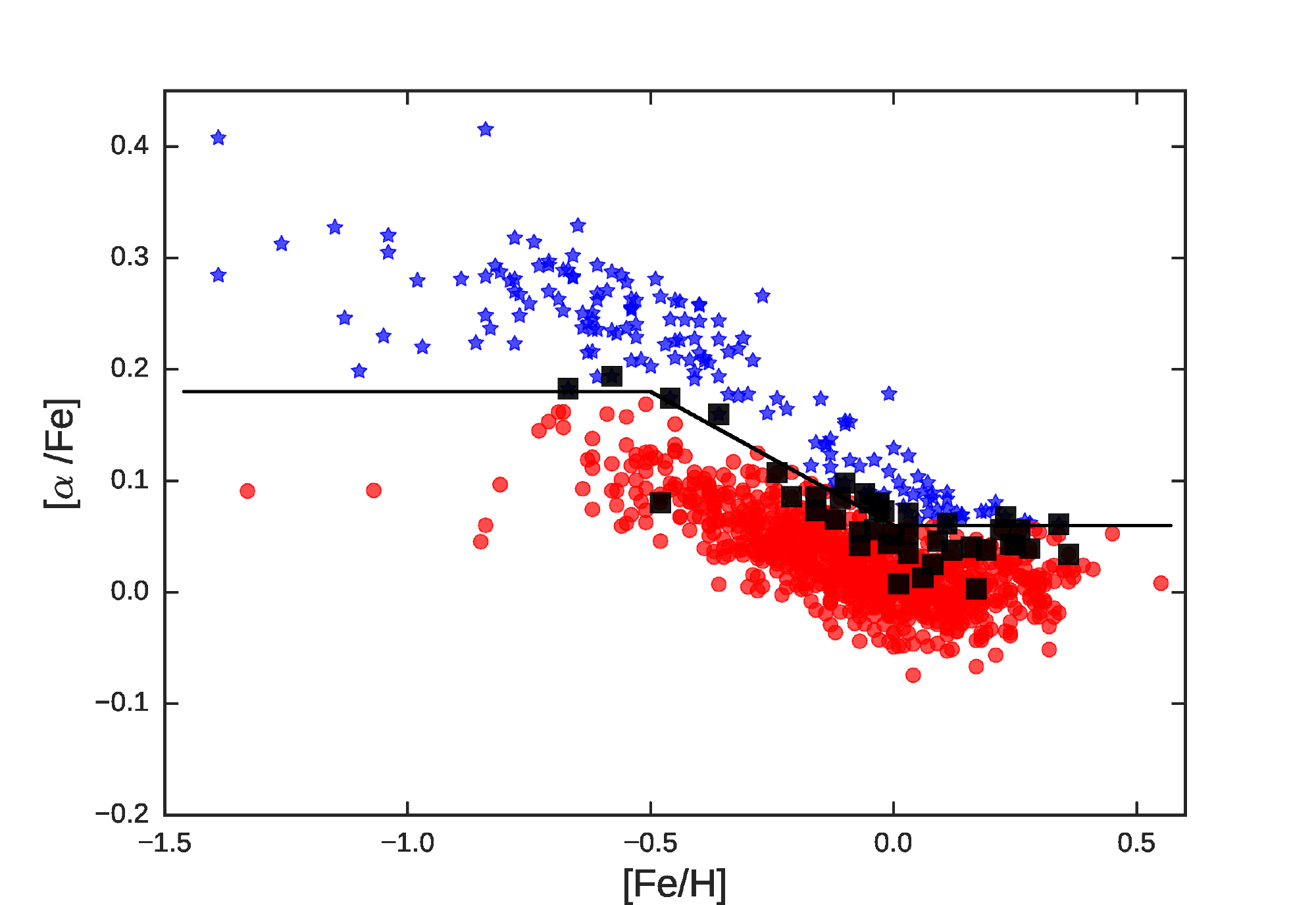}
\caption{Updated abundances of $\alpha$ elements (mean abundance of Si, Mg and Ti) with the new stellar parameters presented in this work. Thin disk stars (red circles) and thick disk stars (blue stars) are chemically separated by their $\alpha$ content (thick line). The \textit{h$\alpha$mr} stars are the prolongation of thick disk stars at [Fe/H]\,$>$\,--0.2\,dex. The stars with a different classification respect to \citet{adibekyan11} are shown as black squares.} 
\label{sep_new}
\end{figure}

\subsection{Revised abundances for refractory elements and \textit{h$\alpha$mr} stars}
In the following sections we will evaluate the behaviour of the previously described elements for the different populations in the Galaxy: thin disk, thick disk and the \textit{h$\alpha$mr} population. We note that stars with [Fe/H]\,$>$\,--0.2\,dex and showing enhancement in $\alpha$-element abundances were first classified as members of a \textit{h$\alpha$mr} population in \citet{adibekyan11, adibekyan13}. Since in this work we are presenting updated values of \logg\ and \teff\ for many stars of our sample we have to re-derive the abundances of refractory elements, specially those of MgI, SiI, TiI and TiII in order to have a coherent definition of \textit{h$\alpha$mr} stars. Moreover we have improved our linelist, using Van der Waals damping constants, and the final abundances for each element are obtained from the WM of all the available individual abundances. In Fig. \ref{sep_new} we can see the new $\alpha$ abundances in our sample and which stars have now a different classification. Most of these stars (24 out of 43) have \teff $<$ 5200K and their abundance change can be explained by the change of \teff\ since \logg\ variations hardly affect neutral species. Nevertheless, the difference in alpha abundances between the old and updated values is lower than 0.05\,dex. \\

In detail, there are 10 stars that before belonged to the thin disk population and now they are considered as \textit{h$\alpha$mr}. Six of these stars have \teff $<$ 5200K and the difference in [$\alpha$/Fe] is 0.004\,$\pm$\,0.005\,dex (in the sense new-old) for the 10 stars. On the other hand, 25 stars considered before as \textit{h$\alpha$mr} now belong to the thin disk. Eleven of these stars are cool and the difference in [$\alpha$/Fe] is --0.047\,$\pm$\,0.021\,dex for the 25 stars. Finally, 50 stars remain classified as \textit{h$\alpha$mr} and in total we have 60 stars belonging to this population. For thin disk stars at [Fe/H]\,$\geq$\,--0.2\,dex, $<$[$\alpha$/Fe]$>$\,=\,0.014\,$\pm$\,0.028\,dex meanwhile for \textit{h$\alpha$mr}, $<$[$\alpha$/Fe]$>$\,=\,0.095\,$\pm$\,0.029\,dex. Despite the new classification of some of our stars the separation between both populations still exists and the difference in [$\alpha$/Fe] is above the errors (the average error of [$\alpha$/Fe] are 0.037, 0.017 and 0.031\,dex for the groups of cool, solar and hot stars as defined in subsection 3.1). The updates abundances for these elements are also provided in an electronic table.

\section{Results and discussion}

\subsection{[X/Fe] ratios for different stellar populations}

\begin{figure*}
\centering
\includegraphics[width=19.0cm]{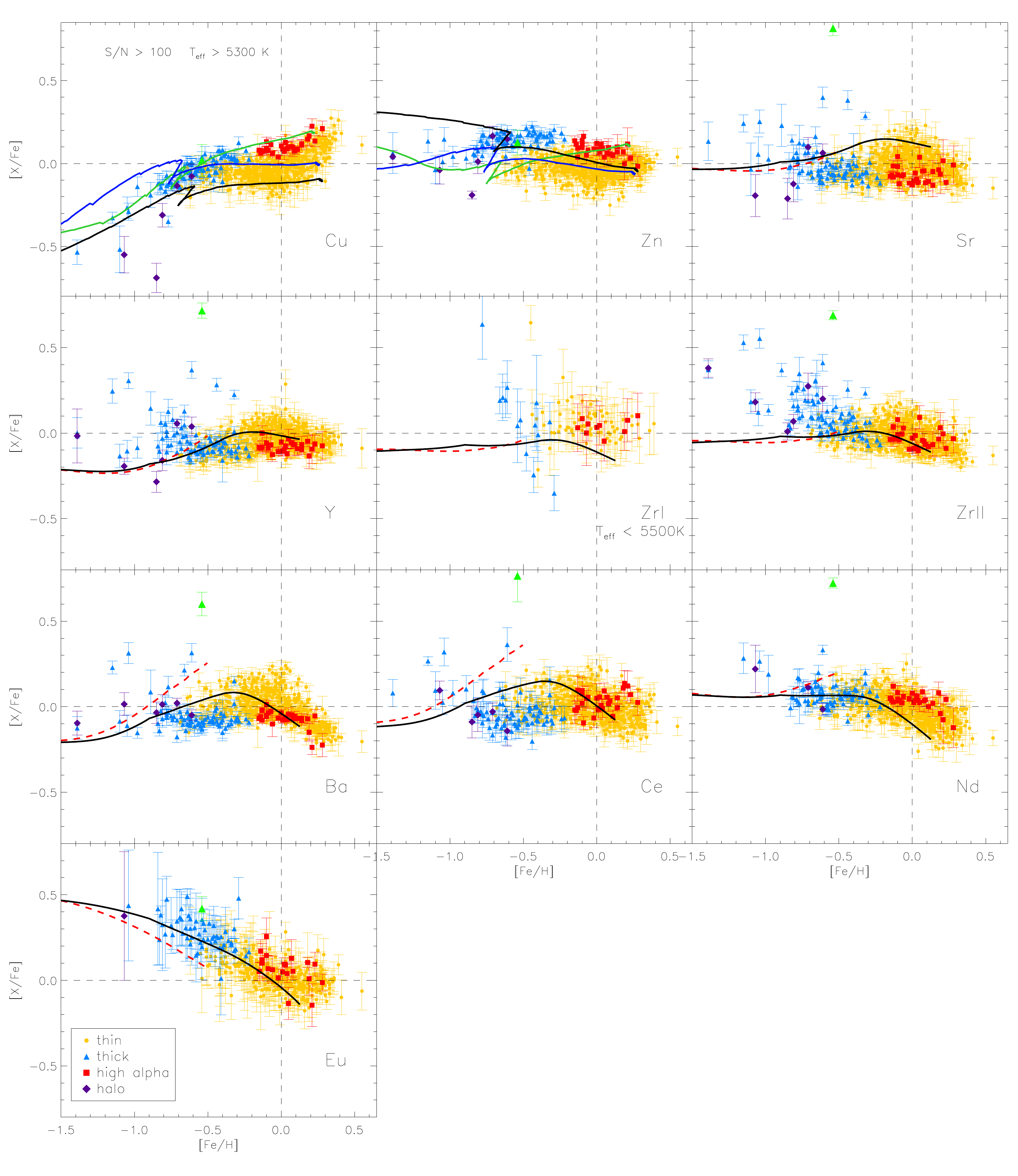}
\caption{Final [X/Fe] ratios as a function of [Fe/H] for stars with \teff $>$ 5300\,K and S/N $>$ 100. The different stellar populations are depicted with different colours and symbols as explained in the legend. The green bigger triangle is the s-enriched star HD11397. For Cu and Zn we overplot the GCE models 1, 4 and 5 (green, blue and black lines, respectively) from \citet{romano10}. For the rest of the elements we show the GCE models from \citet{bisterzo17} for the thin disk (black lines) and the thick disk (red dashed lines).} 
\label{hot_XFe_Fe_errors}
\end{figure*}

\begin{figure*}
\centering
\includegraphics[width=19.0cm]{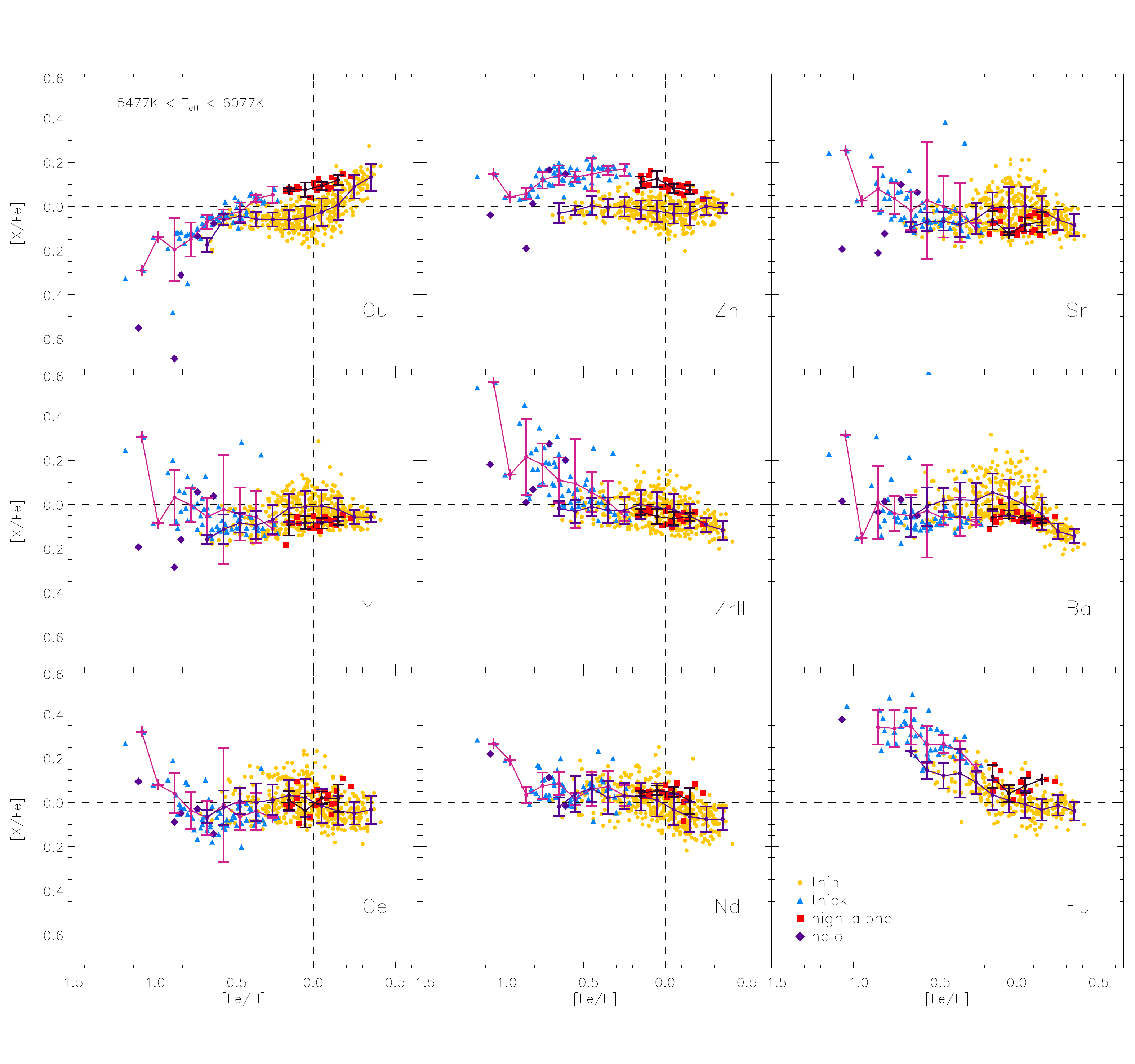}
\caption{Final [X/Fe] ratios as a function of [Fe/H] for stars with \teff$_{\odot}$\,$\pm$\,300\,K. The different stellar populations are depicted with different colours and symbols as explained in the legend. The mean abundances in each metallicity bin of 0.1\,dex are shown together with the standard deviation.} 
\label{solar_XFe_Fe}
\end{figure*}

\begin{center}
\begin{table*}
\caption{Results from the K-S tests for the different [X/Fe] abundance ratios.}
\centering
\label{table_KS}
\begin{tabular}{l|cc|cc|cc|cc}
\hline
\noalign{\medskip} 
 & \multicolumn{4}{c}{Full sample} & \multicolumn{4}{c}{T$_{\odot}$\,$\pm$\,300\,K} \tabularnewline
\noalign{\medskip} 
\hline
\noalign{\medskip} 
 & \multicolumn{2}{c}{Thin-Thick} & \multicolumn{2}{c}{Thin-\textit{h$\alpha$mr}} & \multicolumn{2}{c}{Thin-Thick} & \multicolumn{2}{c}{Thin-\textit{h$\alpha$mr}}\tabularnewline
 & \multicolumn{2}{c}{[Fe/H]\,$<$\,--0.2} & \multicolumn{2}{c}{[Fe/H]\,$\geq$\,--0.2} & \multicolumn{2}{c}{[Fe/H]\,$<$\,--0.2} & \multicolumn{2}{c}{[Fe/H]\,$\geq$\,--0.2}\tabularnewline
\noalign{\medskip} 
\hline
\noalign{\medskip} 
[X/Fe] & p-value & D & p-value & D & p-value & D & p-value & D \tabularnewline  
\noalign{\medskip} 
\hline
\hline
\noalign{\smallskip} 
Cu   & 2.3e-3   & 0.21   & 1.3e-22  &  0.67 & 2.5e-2 & 0.24 & 3.7e-11 & 0.75 \tabularnewline
Zn   & 3.6e-37   & 0.74   & 5.1e-15  & 0.54 & 1.8e-23 & 0.83 & 5.0e-16 & 0.91  \tabularnewline
Sr   & 4.4e-1   &  0.10  & 1.6e-2  &  0.21  & 5.4e-4 & 0.33 & 3.4e-4 & 0.45 \tabularnewline
Y    & 7.4e-2   &  0.15  & 8.3e-8  &   0.39 & 3.4e-3 & 0.29 & 5.8e-7 & 0.58 \tabularnewline
ZrII & 6.8e-10   &  0.38  & 4.8e-2  &  0.18 & 1.8e-8 & 0.49 & 1.5e-2 & 0.33 \tabularnewline
Ba   & 7.1e-22   &  0.57  & 5.5e-8  &  0.39 & 6.4e-8 & 0.48 & 2.1e-4 & 0.46 \tabularnewline
Ce   & 8.3e-6   &  0.29  &  4.8e-3 &  0.23  & 1.2e-3 & 0.31 & 5.5e-3 & 0.37 \tabularnewline
Nd   & 1.3e-3   &  0.25  &  2.2e-8 &  0.44  & 1.5e-2 & 0.26 & 1.0e-8 & 0.66 \tabularnewline
Eu   & 3.6e-8   &  0.67  &  6.0e-6 &  0.41  & 1.3e-6 & 0.76 & 1.2e-6 & 0.71 \tabularnewline
\noalign{\smallskip} 
\hline
\hline
\noalign{\medskip} 
\end{tabular}
\end{table*}
\end{center}

In Fig. \ref{hot_XFe_Fe_errors} we show the [X/Fe] ratios of all our elements for stars hotter than 5300\,K. We decided to study this subsample of hotter stars since at lower \teff\ the abundances present higher uncertainties. We also discarded stars with spectra of S/N\,$<$\,100 to ensure more precise measurements. For this subsample we have 539 thin disk stars, 80 thick disk stars, 28 \textit{h$\alpha$mr} stars and 6 halo stars. We used a chemical criteria based on the [$\alpha$/Fe] ratio to separate the stellar populations as done in \citet{adibekyan11, adibekyan12} except for halo stars, that are selected based on their kinematics. The corresponding [X/Fe] ratios for the full sample are shown in the Appendix, in Fig. \ref{all_XFe_Fe} together with GCE models from \citet{romano10} and \citet{bisterzo17} which will be discussed in the next paragraphs. We can see in this plot that the scatter of abundances is quite high, especially among the thick disk stars. However, this scatter in, in general, higher than the 2-$\sigma$ errors probing that the spread is real (for the coolest stars the typical 1-$\sigma$ errors are around 0.10\,dex, see Table \ref{table_errors}). The abundances of Eu show very high errors, probably overestimated for most of the stars due to the strong effect of the continuum placement uncertainty on the small EWs of the Eu line used here.
We note that the star HD11397 shows very high abundances ([X/Fe]\,$\gtrsim$\,0.7) for the s-process elements Sr, Y, Zr, Ba, Ce and Nd. However, it is not enhanced in the \textit{r}-process element Eu. This star was already discussed in \citet{pompeia08} who reported that it has an abundance profile similar to the mild Ba-stars. We also find three thick disk stars with --0.61$<$\,[Fe/H]\,$<$\,--0.32 (HD126803, CD-436810 and HD28701) that show enhanced abundances of Sr, Y and Zr when compared to other thick disk stars of similar metallicity. Two of them also show enhancement of Ba, Ce and Nd. Although that level of enhancement is not as high as to be considered as \textit{s}-enriched stars.\\

In Figs 3-8 we can see that the most evident trends of abundances with \teff\ take place in cool stars. However, at high temperatures some elements also show slight trends with \teff. This is the case of [Cu/Fe] which hardly presents positive values at \teff\,$\gtrsim$\,6000\,K or [Zn/Fe] that also displays a decreasing trend for the hotter stars. On the other hand [Sr/Fe] tends to increase at higher \teff. Therefore, to avoid the systematic effects associated with the errors on stellar parameters we decided to study the differences in abundances among the different populations only using stars with T$_{\odot}$\,$\pm$\,300\,K as also done in \citep{adibekyan12}. This subsample, shown in Fig. \ref{solar_XFe_Fe} is composed of 328 thin disk stars, 49 thick disk stars, 16 \textit{h$\alpha$mr} stars and 4 halo stars. We also plot the average abundance in each 0.1\,dex metallicity bin for the different components. The population of \textit{h$\alpha$mr} stars (shown with red squares) is well mixed with the thin disk for most of the elements however it shows a clear average enhancement for Cu and Zn following the behaviour of the thick disk stars. Moreover, they have Y and Ba abundances lower on average than thin disk stars. Finally, it is not surprising that they also show higher Eu on average which is mainly produced in SNe II as alpha elements.\\

\textit{\textbf{First elements beyond the iron-peak (Cu, Zn):}} 

Copper and zinc are transition elements between the Fe-peak and the neutron capture elements. Cu was initially thought to be mainly produced by the weak \textit{s}-process \citep{sneden91} but this view has been debated over the years suggesting that SNe Ia would also contribute to their production \citep[e.g.][]{matteucci93,mishenina02}. However, \citet{romano07} ruled out the contribution from SNe Ia and showed that explosive nucleosynthesis in core-collapse SNe is only important at very low metallicities, hence Cu is mostly produced by the weak \textit{s}-process. Moreover, other works claim that only 5\% of Cu is provided by the main-\textit{s} component and the remaining can be explained by SNe II yields from different masses and metallicities, i.e. the weak-\textit{s} component \citep{travaglio04b,bisterzo05}. The weak \textit{s}-process occurs in massive stars during core He and shell C burning, where neutrons are provided by the $^{22}$Ne($\alpha,n$)$^{25}$Mg reaction, and partly produces neutron capture isotopes lighter than A$\sim$90 \citep[e.g.][and refererences therein]{pignatari10}. The weak \textit{s}-process is considered to be of secondary nature, since the neutron source, $^{22}$Ne, is originated from pre-existing CNO nuclei, and thus depends on the initial metallicity of the star. \\

In Fig. \ref{solar_XFe_Fe} we can see that [Cu/Fe] diminishes monotonically for lower metallicities in the thick disk population and the few halo stars of our sample \citep[as already seen in halo stars by][]{sneden91}. This is in agreement with a Cu production mostly by weak \textit{s}-process and a primary contribution by explosive nucleosyntheis in SNe II at low metallicities \citep{romano07}. The maximum Cu for thick disk stars is reached at [Fe/H]\,$\sim$\,--0.4\,dex in agreement with the compilation made by \citet{pignatari10} to then steadily decrease towards lower [Fe/H] and keep more or less constant at --0.4$<$\,[Fe/H]\,$\lesssim$\,--0.2. Thick disk stars have, on average, higher Cu abundances than thin disk stars in the metallicity region --0.5$\lesssim$\,[Fe/H]\,$\lesssim$\,--0.2 as suggested in \citet{reddy06,israelian14,yan15,mikolaitis17}. However, for lower metallicities, although thick disk stars still present higher mean abundances on average, the differences are very small and within the errors. On the contrary, thin disk stars present a very slight increase of abundances for --0.8$<$\,[Fe/H]\,$\lesssim$\,0.1 but at super-solar metallicities there is a steep rise respect to iron which was first reported by \citet{allende04} and further confirmed here with a larger sample. Also, this increase of Cu abundances is found in the large sample of the AMBRE project \citep{mikolaitis17}. The \textit{h$\alpha$mr} population presents the same behaviour as thin disk stars but interestingly the abundances form an upper envelope to the thin disk, making a continuation of the thick disk, and the mean values in each metallicity bin are totally disjunct. To test whether the difference in [Cu/Fe] ratios are significant among the different populations we have performed several Kolmogorov-Smirnov (K-S) tests whose results are shown in Table \ref{table_KS}. In the case of Cu, the K-S test fails to reject the hypothesis that thick disk stars and thin disk stars at [Fe/H]\,$<$\,--0.2 belong to the same population. On the contrary, the K-S test rejects the hypothesis that Cu abundances of high-alpha metal rich stars and thin disk stars (at [Fe/H]\,$\geq$\,--0.2) are drawn from the same parent population with p-values lower than 10$^{-10}$ both for the full sample or the subsample of stars with T$_{\odot}$\,$\pm$\,300\,K. In order to consider the errors of the abundance ratios we have created 1000 samples of randomly selected abundances assuming a gaussian distribution for each [Cu/Fe] value with central value the abundance ratio and sigma equal to the error of the abundance. This test shows that for all the randomly generated samples the p-value of each K-S test is always lower than 2.5$\cdot$10$^{-5}$. We remind that the separation among thin disk stars and \textit{h$\alpha$mr} stars was based on [$\alpha$/Fe] ratios by \citet{adibekyan11} who also showed that the \textit{h$\alpha$mr} are older on average than thin disk stars and have intermediate orbits between the thin and the thick disk stars. Therefore, the tests that we have made here serve to show that \textit{h$\alpha$mr} also have different abundances of other elements apart of $\alpha$ elements but they do not serve as a probe to distinguish them from thin disk stars.\\

In Fig. \ref{hot_XFe_Fe_errors} we show the GCE models using different yields computed by \citet{romano10}. The model that better matches our abundances is Model 1 (green line) which considers the case B yields for normal SNe II from \citet{woosley95}. Models 4 and 5, which consider hypernovae (HNe) yields from \citet{kobayashi09} with different HNe fractions, overestimate the abundances of the thick disk and underestimate the abundances on the thin disk, respectively\footnote{We note the different position of the models in Fig. \ref{hot_XFe_Fe_errors} with respect to \citet{romano10} plots since our solar mean abundances are different from those in \citet{grevesse98}, used in the models of \citet{romano10}.}. 
Moreover, these models are not able to reproduce the increase of Cu at high metallicities. However, Models 4 and 5 seem to better reproduce the Cu trends at very low metallicities as shown in Fig. 16 of \citet{romano10}. Neverthless, these authors warn about the lack of AGB yields in a full range of masses and metallicities to test their effects at low [Fe/H].\\

\begin{figure*}
\centering
\includegraphics[width=19.0cm]{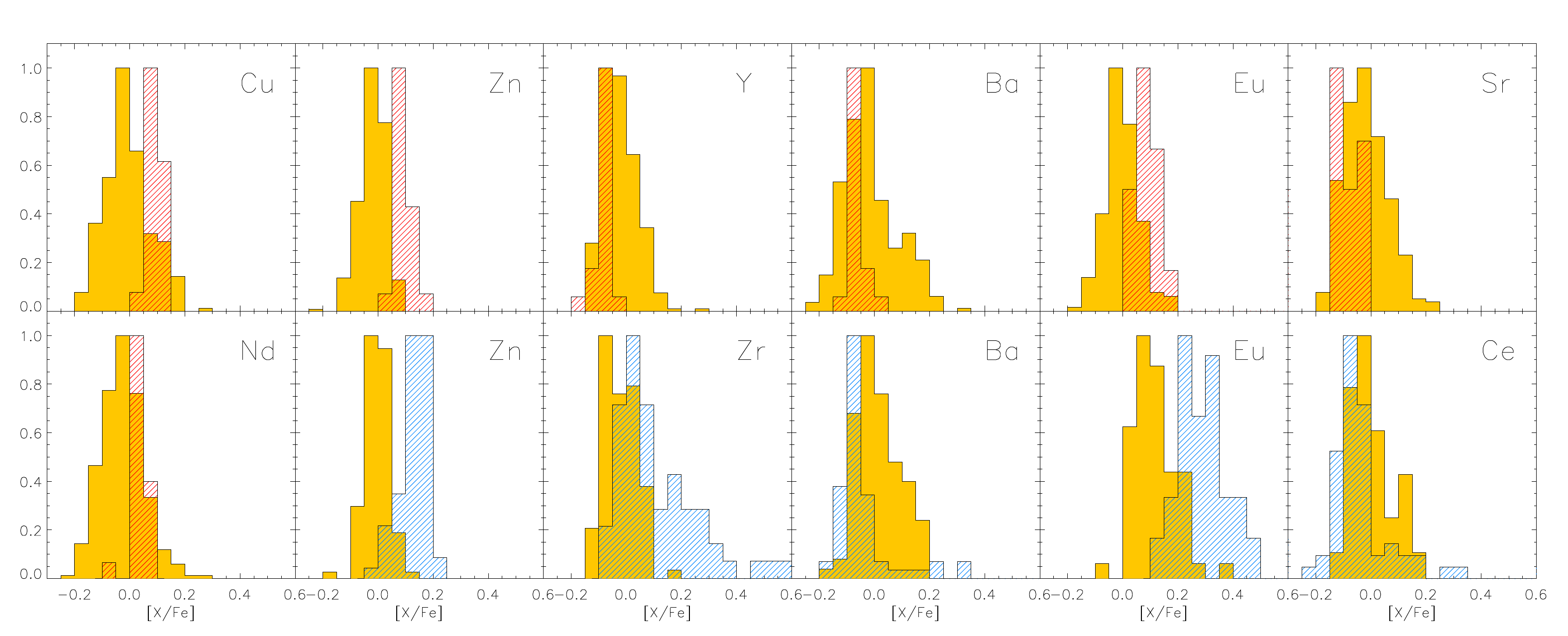}
\caption{Normalized distributions of several [X/Fe] ratios for stars with T$_{\odot}$\,$\pm$\,300\,K. The yellow histograms represent the thin disk stars at [Fe/H]\,$\geq$\,--0.2 dex when compared to \textit{h$\alpha$mr} (red dashed histogram) or at [Fe/H]\,$<$\,--0.2 dex when compared to thick disk stars (blue histogram).} 
\label{histogram}
\end{figure*}

The production of Zn is somehow more complex. Half of solar Zn is in the isotope $^{64}$Zn which is produced by $\alpha$ freezout in neutrino winds during supernova explosions of massive stars while all the other Zn isotopes are produced by the weak \textit{s}-process \citep[e.g.][and refererences therein]{bisterzo05}. The [Zn/Fe] ratios resemble somehow the behaviour of alpha elements (with a flattening of abundances around solar metallicity caused by the well-known iron contribution from SNe Ia and clearly separated thin and thick disk), however the growth of abundances towards lower [Fe/H] is very slight and it does not reach as high values as alpha elements in our [Fe/H] range \citep[e.g. $\alpha$/Fe$\,>$\,0.3\,dex at metallicity around --1\,dex in][and Fig. \ref{sep_new}]{adibekyan12}. Moreover, thin disk stars present nearly flat Zn abundances meanwhile alpha elements increase for lower metallicities. A maximum value of $\sim$ 0.2\,dex is observed for stars with --1.0\,$<$\,[Fe/H]\,$<$\,--0.5 and then slightly decrease for lower metallicities as reported by \citet{saito09}. The work by \citet{mikolaitis17} also shows this slight decrease though it starts at lower [Fe/H]. Although thick disk stars form kind of a plateau, the maximum Zn in the thick disk seems to be observed around metallicity --0.5\,dex as reported to Cu, but drops at higher metallicities in contrast with Cu. \citet{jonay10,bensby14,mikolaitis17} and \citet{duffau17} also found a somewhat decreasing trend with metallicity. At very low metallicities Zn abundances continue to increase, reaching values of [Zn/Fe]\,$\sim$\,0.5\,dex \citep[e.g.][]{saito09,romano10}. This overall trend can be explained by chemical-evolution models where Zn is produced by SNe II, HNe, and SNe Ia with various metallicities \citep{kobayashi09,saito09} although models cannot match well the behaviour at low metallicities.\\

In Fig. \ref{solar_XFe_Fe} we can also observe a very well separated thick disk in all [Fe/H] bins as first noted by \citet{bensby03} with a smaller sample and further confirmed by \citet{mikolaitis17}. Interestingly, the \textit{h$\alpha$mr} stars present also high Zn abundances when compared to thin disk stars, a fact in agreement with the $\alpha$-kind behaviour of Zn abundances. The K-S tests for the full sample and the solar stars reject the hypothesis that [Zn/Fe] ratios for the thin disk and \textit{h$\alpha$mr} stars are drawn from the same population (see Table \ref{table_KS}). A similar result is obtained for the comparison between thick disk and thin disk stars with even lower p-values. Also, the random generated samples always give p-values lower than 10$^{-5}$ and 10$^{-14}$ for the comparison between thin-\textit{h$\alpha$mr} and thin-thick, respectively. For a more visual comparison we present the distribution of some abundance ratios for the stars with T$_{\odot}$\,$\pm$\,300\,K in Fig. \ref{histogram}. Finally, in Fig. \ref{hot_XFe_Fe_errors} we also show the same models from \citet{romano10} as for Cu. Models 4 and 5 (those including HNe yields) reproduce better our abundances and the general lowering trend from the thick disk to the thin disk. However, in their Fig. 16, \citet{romano10} show that these models do not work well at low metallicities pointing to the necessity of increasing the Zn yields from metal-poor core-collapse SNe.\\

\begin{figure}
\centering
\includegraphics[width=9cm]{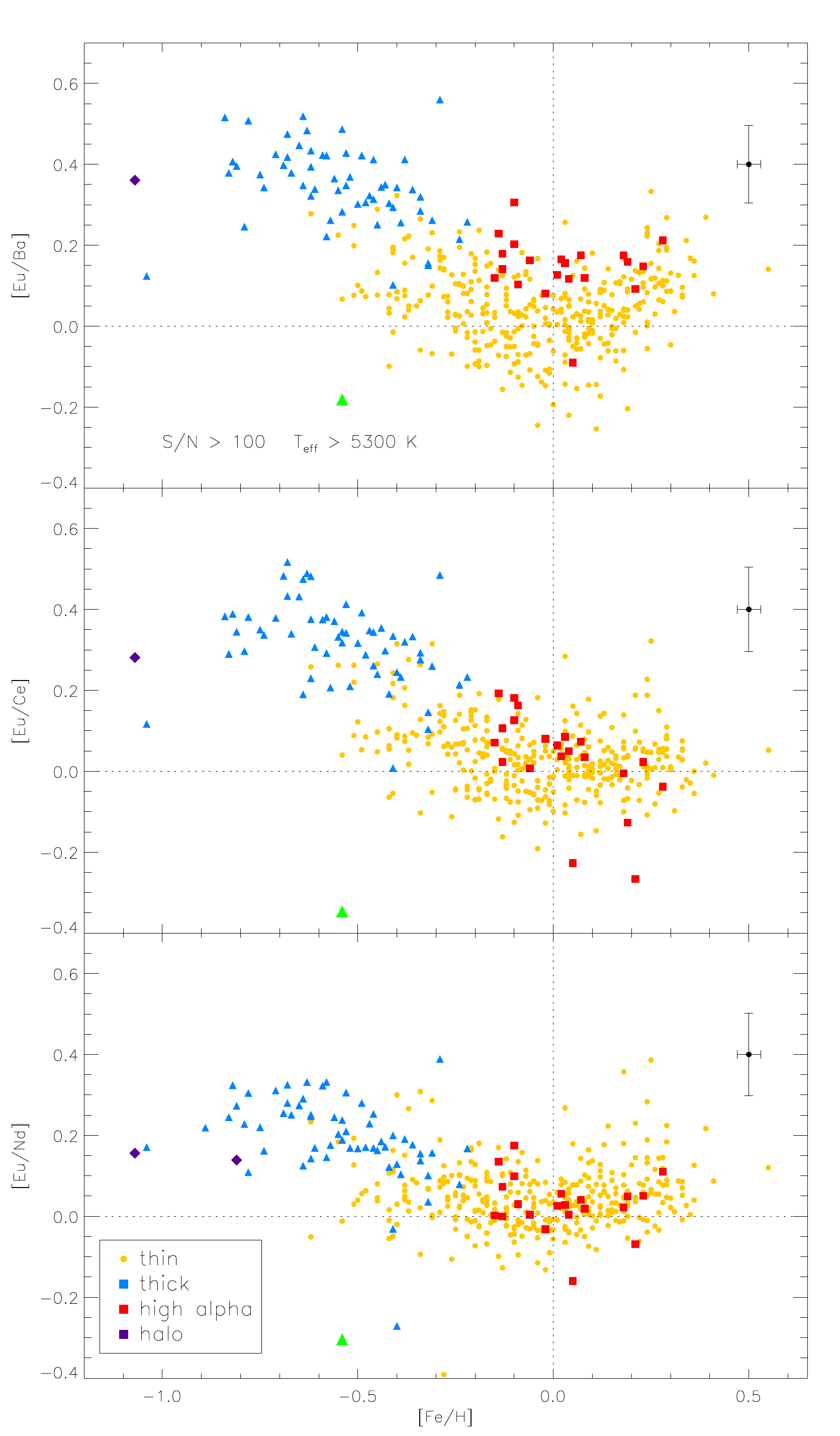}
\caption{Abundance ratios between the \textit{r}-process element Eu and the heavy-\textit{s} elements as a function of [Fe/H] for stars with \teff $>$ 5300\,K and S/N $>$ 100. Symbols as in Fig. \ref{hot_XFe_Fe_errors}.} 
\label{EuBa}
\end{figure}

\textit{\textbf{Light \textit{s}-process elements (Sr, Y, Zr):}} 

Light \textit{s}-process elements (corresponding to the first peak) are mainly produced by AGB stars through the main \textit{s}-process, where the required free neutrons are supplied mainly by the reaction $^{13}$C($\alpha,n$)$^{16}$O and to a lower extent by the reaction $^{22}$Ne($\alpha,n$)$^{25}$Mg. However, some studies point to a non-negligible contribution of the weak \textit{s}-process to the production of these elements \citep[30\% as shown in ][]{pignatari10}. It is important to note that there is a lack of \textit{s}-process yields calculations of AGB stars of different masses at different metallicities. Together with some uncertainties such as the convection treatment, stellar rotation, mass loss or the size of the $^{13}$C pocket; it makes the development of GCE models a difficult task \citep{karakas16a}. In Fig. \ref{solar_XFe_Fe} we can see that these three elements behave in a different way despite belonging to the same group. Zr shows a clear increasing trend for lower metallicities including halo stars, as also observed in previous works \citep[e.g.][]{jonay10,mishenina13,battistini16,zhao16} meanwhile Y and Sr present a flatter trend and most of thick disk stars have subsolar abundances, similar to \citet{bensby14,zhao16}, respectively. Thick disk stars tend to increase their Y and Sr abundances from $\sim$\,--0.1\,dex at [Fe/H]\,=\,--0.2 to $\sim$\,0.1\,dex at [Fe/H]\,=\,--1, but the increase is very slight when compared to Zr. Thin disk stars present a slight rise for Sr and Y towards a maximum at [Fe/H]\,$\sim$\,0.0\,dex to then decrease at super-solar metallicities while for \ion{Zr}{II} they present a flat trend till solar metallicity to then also decrease at [Fe/H]\,$\geq$\,0\,dex. The group of \textit{h$\alpha$mr} stars also present subsolar abundances along [Fe/H], as thick disk stars, very clear for Y abundances.The K-S test for [Y/Fe] ratios provide p-values lower than 10$^{-6}$ for the comparison of thin disk stars and \textit{h$\alpha$mr} (see Table \ref{table_KS}). The K-S tests for the random generated samples considering gaussian distributions always give p-values lower than 0.05 demonstrating that Y abundances for both populations probably do not come from the same parent population. For the case of [Sr/Fe] we obtain p-values lower than 0.05 in 98.7\% of the tests although we note that this percentage decreases significantly when considering the full sample. In contrast, Zr abundances for \textit{h$\alpha$mr} stars are well mixed with those of thin disk stars although they present a rather flat trend opposite to the lowering trend of thin disk stars. Finally, the abundances of Zr for thick disk stars are clearly higher than for thin disk stars and for all the 1000 random generated samples we can discard the hypothesis that both groups are drawn from the same population. To evaluate the behaviour of \ion{Zr}{I} abundances we should look at the full sample of cool stars in Fig. \ref{all_XFe_Fe}. Although the dispersion is higher we still can see the increasing trend of abundances as [Fe/H] decreases, similar to the behaviour of \ion{Zr}{II}. This could imply that at lower metallicities a non-negligible contribution from massive stars to Zr took place (see further discussion in next subsection). \\

In Fig. \ref{hot_XFe_Fe_errors} we also plot the GCE models for Y and Zr computed by \citet{bisterzo17} and for Sr (Bisterzo et al., private communication). These models have been built considering the contributions of \textit{r}-process, \textit{s}-process and Lighter Element Primary Process (LEPP, this will be further discussed in Sect. 4.3)\footnote{We note the different position of the models in Fig. \ref{hot_XFe_Fe_errors} with respect to \citet{bisterzo17} plots since our solar mean abundances are different from those in \citet{lodders09}, used in the models of \citet{bisterzo17}.}. The \textit{r}-process contribution is only important for stars with lower metallicities as shown here. For the \textit{s}-process contribution these authors considered AGB yields with a standard $^{13}$C pocket choice. In general, these models reproduce well our observations for thin disk stars although our maximum abundances (taking place around solar metallicity) are delayed with respect to the models (where the maximum is observed at [Fe/H]\,$\sim$\,--0.25\,dex. This delay is probably caused by the uncertainties on SNe Ia yields (which affect Fe abundances) and by different assumptions on the Star Formation Rate (SFR) of the models (which determine the delay of AGB stars contribution to the elements discussed here) as pointed by \citet{bisterzo17}. On the other hand, thick disk models do not match our observations, especially for Zr, where the model is basically flat but our [Zr/Fe] ratios clearly increase towards lower metallicities. \citet{bisterzo17} also explore the effect of different $^{13}$C pocket choices for Y abundances but the differences are not very high with respect to the standard case in our metallicity range. \\

\textit{\textbf{Heavy \textit{s}-process elements (Ba, Ce, Nd):}}

These elements correspond to the second peak of s-process production, also known as heavy \textit{s}-process elements. The main component of the \textit{s}-process, that is, the production by AGB stars, dominates the synthesis of these elements.
In Fig. \ref{solar_XFe_Fe} we can see a clear fall of Ba abundances for thin disk stars from a maximum [Ba/Fe]\,$\sim$\,0.25\,dex at solar metallicity to [Ba/Fe]\,$<$\,--0.2 at super-solar metallicities, similar to the values reported by \citet{israelian14,bensby14}. This decrease of abundances at high metallicities is also observed for Nd but not for Ce. The decrease of Ba abundances at super-solar metallicities seems to be at odds with the overabundances found in young metal-rich clusters by \citet{dorazi09}\citep[see also Fig. 4 of][]{maiorca12} although not all our metal-rich stars are young\footnote{In a forthcoming work we will discuss the abundance trends with age}. Ba and Ce are mainly \textit{s}-process in solar material ($>$80\%) and they show a very slight reduction of abundances from the maximum towards lower metallicities for the thin disk population. The work by \citet{mishenina13} reports a similar behaviour for Ce abundances, meanwhile other authors show a more flatten trend \citep[e.g.][]{reddy06,battistini16} and \citet{allende04} present a strong increasing trend as [Fe/H] decreases. On the other hand, Nd, which is produced in a similar proportion by \textit{r}- and \textit{s}-processes presents a slight increasing trend as [Fe/H] drops for thin disk stars, also reported by \citet{allende04,jonay13,mishenina13,battistini16} which reflects the important contribution of massive stars to this element. The thick disk stars present mostly sub-solar abundances of Ba and Ce whereas Nd is super-solar and also increases towards lower metallicities. Interestingly, the \textit{h$\alpha$mr} stars show clear subsolar abundances for Ba, as a continuation of the thick disk, resembling the behaviour of Y. However, for Nd, they present higher abundances on average when compared to the thin disk. We have also performed K-S tests which show that the abundances of thick and thin disk stars are probably drawn from different parent populations (see Table \ref{histogram}) for the case of Ba (all K-S tests for the random samples give p-values lower than 0.01) and for Ce (99.8\% of the tests provide p-values lower than 0.05). When comparing the \textit{h$\alpha$mr} stars with the thin disk stars at [Fe/H]\,$>$\,--0.2 we found that 99.8\% of the K-S tests applied to the random generated samples have a p-value lower than 0.05. \\

In Fig. \ref{hot_XFe_Fe_errors} we overplot GCE models of Ba by \citet{bisterzo17} and Ce, Nd (Bisterzo et al., private communication). These models have the same ingredients as for light-\textit{s} elements. Models for Ba and Ce abundances match quite well the observations of the thin disk but here we observe the same delay in the maximum abundances as happens for Sr, Y and Zr. Our maximum abundances occur at solar metallicity but in the models, Ba and Ce peak at [Fe/H]\,$\sim$\,--0.35\,dex. On the other hand, Nd is better reproduce although the models keep flat at [Fe/H]\,$<$\,--0.3\,dex despite the abundances slightly increase towards lower metallicities\footnote{We have shifted the models of Nd by +0.15\,dex in Fig. \ref{hot_XFe_Fe_errors} since our solar abundance is much higher than the solar reference of \citet{bisterzo17} and the models would look too low when compared with our data.}. As happens for light-\textit{s} elements, any of the models for the thick disk seem to reproduce our observations. However, we have to consider the lack of stars at [Fe/H]\,$<$\,--0.8\,dex in our sample. By filling these region with more stars we might have different trends for the thick disk than observed here.\\

\textit{\textbf{The \textit{r}-process element Eu:}} 

Eu is the only element of our work with a dominant contribution by the \textit{r}-process. Its steep increase in abundance towards low metallicities resembles the behaviour of $\alpha$-elements. Therefore, due to the unknown mechanisms owing to Eu production, GCE models have considered that it is synthetized by a primary process in massive stars exploding as SNe II \citep[e.g.][]{travaglio99,bisterzo17}. This trend has been extensively studied in the literature with qualitative good agreement among different authors. We find a maximum [Eu/Fe] value of $\sim$\,0.5\,dex at [Fe/H]\,$\sim$\,--0.8\,dex to then monotonically decrease towards higher metallicities and become flat at [Fe/H]\,$>$\,0\,dex. All the K-S tests reject the hypothesis that Eu abundances in thick disk stars and thin disk stars are drawn from the same population (p-values always lower than 0.02). In Fig. \ref{histogram} we can appreciate how well separated are both populations. Also, 97.2\% of the tests provide p-values lower than 0.05 when comparing the populations of thin disk stars and \textit{h$\alpha$mr} stars, since these last group presents higher Eu abundances, as can be expected due the $\alpha$-like behaviour of Eu. The GCE models by \citet{bisterzo17} shown in Fig. \ref{hot_XFe_Fe_errors} match very well our observations although they seem to underestimate the abundances of the thick disk stars\footnote{We have shifted the models of Eu by +0.1\,dex in Fig. \ref{hot_XFe_Fe_errors} since our solar abundance is quite higher than the solar reference of \citet{bisterzo17} and the models would look too low when compared with our data.}.

\subsection{Relative contribution of \textit{r}- and \textit{s}-process elements along the chemical history of the Galaxy.}

Since the \textit{s}- and \textit{r}-processes are associated with stars of different masses and metallicities that eject their material to the ISM at different moments of the evolution of the Galaxy, their contribution to heavy element production varies with time. Therefore, a way to disentangle the contribution of each process is to check the behaviour of abundance ratios of different kind of elements. For example, the [Ba/Eu] ratio has been extensively used to unveil whether the \textit{s}- or \textit{r}-process dominated the nucleosynthesis at a given moment of the evolution of the Galaxy. Early studies on very metal-poor stars reported the enrichment of Eu when compared to other \textit{n}-capture elements \citep{spite78} pointing to the importance of the \textit{r}-process for very old stars. At very low metallicities the \textit{r}-process is expected to dominate the production of heavy elements since massive stars were the first to explode as core-collapse SNe and enrich the ISM with their material, before the AGB stars contributed to the main \textit{s}-process. In Fig. \ref{EuBa} we can see how the ratios of Eu versus other heavy-\textit{s} elements increase towards lower metallicities. Since the contribution from \textit{r}-process to Ba and Ce is lower ($<$\,20\%) than to Nd ($\sim$\,45\%) the ratios of [Eu/Ba] and [Eu/Ce] reach values up to $\sim$\,0.5\,dex while the maximum of [Eu/Nd] is $\sim$\,0.35\,dex. The minimum of these ratios is observed around solar metallicity to then grow again towards super-solar metallicities for Ba and Nd, due to the continuous decrease of these heavy-\textit{s} elements as the metallicity increases. \\

\begin{figure}
\centering
\includegraphics[width=9cm]{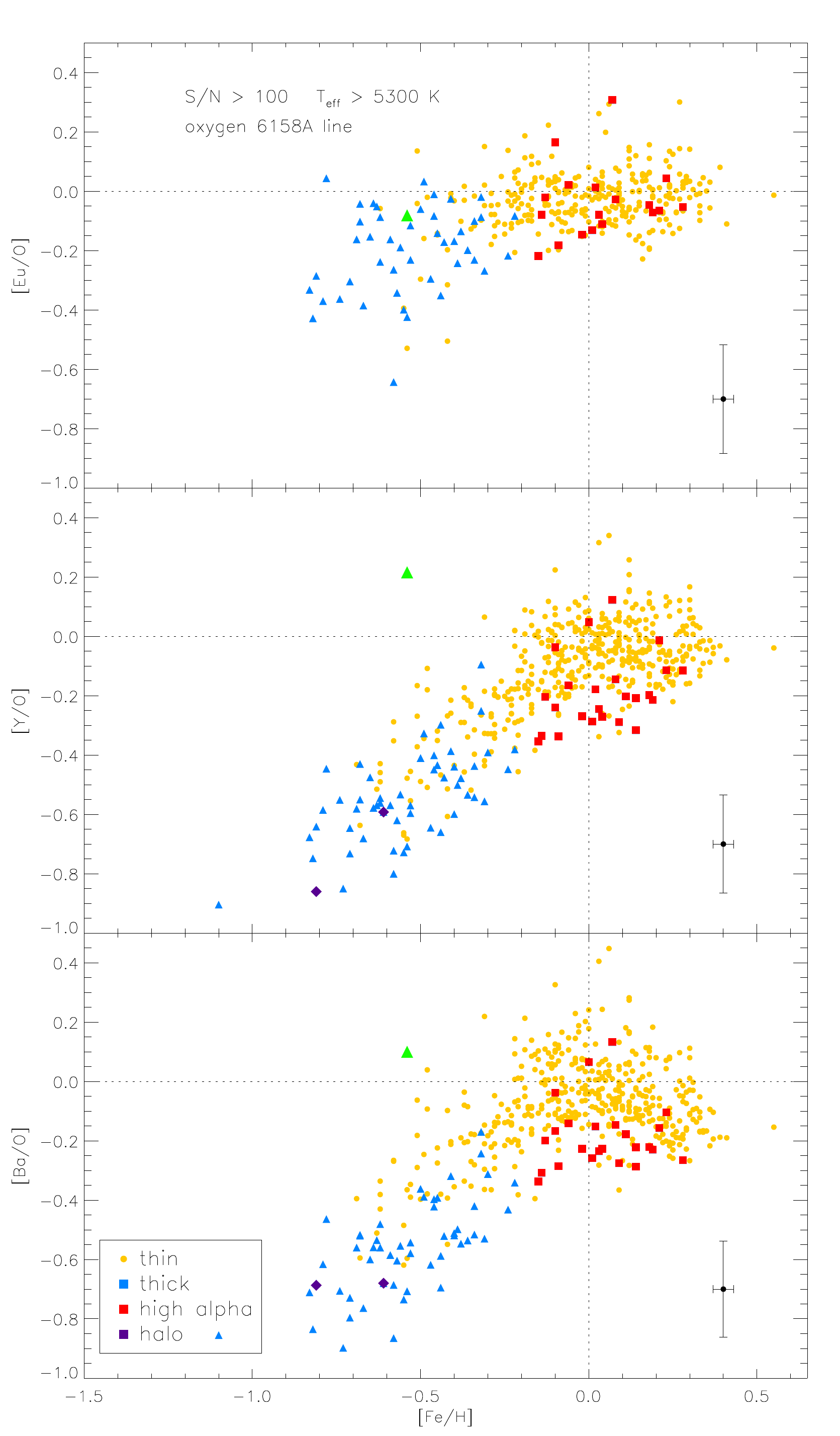}
\caption{Abundance ratios between the $\alpha$ element O and the \textit{n}-capture elements Eu, Ba and Y as a function of [Fe/H] for stars with \teff $>$ 5300\,K and S/N $>$ 100. Symbols as in Fig. \ref{hot_XFe_Fe_errors}.} 
\label{OEu}
\end{figure}

\begin{figure}
\centering
\includegraphics[width=9cm]{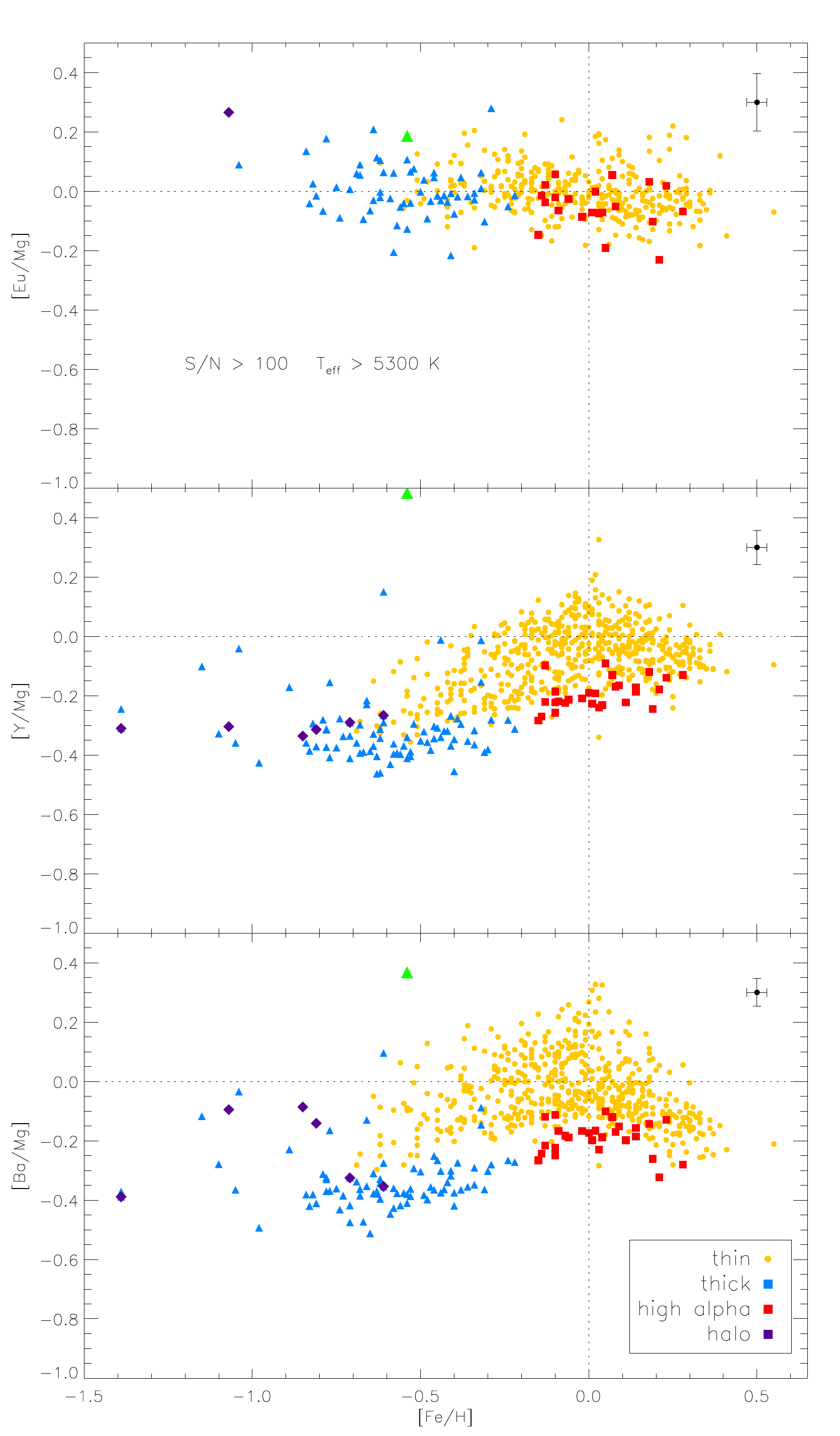}
\caption{Abundance ratios between the $\alpha$ element Mg and the \textit{n}-capture elements Eu, Ba and Y as a function of [Fe/H] for stars with \teff $>$ 5300\,K and S/N $>$ 100. Symbols as in Fig. \ref{hot_XFe_Fe_errors}.} 
\label{MgEu}
\end{figure}

Also, by studying different ratios we can get information about the masses of the progenitors enriching the ISM at the formation time of our stars. The work by \citet{travaglio99} showed that the best progenitors to reproduce the \textit{r}-process contribution to the enrichment of the Galaxy are SNe II from stars with masses 8-10\,M$_{\odot}$. On the other hand, more massive SNe II of M\,$>$\,15\,M$_{\odot}$ enriched the ISM with oxygen at earlier times since those massive stars evolve faster. As a consequence we can observe that the ratios of \textit{r}-process elements with respect to oxygen are negative for low metallicities. In Fig. \ref{OEu} we show the ratios between Eu-Y-Ba and O using the abundances of the oxygen line at 6158\AA\ derived for the same sample by \citet{bertrandelis15}. Since we do not have very metal poor stars we cannot observe the behaviour of Y and Ba at very low metallicities, where they are considered to be mainly produced by the \textit{r}- and not the \textit{s}-process. We can see how [Eu/O] has a less steep decline towards lower metallicities when compared to Ba and Y. This is because Eu is a pure \textit{r}-process whereas Y and Ba at [Fe/H]$\sim$\,--1\,dex are mainly produced by AGB stars which evolve slower than the progenitors of Eu and present even a longer delay with respect to the more massive progenitors of oxygen. At this point it is also interesting to compare our heavy elements with Mg, another $\alpha$ element, using the rederived abundances in this work. In Fig. \ref{MgEu} we can see the same decreasing trends towards lower metallicities for Y and Ba, but less steep than when comparing oxygen, while [Eu/Mg] is mostly flat. This might be explained by increasing O/Mg yields for higher mass SNe progenitors \citep[e.g.][and references therein]{woosley95,mcwilliam08}. Thus, the production of oxygen would start earlier in the Galaxy producing higher [O/Mg] at lower [Fe/H]. Moreover, the [Eu/Mg] is mostly flat suggesting that these two elements receive an important contribution from SNe progenitors of similar masses but less massive than oxygen progenitors as explained above. However, the study of \citet{mcwilliam08} discarded the possibility of having increasing O/Mg yields for higher mass SNe progenitors, since that would imply a metallicity-dependent Initial Mass Function (IMF), that is, an increase of the fraction of low mass SNe at higher [Fe/H]. Instead, they proposed that a metallicity-dependent modulation of the SNe O/Mg ratio can perfectly explain the behaviour of this ratio both in the disk and in the bulge. In Fig. \ref{MgEu} it is also interesting to see the well-defined separation of thick disk and \textit{h$\alpha$mr} stars with respect to the thin disk group for [Y/Mg] and [Ba/Mg].\\

\begin{figure}
\centering
\includegraphics[width=9.2cm]{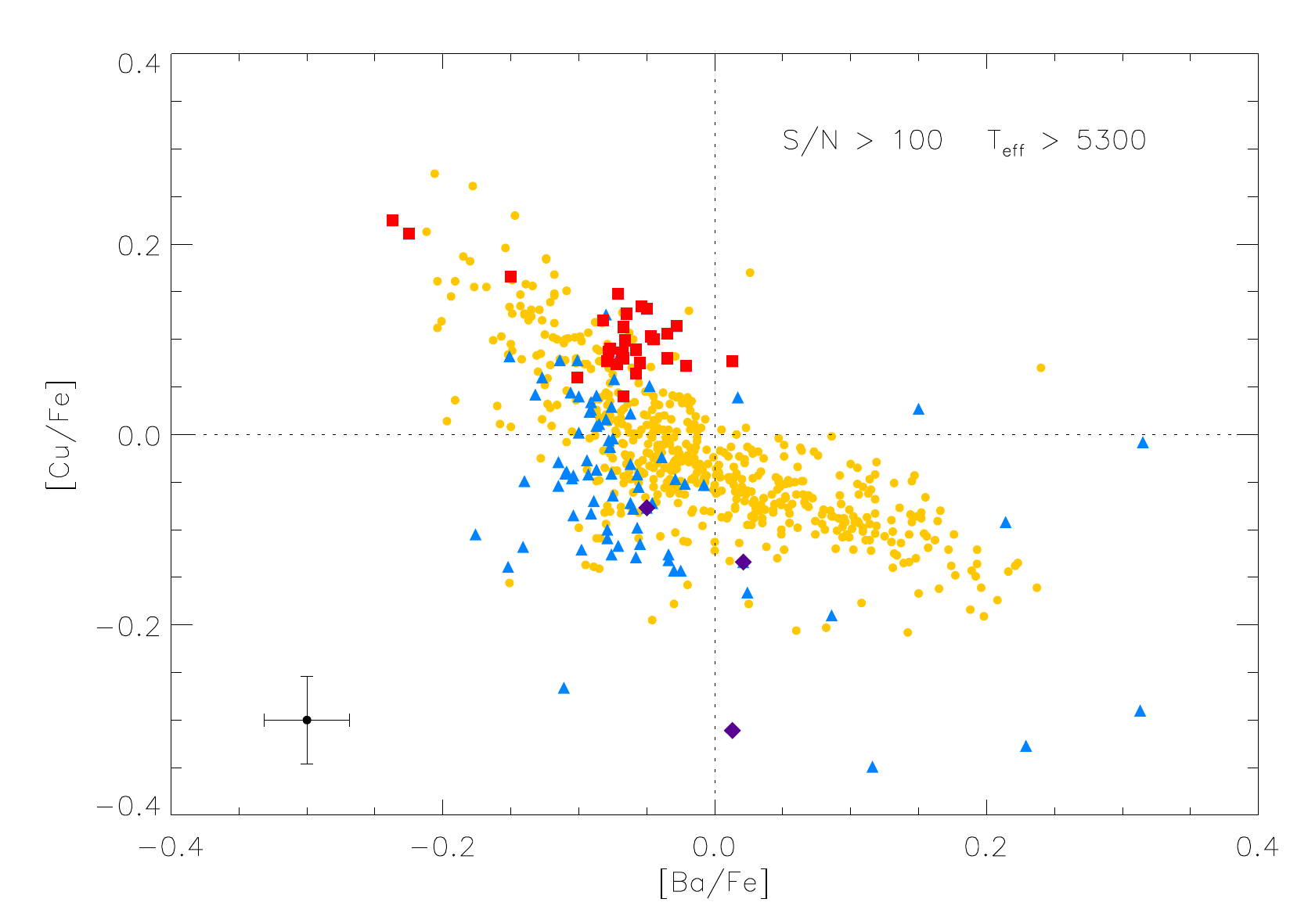}
\caption{[Cu/Fe] as a function of [Ba/Fe] for stars with \teff $>$ 5300\,K and S/N $>$ 100. Symbols as in Fig. \ref{hot_XFe_Fe_errors}.} 
\label{CuBa}
\end{figure}

In Fig. \ref{CuBa} we show the correlation of [Cu/Fe] with [Ba/Fe] abundances which is quite obvious for thin disk stars but does not seem to hold for the other populations. This correlation was first found by \citet{castro99} using a small sample of metal rich stars in the Ursa Major moving group. They pointed to a possible anticorrelation between Cu and \textit{s}-process elements maybe caused by Cu acting as seed for \textit{n}-capture elements. However, we do not find such a tight correlation with any of the other heavy-\textit{s} or light-\textit{s} elements. Later, the work by \citet{allen11} did not find that Cu-Ba correlation in a sample of barium and normal stars while Zn showed increasing trends with Ba, Sr, Y, Nd, and more tight increasing trends with Eu, and other \textit{r}-elements (Gd and Dy). Those authors conclude that the \textit{r}-process is contributing to Zn production with a higher proportion than to Cu for those stars. Contrary to those trends with \textit{s}-elements in \citet{allen11}, our thin disk stars present lower Zn abundances as Ba and especially Sr increase, but in a less tight way as Cu. On the other hand, we also find slightly higher abundances of Zn as Eu increases meanwhile Cu present a flatten trend suggesting that Zn receives a major contribution from SNe II. This is in agreement with current nucelosynthesis models where Zn receives an important contribution of neutrino winds during SNe explosions of massive stars. Nevertheless, the dispersion in these correlations is too high to extract any firm conclusion. \\

\begin{figure}
\centering
\includegraphics[width=9.2cm]{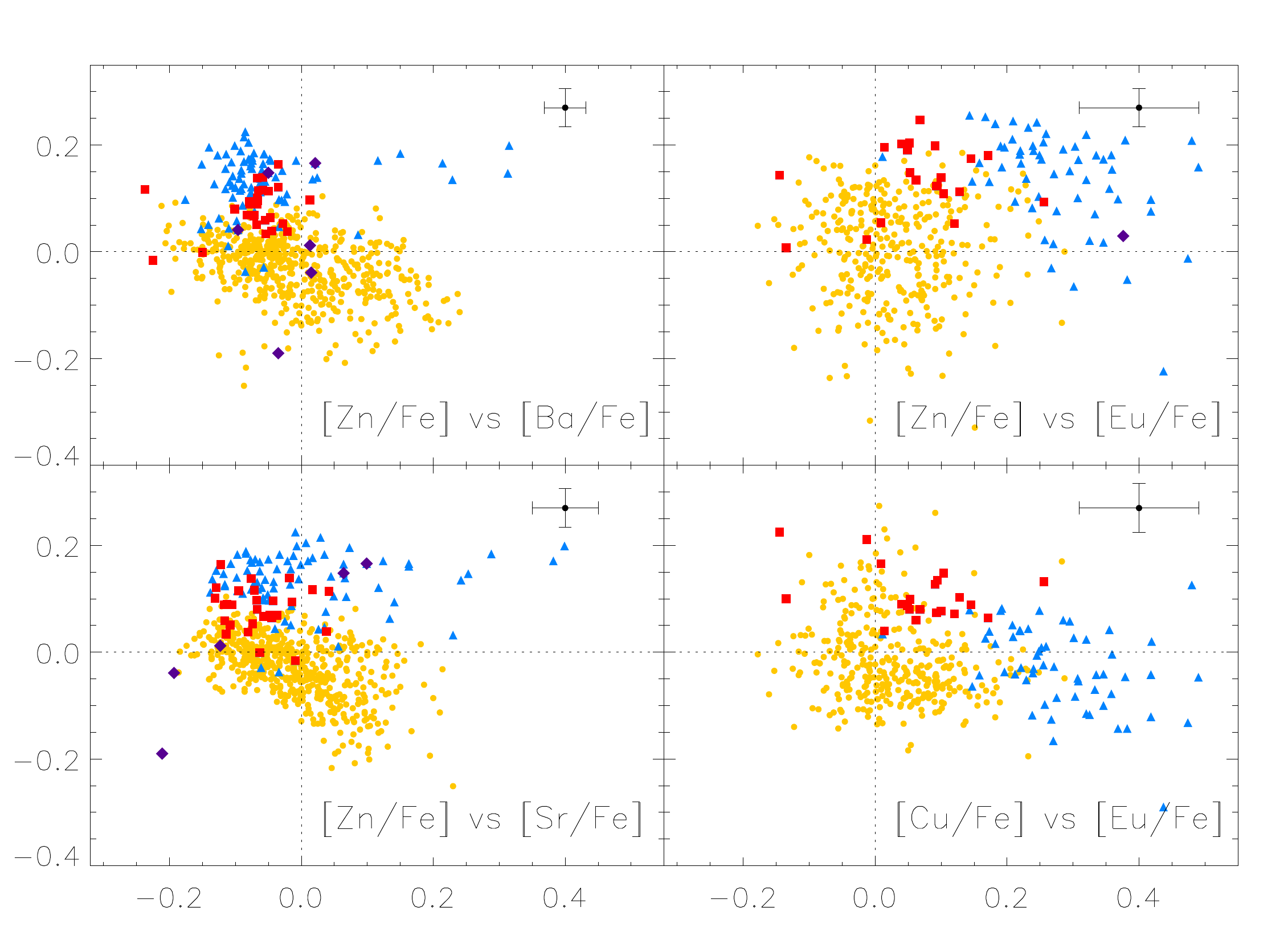}
\caption{Several abundances ratios for stars with \teff $>$ 5300\,K and S/N $>$ 100. Symbols as in Fig. \ref{hot_XFe_Fe_errors}.} 
\label{CuZn}
\end{figure}

\subsection{The ratio of heavy-\textit{s} to light-\textit{s} elements}

The main component of the \textit{s}-process produced in low-mass AGB stars (1.3\,M$_{\odot}$\,$\lesssim$\,M\,$\lesssim$\,3.5\,M$_{\odot}$) reproduces most of the \textit{s}-only isotopes in the solar system. The \textit{n}-capture takes place in the He intershell where $^{56}$Fe nuclei are fed by neutrons produced in the reaction $^{13}$C($\alpha,n$)$^{16}$O during the interpulse phase of thermally-pulsing AGB stars (TP-AGB)\citep[e.g.][]{travaglio04a}. This reaction releases a low neutron density ($<$10$^{8}$ n/cm$^{3}$) and needs a lower temperature than the $^{22}$Ne($\alpha,n$)$^{25}$Mg reaction \cite[e.g.][]{travaglio04a,fishlock14,bisterzo16} which is only partially activated during the convective thermal pulses in this kind of stars. Also, the elements produced in this way may depend on the size of the $^{13}$C pocket \citep[e.g.][]{bisterzo14}, a $^{13}$C rich region created by proton capture of $^{12}$C during the different third dredge-up mixing episodes. Light-\textit{s} elements are created first and as the neutron exposure increases the second-peak group is created. For more massive AGB stars, M\,$\gtrsim$\,4-8\,M$_{\odot}$, (and at halo metallicities) there is a higher amount of primary $^{22}$Ne which sets off the second reaction meanwhile the neutron source by the $^{13}$C reaction is lower due to the smaller $^{13}$C pocket size of these stars. In the $^{22}$Ne raction the density of neutrons is much higher ($>$10$^{11}$ n/cm$^{3}$), triggering an overproduction of neutron-rich isotopes such as $^{86}$Kr, $^{87}$Rb $^{96}$Zr. However, the contribution of intermediate-mass AGB stars to heavy-\textit{s} elements is not significant because their He-intershell is one order of magnitude smaller than in low-mass AGB stars with an uncertain formation of the $^{13}$C pocket and a less efficient third dredge-up \citep[e.g.][]{bisterzo16}. Moreover, the neutron exposure of the $^{22}$Ne reaction is lower, and thus a smaller quantity of \textit{s}-elements is expected, especially the heavier ones \citep[e.g.][]{cristallo15b}. Nevertheless, intermediate-mass AGB stars can supply up to 10\% to light-\textit{s} elements \citep{travaglio04a}. \\

\begin{figure}
\centering
\includegraphics[width=9cm]{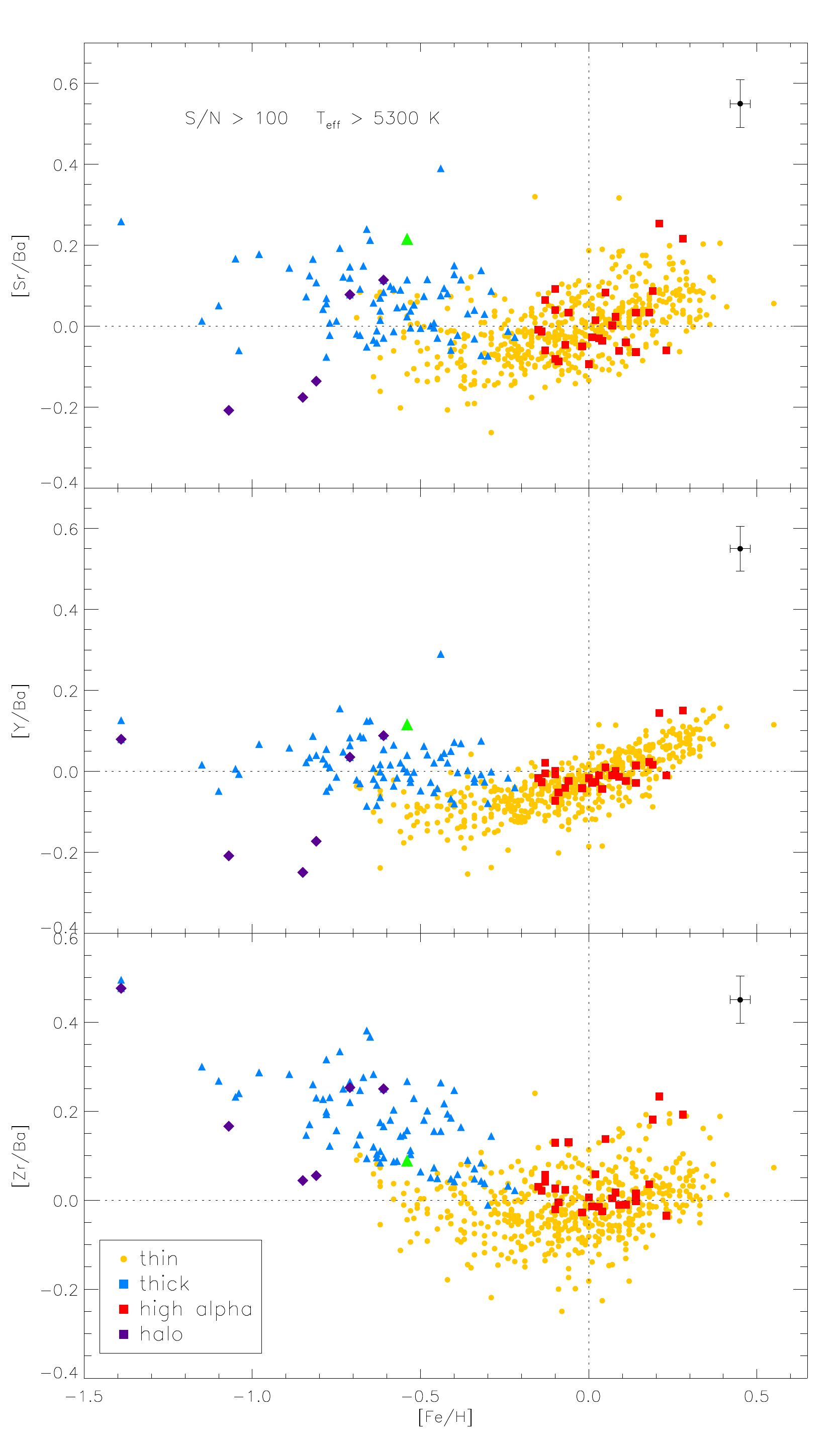}
\caption{Abundance ratios of light-\textit{s} elements respect to the heavy-\textit{s} element Ba as a function of [Fe/H] for stars with \teff $>$ 5300\,K and S/N $>$ 100. Symbols as in Fig. \ref{hot_XFe_Fe_errors}.} 
\label{Ba_lights}
\end{figure}

Low-mass AGB stars produce a higher proportion of heavy-\textit{s} respect to light-\textit{s} elements (evaluated with the ratio [\textit{hs/ls}]) at solar metallicities and below \citep[e.g.][]{karakas16b}. This fact can be seen in Fig \ref{Ba_lights} where we show the ratios of the three light-\textit{s} elements with respect to Ba, a heavy-\textit{s} element. At metallicities lower than solar, thin disk stars show negative ratios, that is, they have higher abundances of the heavy-\textit{s} element Ba. However, for a given [Fe/H], thick disk stars have less Ba with respect to light-\textit{s} (lower [\textit{hs/ls}]) than thin disk stars. This is probably caused because thick disk stars are older and at their time of formation less low-mass AGB stars (which evolve slower than intermediate-mass AGBs) have contributed to the ISM enrichment where they were formed. On the other hand, at super-solar metallicities light-\textit{s} elements with respect to Ba show an increasing trend with metallicity, very clear for [Y/Ba]. This might be caused by the contribution of metal rich AGB stars which provide higher abundances of light-\textit{s} elements such as Y and Sr than heavy-\textit{s} elements like Ba or Ce \citep[see Fig. 10 in ][]{karakas16b}.\\

\begin{figure}
\centering
\includegraphics[width=9.2cm]{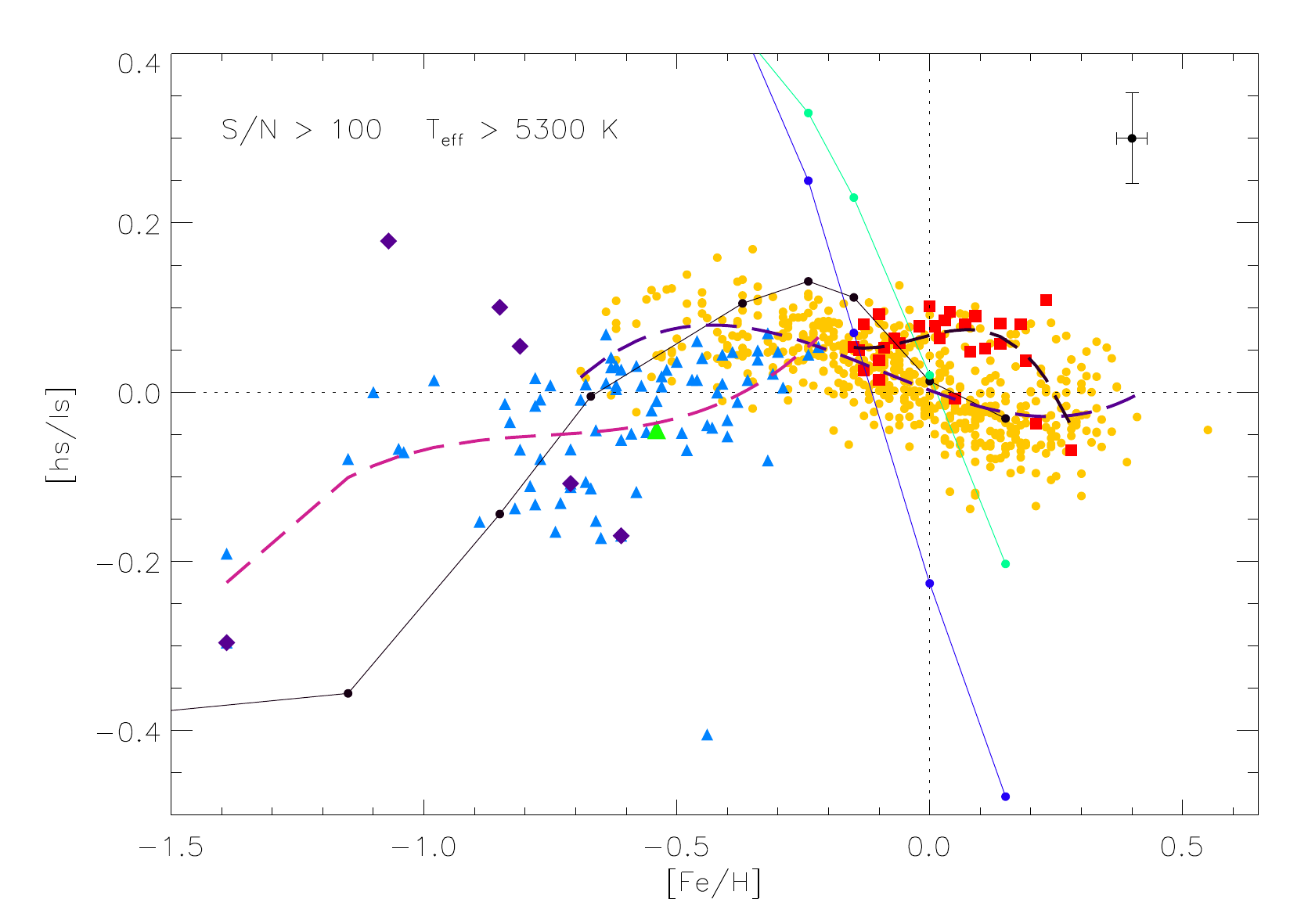}
\caption{Heavy-s to light-\textit{s} ratios as a function of [Fe/H] for stars with \teff $>$ 5300\,K and S/N $>$ 100. Symbols as in Fig. \ref{hot_XFe_Fe_errors}. The blue, green and black lines are AGB models of 2\,M$_{\odot}$, 3\,M$_{\odot}$ and 6\,M$_{\odot}$, respectively, from \citet{cristallo15b}. The long-dashed lines are polynomial fits to the different populations.} 
\label{hsls}
\end{figure}

This behaviour can also be observed by evaluating the ratio of heavy-\textit{s} to light-\textit{s} in Fig. \ref{hsls}. In this plot \textit{hs} is the average abundance of Ba, Ce and Nd\footnote{We note that a similar plot can be obtained if only Ba and Ce are considered as heavy-\textit{s}, with [\textit{hs/ls}] ratios slightly shifted towards lower values for thick disk stars. In this case, only a very small fraction of thick disk stars would show positive [\textit{hs/ls}] values.} while \textit{ls} is the average abundance of Sr, Y and Zr. This ratio is rather flat for thin disk stars at low [Fe/H] but has a maximum at [Fe/H]$\sim$\,--0.4\,dex where it starts to decline towards higher metallicities. It is also at [Fe/H]\,$\sim$\,--0.5\,dex where the production of light-\textit{s} elements such as Y in low-mass AGB stars begins to be higher than Ba as the metallicity increases \citep[see Fig. 1 from ][]{travaglio04a}. As the metallicity decreases there is a higher amount of neutrons available per Fe seed, hence the higher neutron density allows for the build-up of heavier elements and the [\textit{hs/ls}] ratio is expected to increase as the metallicity diminishes \cite[e.g.][]{busso99}. As the metallicity further decreases, the [\textit{hs/ls}] ratio reaches a maximum around [Fe/H]\,=\,--1\,dex (only one of our halo stars shows a high value at this metallicity) due to the progressive build-up of the third s-peak at Pb, which becomes dominant over the \textit{ls} and \textit{hs} production \citep[e.g.][]{travaglio04a}. \\

The rise of [\textit{hs/ls}] towards lower metallicities is very clear for thin disk stars, however, thick disk stars present the opposite trend, decreasing [\textit{hs/ls}] for decreasing [Fe/H]. This fact suggests that low-mass AGB stars are not the only important contribution to \textit{s}-elements of thick disk stars. Interestingly, most of the \textit{h$\alpha$mr} stars seem to follow the trend of thick disk stars (that starts from subsolar ratios up to [Fe/H]\,$\sim$\,--0.4\,dex to continue with supersolar ratios as the metallicity increases) and they present higher [\textit{hs/ls}] on average than thin disk stars at super-solar metallicities. The overabundance of light-\textit{s} elements with respect to heavy-\textit{s} elements was noted by \citet{travaglio04a} but at lower metallicities than the values found here for our thick disk stars. Also, in the work of \citet{battistini16} they find high abundances of Sr and Zr at [Fe/H]$\sim$\,--0.5\,dex as in this work. This overabundance of light-\textit{s} elements could not be atributed to \textit{r}-process or the standard weak \textit{s}-process so \citet{travaglio04a} proposed that an extra source of primary \textit{s}-process production was contributing to light-\textit{s} elements at low metallicities and they called it Lighter Element Primary Process (LEPP). In recent works, based on the analysis of \textit{s}-only isotopes, \citet{bisterzo14,bisterzo16} also claimed the existence of an unknown \textit{s}-process contribution to explain the solar abundances of light-\textit{s} elements. In addition, \citet{pignatari13} proposed that a so-called \textit{cs}-component ('cold' C-burning component) could produce the missing light-\textit{s} elements through the $^{13}$C($\alpha,n$)$^{16}$O reaction in massive stars (opposite to the expected $^{22}$Ne source), explaining (at least partially) the LEPP signature proposed by \citet{travaglio04a}. On the other hand, by increasing the size of the $^{13}$C pocket the models of \citep{trippella14} would not require the contribution from a LEPP. Moreover, \citet{cristallo15a} found that the LEPP is not completely necessary to explain the solar composition of \textit{s}-only isotopes when considering other aspects in their models such as rotation or star formation rate uncertainties. Indeed, our [\textit{hs/ls}] trend for thick disk stars resembles that of intermediate-mass AGB yields (5-6\,M$_{\odot}$) by \citet{cristallo15b}, thus suggesting that the s-process production of our thick disk stars is dominated by intermediate-mass instead of low-mass AGB stars. On the contrary, this missing component of light-\textit{s} elements might be explained by the contribution of massive stars which can reach up to 30\% as shown in \citet{pignatari10}. In a recent work, \cite{bisterzo17} reports that by including the contribution of fast-rotating metal-poor stars to the weak \textit{s}-process \citep[using the yields from][]{frischknecht16} the solar abundances of Sr-Y-Zr can be increased, which could partially account to the solar LEPP. However, the inclusion of these yields has a major effect on the abundances of light-\textit{s} at very low metallicities, as previously reported by \citet{cescutti13}. Nevertheless, \cite{bisterzo17} point to the necessity of a combination of different \textit{r}- and \textit{s}-process to totally account for the solar and the metal-poor LEPP. The study of other populations in the Galaxy and the late major improvements in GCE models will help to understand this issue in the future.\\


\section{Summary}

In this work we present a detailed and homogeneous study of chemical abundances of Cu, Zn, Sr, Y, Zr, Ba, Ce, Nd and Eu for the HARPS-GTO sample which complements our previous studies on Li, C, O, iron-peak and $\alpha$ elements in a homogeneous way. One of the advantages of this high quality volume-limited sample is that it allows to study the GCE at high metallicities since it contains a high number of metal-rich stars. The purpose of this work is to evaluate the production and evolution of neutron-capture elements in the different populations of the Galaxy: thin disk, thick disk and \textit{h$\alpha$mr} stars. Our main discoveries and conclusions are the following:

\begin{itemize}
 \item For the thin disk population we find mostly flat trends up to solar metallicity for Zr, Ce and Nd meanwhile Sr, Y and Ba show slightly rising abundances as [Fe/H] increases with maximum values around solar metallicity. Then, the trend is changed and the abundances decrease for higher metallicities, with Zr, Ba and Nd showing steeper declines. On the contrary, Cu presents a a flat trend up to solar metallicity to then continuously increase reaching a maximum [Cu/Fe]\,$\sim$\,0.3 at [Fe/H]\,$\sim$\,0.4\,dex meanwhile Zn remains practically flat at all [Fe/H] and Eu shows a continuous lowering trend as [Fe/H] increases. When comparing our observations of heavy-\textit{s} and light-\textit{s} elements with GCE models from  \citet{bisterzo17} we observe that the general trends of thin disk stars are well reproduced but our maximum abundances are delayed by 0.2-0.3\,dex with respect to the models. On the other hand, Cu and Zn increases along [Fe/H] for thin disk stars can be reproduced by the models of \citet{romano10} considering normal SNe II yields.\\
 
 \item Thick disk stars present disjunct abundances for Zn and Eu, resembling the behaviour of $\alpha$ elements, although Zn remains mostly flat in our [Fe/H] range. Eu and Zr present the steeper lowering trends of abundances towards higher metallicities whereas Cu shows a steep increase as [Fe/H] rises. The abundances of Y and Ba for thick disk stars are mainly subsolar, Ce and Sr present close to solar abundances on average and Zn, Zr Nd and Eu show super-solar abundances for most of the thick disk stars. The K-S tests reject the hypothesis that the abundances of thick disk stars and thin disk stars at [Fe/H]\,$<$\,--0.2\,dex belong to the same parent population for Zn, Zr, Ba and Eu. In general, none of the GCE models tested here can match the observations of thick stars except for Cu. However, we note that the low number of thick disk stars at [Fe/H]\,$<$\,--0.8\,dex in our sample is might be producing different trends than expected.\\
 
 \item We find that the \textit{h$\alpha$mr} stars show clear different abundances of Cu, Zn, Y, Ba, Nd and Eu when compared to the thin disk stars further supporting the different chemical enrichment of this population first discovered by \citep{adibekyan11}. Cu, Zn, Nd and Eu are higher when compared to thin disk stars at the same [Fe/H] whereas Ba and Y are lower on average. We have performed K-S tests that confirm this behaviour, with Cu and Zn showing the highest distance among both populations. \\
 
 \item By comparing the pure \textit{r} element Eu with the heavy-\textit{s} elements we can see how at lower metallicities the earlier production of \textit{r}-process elements by massive stars provides high [Eu/Ba] and [Eu/Ce] since Ba and Ce do not receive a major contribution of \textit{r}-process in our metallicity range. On the other hand, the [Eu/Nd] ratio is less steep because 50\% of Nd is produced by the \textit{r}-process. We also find that [Eu/O] ratio is negative at lower metallicities and slightly increase to become flat at [Fe/H]\,$\sim$\,--0\,dex. This trend shows that the SNe progenitor's masses producing O are higher than those producing Eu \citep{travaglio99}. Moreover, the steeper and more negative trends of [Y/O] and [Ba/O] agree with the production of these elements by AGB stars, since they have longer lifetimes than the massive stars contributing to Eu and O.\\
 
 \item We find that for the thin disk stars the ratio [\textit{hs/ls}] shows the expected behaviour, increasing values towards decreasing metallicities. This is a reflection of the higher neutron exposure as the metallicity declines (more neutrons per Fe seed), so for [Fe/H]\,$\gtrsim$\,--0.5\,dex the \textit{s}-process peaks at Sr-Y-Zr, for intermediate metallicities the \textit{s}-process mainly produces the second-peak (Ba-La-Ce-Sm-Nd) and for halo metallicities the high neutron exposure allows for the formation of Pb, the heaviest \textit{s}-process element. However, the thick disk population present negative [\textit{hs/ls}] supporting previous findings of a missing contribution for light-\textit{s} elements at low metallicities \citep{travaglio04a} to explain the higher than expected abundances of elements such as Sr, Y and Zr. Nevertheless, we show that the low [\textit{hs/ls}] of our thick disk might be explained with yields of intermediate-mass AGB stars \citep{cristallo15b} instead of the major expected contribution from low-mass AGB stars. Finally, we also find an increase of light-\textit{s} abundances with respect to heavy-\textit{s} at super-solar metallicities which might be produced by the contribution of metal rich AGB stars \cite{karakas16b}. Interestingly, \textit{h$\alpha$mr} mostly present positive [\textit{hs/ls}] ratios whereas thin disk stars at similar [Fe/H] are spread at positive and negative ratios.\\
\end{itemize}

Our work demonstrates that even in the era of very large surveys, homogeneous and high-precision data of relatively small samples (though probably this is among the largest in its kind) can provide very important insides for our understanding of stellar nucleosynthesis and Galactic Chemical Evolution.

\begin{acknowledgements}
E.D.M., V.Zh.A., N.C.S. and S.G.S. acknowledge the support from Funda\c{c}\~ao para a Ci\^encia e a Tecnologia (FCT) through national funds
and from FEDER through COMPETE2020 by the following grants UID/FIS/04434/2013 \& POCI--01--0145-FEDER--007672, PTDC/FIS-AST/7073/2014 \& POCI--01--0145-FEDER--016880 and PTDC/FIS-AST/1526/2014 \& POCI--01--0145-FEDER--016886. E.D.M. acknowledges the support by the fellowship SFRH/BPD/76606/2011 funded by FCT (Portugal) and by the Investigador FCT contract IF/00849/2015. V.Zh.A., N.C.S. and S.G.S. also acknowledge the support from FCT through Investigador FCT contracts IF/00650/2015, IF/00169/2012/CP0150/CT0002 and IF/00028/2014/CP1215/CT0002 funded by FCT (Portugal) and POPH/FSE (EC). J.I.G.H. acknowledges financial support from the Spanish Ministry of Economy and Competitiveness (MINECO) under the 2013 Ram\'on y Cajal program MINECO RYC--2013--14875, and the Spanish ministry project MINECO AYA2014--56359-P.\\

We thank Sara Bisterzo and Donatella Romano for kindly providing their GCE models. This research has made use of the SIMBAD database operated at CDS, Strasbourg (France), the IRAF facility and the VALD3 database. Finally, we are sincerely grateful to the anonymous referee, whose careful review helped to improve this paper.

\end{acknowledgements}

\bibliographystyle{aa}
\bibliography{edm_bibliography}

\begin{thebibliography}{98}
\expandafter\ifx\csname natexlab\endcsname\relax\def\natexlab#1{#1}\fi

\bibitem[{{Adibekyan} {et~al.}(2015){Adibekyan}, {Figueira}, {Santos}, {Sousa},
  {Faria}, {Delgado-Mena}, {Oshagh}, {Tsantaki}, {Hakobyan}, {Gonz{\'a}lez
  Hern{\'a}ndez}, {Su{\'a}rez-Andr{\'e}s}, \& {Israelian}}]{adibekyan15c}
{Adibekyan}, V., {Figueira}, P., {Santos}, N.~C., {et~al.} 2015, \aap, 583, A94

\bibitem[{{Adibekyan} {et~al.}(2013){Adibekyan}, {Figueira}, {Santos},
  {Hakobyan}, {Sousa}, {Pace}, {Delgado Mena}, {Robin}, {Israelian}, \&
  {Gonz{\'a}lez Hern{\'a}ndez}}]{adibekyan13}
{Adibekyan}, V.~Z., {Figueira}, P., {Santos}, N.~C., {et~al.} 2013, \aap, 554,
  A44

\bibitem[{{Adibekyan} {et~al.}(2011){Adibekyan}, {Santos}, {Sousa}, \&
  {Israelian}}]{adibekyan11}
{Adibekyan}, V.~Z., {Santos}, N.~C., {Sousa}, S.~G., \& {Israelian}, G. 2011,
  \aap, 535, L11

\bibitem[{{Adibekyan} {et~al.}(2012){Adibekyan}, {Sousa}, {Santos}, {Delgado
  Mena}, {Gonz{\'a}lez Hern{\'a}ndez}, {Israelian}, {Mayor}, \&
  {Khachatryan}}]{adibekyan12}
{Adibekyan}, V.~Z., {Sousa}, S.~G., {Santos}, N.~C., {et~al.} 2012, \aap, 545,
  A32

\bibitem[{{Allen} \& {Porto de Mello}(2011)}]{allen11}
{Allen}, D.~M. \& {Porto de Mello}, G.~F. 2011, \aap, 525, A63

\bibitem[{{Allende Prieto} {et~al.}(2004){Allende Prieto}, {Barklem},
  {Lambert}, \& {Cunha}}]{allende04}
{Allende Prieto}, C., {Barklem}, P.~S., {Lambert}, D.~L., \& {Cunha}, K. 2004,
  \aap, 420, 183

\bibitem[{{Arlandini} {et~al.}(1999){Arlandini}, {K{\"a}ppeler}, {Wisshak},
  {Gallino}, {Lugaro}, {Busso}, \& {Straniero}}]{arlandini99}
{Arlandini}, C., {K{\"a}ppeler}, F., {Wisshak}, K., {et~al.} 1999, \apj, 525,
  886

\bibitem[{{Astraatmadja} \& {Bailer-Jones}(2016)}]{Astraatmadja2016}
{Astraatmadja}, T.~L. \& {Bailer-Jones}, C.~A.~L. 2016, \apj, 832, 137

\bibitem[{{Barbuy} {et~al.}(2015){Barbuy}, {Fria{\c c}a}, {da Silveira},
  {Hill}, {Zoccali}, {Minniti}, {Renzini}, {Ortolani}, \&
  {G{\'o}mez}}]{barbuy15}
{Barbuy}, B., {Fria{\c c}a}, A.~C.~S., {da Silveira}, C.~R., {et~al.} 2015,
  \aap, 580, A40

\bibitem[{{Battistini} \& {Bensby}(2016)}]{battistini16}
{Battistini}, C. \& {Bensby}, T. 2016, \aap, 586, A49

\bibitem[{{Bensby} {et~al.}(2003){Bensby}, {Feltzing}, \&
  {Lundstr{\"o}m}}]{bensby03}
{Bensby}, T., {Feltzing}, S., \& {Lundstr{\"o}m}, I. 2003, \aap, 410, 527

\bibitem[{{Bensby} {et~al.}(2014){Bensby}, {Feltzing}, \& {Oey}}]{bensby14}
{Bensby}, T., {Feltzing}, S., \& {Oey}, M.~S. 2014, \aap, 562, A71

\bibitem[{{Bertran de Lis} {et~al.}(2015){Bertran de Lis}, {Delgado Mena},
  {Adibekyan}, {Santos}, \& {Sousa}}]{bertrandelis15}
{Bertran de Lis}, S., {Delgado Mena}, E., {Adibekyan}, V.~Z., {Santos}, N.~C.,
  \& {Sousa}, S.~G. 2015, \aap, 576, A89

\bibitem[{{Bessell} {et~al.}(1998){Bessell}, {Castelli}, \& {Plez}}]{bessell98}
{Bessell}, M.~S., {Castelli}, F., \& {Plez}, B. 1998, \aap, 333, 231

\bibitem[{{Bisterzo} {et~al.}(2005){Bisterzo}, {Pompeia}, {Gallino},
  {Pignatari}, {Cunha}, {Heger}, \& {Smith}}]{bisterzo05}
{Bisterzo}, S., {Pompeia}, L., {Gallino}, R., {et~al.} 2005, Nuclear Physics A,
  758, 284

\bibitem[{{Bisterzo} {et~al.}(2014){Bisterzo}, {Travaglio}, {Gallino},
  {Wiescher}, \& {K{\"a}ppeler}}]{bisterzo14}
{Bisterzo}, S., {Travaglio}, C., {Gallino}, R., {Wiescher}, M., \&
  {K{\"a}ppeler}, F. 2014, \apj, 787, 10

\bibitem[{{Bisterzo} {et~al.}(2016){Bisterzo}, {Travaglio}, {Wiescher},
  {Gallino}, {K{\"o}ppeler}, {Straniero}, {Cristallo}, {Imbriani},
  {G{\"o}rres}, \& {deBoer}}]{bisterzo16}
{Bisterzo}, S., {Travaglio}, C., {Wiescher}, M., {et~al.} 2016, Journal of
  Physics Conference Series, 665, 012023

\bibitem[{{Bisterzo} {et~al.}(2017){Bisterzo}, {Travaglio}, {Wiescher},
  {K{\"a}ppeler}, \& {Gallino}}]{bisterzo17}
{Bisterzo}, S., {Travaglio}, C., {Wiescher}, M., {K{\"a}ppeler}, F., \&
  {Gallino}, R. 2017, \apj, 835, 97

\bibitem[{{Bressan} {et~al.}(2012){Bressan}, {Marigo}, {Girardi}, {Salasnich},
  {Dal Cero}, {Rubele}, \& {Nanni}}]{bressan12}
{Bressan}, A., {Marigo}, P., {Girardi}, L., {et~al.} 2012, \mnras, 427, 127

\bibitem[{{Burbidge} {et~al.}(1957){Burbidge}, {Burbidge}, {Fowler}, \&
  {Hoyle}}]{burbidge57}
{Burbidge}, E.~M., {Burbidge}, G.~R., {Fowler}, W.~A., \& {Hoyle}, F. 1957,
  Reviews of Modern Physics, 29, 547

\bibitem[{{Busso} {et~al.}(1999){Busso}, {Gallino}, \& {Wasserburg}}]{busso99}
{Busso}, M., {Gallino}, R., \& {Wasserburg}, G.~J. 1999, \araa, 37, 239

\bibitem[{{Castro} {et~al.}(1999){Castro}, {Porto de Mello}, \& {da
  Silva}}]{castro99}
{Castro}, S., {Porto de Mello}, G.~F., \& {da Silva}, L. 1999, \mnras, 305, 693

\bibitem[{{Cayrel}(1988)}]{cayrel88}
{Cayrel}, R. 1988, in IAU Symposium, Vol. 132, The Impact of Very High S/N
  Spectroscopy on Stellar Physics, ed. G.~{Cayrel de Strobel} \& M.~{Spite},
  345

\bibitem[{{Cescutti} {et~al.}(2013){Cescutti}, {Chiappini}, {Hirschi},
  {Meynet}, \& {Frischknecht}}]{cescutti13}
{Cescutti}, G., {Chiappini}, C., {Hirschi}, R., {Meynet}, G., \&
  {Frischknecht}, U. 2013, \aap, 553, A51

\bibitem[{{Chabrier}(2001)}]{chabrier01}
{Chabrier}, G. 2001, \apj, 554, 1274

\bibitem[{{Cowan} \& {Thielemann}(2004)}]{cowan04}
{Cowan}, J.~J. \& {Thielemann}, F.-K. 2004, Physics Today, 57, 47

\bibitem[{{Cristallo} {et~al.}(2015{\natexlab{a}}){Cristallo}, {Abia},
  {Straniero}, \& {Piersanti}}]{cristallo15a}
{Cristallo}, S., {Abia}, C., {Straniero}, O., \& {Piersanti}, L.
  2015{\natexlab{a}}, \apj, 801, 53

\bibitem[{{Cristallo} {et~al.}(2015{\natexlab{b}}){Cristallo}, {Straniero},
  {Piersanti}, \& {Gobrecht}}]{cristallo15b}
{Cristallo}, S., {Straniero}, O., {Piersanti}, L., \& {Gobrecht}, D.
  2015{\natexlab{b}}, \apjs, 219, 40

\bibitem[{{da Silva} {et~al.}(2006){da Silva}, {Girardi}, {Pasquini},
  {Setiawan}, {von der L{\"u}he}, {de Medeiros}, {Hatzes}, {D{\"o}llinger}, \&
  {Weiss}}]{dasilva06_param}
{da Silva}, L., {Girardi}, L., {Pasquini}, L., {et~al.} 2006, \aap, 458, 609

\bibitem[{{Delgado Mena} {et~al.}(2015){Delgado Mena}, {Bertr{\'a}n de Lis},
  {Adibekyan}, {Sousa}, {Figueira}, {Mortier}, {Gonz{\'a}lez Hern{\'a}ndez},
  {Tsantaki}, {Israelian}, \& {Santos}}]{delgado15}
{Delgado Mena}, E., {Bertr{\'a}n de Lis}, S., {Adibekyan}, V.~Z., {et~al.}
  2015, \aap, 576, A69

\bibitem[{{Delgado Mena} {et~al.}(2014){Delgado Mena}, {Israelian},
  {Gonz{\'a}lez Hern{\'a}ndez}, {Sousa}, {Mortier}, {Santos}, {Adibekyan},
  {Fernandes}, {Rebolo}, {Udry}, \& {Mayor}}]{delgado14}
{Delgado Mena}, E., {Israelian}, G., {Gonz{\'a}lez Hern{\'a}ndez}, J.~I.,
  {et~al.} 2014, \aap, 562, A92

\bibitem[{{D'Orazi} {et~al.}(2009){D'Orazi}, {Magrini}, {Randich}, {Galli},
  {Busso}, \& {Sestito}}]{dorazi09}
{D'Orazi}, V., {Magrini}, L., {Randich}, S., {et~al.} 2009, \apjl, 693, L31

\bibitem[{{Duffau} {et~al.}(2017){Duffau}, {Caffau}, {Sbordone}, {Bonifacio},
  {Andrievsky}, {Korotin}, {Babusiaux}, {Salvadori}, {Monaco}, {Francois},
  {Skuladottir}, {Bragaglia}, {Donati}, {Spina}, {Gallagher}, {Ludwig},
  {Christlieb}, {Hansen}, {Mott}, {Steffen}, {Zaggia}, {Blanco-Cuaresma},
  {Calura}, {Friel}, {Jimenez-Esteban}, {Koch}, {Magrini}, {Pancino}, {Tang},
  {Tautvaisiene}, {Vallenari}, {Hawkins}, {Gilmore}, {Randich}, {Feltzing},
  {Bensby}, {Flaccomio}, {Smiljanic}, {Bayo}, {Carraro}, {Casey}, {Costado},
  {Damiani}, {Franciosini}, {Hourihane}, {Jofre}, {Lardo}, {Lewis},
  {Morbidelli}, {Sousa}, \& {Worley}}]{duffau17}
{Duffau}, S., {Caffau}, E., {Sbordone}, L., {et~al.} 2017, ArXiv e-prints

\bibitem[{{Fishlock} {et~al.}(2014){Fishlock}, {Karakas}, {Lugaro}, \&
  {Yong}}]{fishlock14}
{Fishlock}, C.~K., {Karakas}, A.~I., {Lugaro}, M., \& {Yong}, D. 2014, \apj,
  797, 44

\bibitem[{{Flower}(1996)}]{flower96}
{Flower}, P.~J. 1996, \apj, 469, 355

\bibitem[{{Frischknecht} {et~al.}(2016){Frischknecht}, {Hirschi}, {Pignatari},
  {Maeder}, {Meynet}, {Chiappini}, {Thielemann}, {Rauscher}, {Georgy}, \&
  {Ekstr{\"o}m}}]{frischknecht16}
{Frischknecht}, U., {Hirschi}, R., {Pignatari}, M., {et~al.} 2016, \mnras, 456,
  1803

\bibitem[{{Gilmore} {et~al.}(2012){Gilmore}, {Randich}, {Asplund}, {Binney},
  {Bonifacio}, {Drew}, {Feltzing}, {Ferguson}, {Jeffries}, {Micela}, \&
  et~al.}]{gilmore12}
{Gilmore}, G., {Randich}, S., {Asplund}, M., {et~al.} 2012, The Messenger, 147,
  25

\bibitem[{{Gonz{\'a}lez Hern{\'a}ndez} {et~al.}(2013){Gonz{\'a}lez
  Hern{\'a}ndez}, {Delgado-Mena}, {Sousa}, {Israelian}, {Santos}, {Adibekyan},
  \& {Udry}}]{jonay13}
{Gonz{\'a}lez Hern{\'a}ndez}, J.~I., {Delgado-Mena}, E., {Sousa}, S.~G.,
  {et~al.} 2013, \aap, 552, A6

\bibitem[{{Gonz{\'a}lez Hern{\'a}ndez} {et~al.}(2010){Gonz{\'a}lez
  Hern{\'a}ndez}, {Israelian}, {Santos}, {Sousa}, {Delgado-Mena}, {Neves}, \&
  {Udry}}]{jonay10}
{Gonz{\'a}lez Hern{\'a}ndez}, J.~I., {Israelian}, G., {Santos}, N.~C., {et~al.}
  2010, \apj, 720, 1592

\bibitem[{{Grevesse} \& {Sauval}(1998)}]{grevesse98}
{Grevesse}, N. \& {Sauval}, A.~J. 1998, \ssr, 85, 161

\bibitem[{{Heijmans} {et~al.}(2012){Heijmans}, {Asplund}, {Barden}, {Birchall},
  {Carollo}, {Bland-Hawthorn}, {Brzeski}, {Case}, {Churilov}, {Colless},
  {Dean}, {De Silva}, {Farrell}, {Fiegert}, {Freeman}, {Gers}, {Goodwin},
  {Gray}, {Heald}, {Heng}, {Jones}, {Kobayashi}, {Klauser}, {Kondrat},
  {Lawrence}, {Lee}, {Mathews}, {Mayfield}, {Miziarski}, {Monnet}, {Muller},
  {Pai}, {Patterson}, {Penny}, {Orr}, {Sheinis}, {Shortridge}, {Smedley},
  {Smith}, {Stafford}, {Staszak}, {Vuong}, {Waller}, {Whittard}, {Wylie de
  Boer}, {Xavier}, {Zheng}, {Zhelem}, \& {Zucker}}]{heijmans12}
{Heijmans}, J., {Asplund}, M., {Barden}, S., {et~al.} 2012, in \procspie, Vol.
  8446, Ground-based and Airborne Instrumentation for Astronomy IV, 84460W

\bibitem[{{Heiter} {et~al.}(2015){Heiter}, {Lind}, {Asplund}, {Barklem},
  {Bergemann}, {Magrini}, {Masseron}, {Mikolaitis}, {Pickering}, \&
  {Ruffoni}}]{heiter15}
{Heiter}, U., {Lind}, K., {Asplund}, M., {et~al.} 2015, \physscr, 90, 054010

\bibitem[{{Israelian} {et~al.}(2014){Israelian}, {Bertran de Lis}, {Delgado
  Mena}, \& {Adibekyan}}]{israelian14}
{Israelian}, G., {Bertran de Lis}, S., {Delgado Mena}, E., \& {Adibekyan},
  V.~Z. 2014, \memsai, 85, 265

\bibitem[{{Kappeler} {et~al.}(1989){Kappeler}, {Beer}, \&
  {Wisshak}}]{kappeler89}
{Kappeler}, F., {Beer}, H., \& {Wisshak}, K. 1989, Reports on Progress in
  Physics, 52, 945

\bibitem[{{K{\"a}ppeler} {et~al.}(2011){K{\"a}ppeler}, {Gallino}, {Bisterzo},
  \& {Aoki}}]{kappeler11}
{K{\"a}ppeler}, F., {Gallino}, R., {Bisterzo}, S., \& {Aoki}, W. 2011, Reviews
  of Modern Physics, 83, 157

\bibitem[{{Karakas}(2016)}]{karakas16a}
{Karakas}, A.~I. 2016, \memsai

\bibitem[{{Karakas} \& {Lugaro}(2016)}]{karakas16b}
{Karakas}, A.~I. \& {Lugaro}, M. 2016, \apj

\bibitem[{{Kobayashi} \& {Nomoto}(2009)}]{kobayashi09}
{Kobayashi}, C. \& {Nomoto}, K. 2009, \apj, 707, 1466

\bibitem[{{Kurucz}(1993)}]{kurucz}
{Kurucz}, R. 1993, ATLAS9 Stellar Atmosphere Programs and 2 km/s grid.~Kurucz
  CD-ROM No.~13.~ Cambridge, Mass.: Smithsonian Astrophysical Observatory,
  1993., 13

\bibitem[{{Lindegren} \& {Feltzing}(2013)}]{lindregen13}
{Lindegren}, L. \& {Feltzing}, S. 2013, \aap, 553, A94

\bibitem[{{Lo Curto} {et~al.}(2010){Lo Curto}, {Mayor}, {Benz}, {Bouchy},
  {Lovis}, {Moutou}, {Naef}, {Pepe}, {Queloz}, {Santos}, {Segransan}, \&
  {Udry}}]{locurto}
{Lo Curto}, G., {Mayor}, M., {Benz}, W., {et~al.} 2010, \aap, 512, A48

\bibitem[{{Lodders} {et~al.}(2009){Lodders}, {Palme}, \& {Gail}}]{lodders09}
{Lodders}, K., {Palme}, H., \& {Gail}, H.-P. 2009, Landolt B{\"o}rnstein

\bibitem[{{Maiorca} {et~al.}(2012){Maiorca}, {Magrini}, {Busso}, {Randich},
  {Palmerini}, \& {Trippella}}]{maiorca12}
{Maiorca}, E., {Magrini}, L., {Busso}, M., {et~al.} 2012, \apj, 747, 53

\bibitem[{{Matteucci} {et~al.}(1993){Matteucci}, {Raiteri}, {Busson},
  {Gallino}, \& {Gratton}}]{matteucci93}
{Matteucci}, F., {Raiteri}, C.~M., {Busson}, M., {Gallino}, R., \& {Gratton},
  R. 1993, \aap, 272, 421

\bibitem[{{Mayor} {et~al.}(2003){Mayor}, {Pepe}, {Queloz}, {Bouchy},
  {Rupprecht}, {Lo Curto}, {Avila}, {Benz}, {Bertaux}, {Bonfils}, {Dall},
  {Dekker}, {Delabre}, {Eckert}, {Fleury}, {Gilliotte}, {Gojak}, {Guzman},
  {Kohler}, {Lizon}, {Longinotti}, {Lovis}, {Megevand}, {Pasquini}, {Reyes},
  {Sivan}, {Sosnowska}, {Soto}, {Udry}, {van Kesteren}, {Weber}, \&
  {Weilenmann}}]{mayor03}
{Mayor}, M., {Pepe}, F., {Queloz}, D., {et~al.} 2003, The Messenger, 114, 20

\bibitem[{{McWilliam} {et~al.}(2008){McWilliam}, {Matteucci}, {Ballero},
  {Rich}, {Fulbright}, \& {Cescutti}}]{mcwilliam08}
{McWilliam}, A., {Matteucci}, F., {Ballero}, S., {et~al.} 2008, \aj, 136, 367

\bibitem[{{Mikolaitis} {et~al.}(2017){Mikolaitis}, {de Laverny},
  {Recio-Blanco}, {Hill}, {Worley}, \& {de Pascale}}]{mikolaitis17}
{Mikolaitis}, {\v S}., {de Laverny}, P., {Recio-Blanco}, A., {et~al.} 2017,
  \aap, 600, A22

\bibitem[{{Mishenina} {et~al.}(2002){Mishenina}, {Kovtyukh}, {Soubiran},
  {Travaglio}, \& {Busso}}]{mishenina02}
{Mishenina}, T.~V., {Kovtyukh}, V.~V., {Soubiran}, C., {Travaglio}, C., \&
  {Busso}, M. 2002, \aap, 396, 189

\bibitem[{{Mishenina} {et~al.}(2013){Mishenina}, {Pignatari}, {Korotin},
  {Soubiran}, {Charbonnel}, {Thielemann}, {Gorbaneva}, \&
  {Basak}}]{mishenina13}
{Mishenina}, T.~V., {Pignatari}, M., {Korotin}, S.~A., {et~al.} 2013, \aap,
  552, A128

\bibitem[{{Mortier} {et~al.}(2014){Mortier}, {Sousa}, {Adibekyan},
  {Brand{\~a}o}, \& {Santos}}]{mortier14}
{Mortier}, A., {Sousa}, S.~G., {Adibekyan}, V.~Z., {Brand{\~a}o}, I.~M., \&
  {Santos}, N.~C. 2014, ArXiv e-prints

\bibitem[{{Pignatari} {et~al.}(2010){Pignatari}, {Gallino}, {Heil}, {Wiescher},
  {K{\"a}ppeler}, {Herwig}, \& {Bisterzo}}]{pignatari10}
{Pignatari}, M., {Gallino}, R., {Heil}, M., {et~al.} 2010, \apj, 710, 1557

\bibitem[{{Pignatari} {et~al.}(2013){Pignatari}, {Hirschi}, {Wiescher},
  {Gallino}, {Bennett}, {Beard}, {Fryer}, {Herwig}, {Rockefeller}, \&
  {Timmes}}]{pignatari13}
{Pignatari}, M., {Hirschi}, R., {Wiescher}, M., {et~al.} 2013, \apj, 762, 31

\bibitem[{{Pomp{\'e}ia} \& {Allen}(2008)}]{pompeia08}
{Pomp{\'e}ia}, L. \& {Allen}, D.~M. 2008, \aap, 488, 723

\bibitem[{{Prantzos} {et~al.}(1990){Prantzos}, {Hashimoto}, \&
  {Nomoto}}]{prantzos90}
{Prantzos}, N., {Hashimoto}, M., \& {Nomoto}, K. 1990, \aap, 234, 211

\bibitem[{{Prochaska} {et~al.}(2000){Prochaska}, {Naumov}, {Carney},
  {McWilliam}, \& {Wolfe}}]{prochaska00}
{Prochaska}, J.~X., {Naumov}, S.~O., {Carney}, B.~W., {McWilliam}, A., \&
  {Wolfe}, A.~M. 2000, \aj, 120, 2513

\bibitem[{{Raiteri} {et~al.}(1993){Raiteri}, {Gallino}, {Busso}, {Neuberger},
  \& {Kaeppeler}}]{raiteri93}
{Raiteri}, C.~M., {Gallino}, R., {Busso}, M., {Neuberger}, D., \& {Kaeppeler},
  F. 1993, \apj, 419, 207

\bibitem[{{Reddy} {et~al.}(2006){Reddy}, {Lambert}, \& {Allende
  Prieto}}]{reddy06}
{Reddy}, B.~E., {Lambert}, D.~L., \& {Allende Prieto}, C. 2006, \mnras, 367,
  1329

\bibitem[{{Reddy} {et~al.}(2003){Reddy}, {Tomkin}, {Lambert}, \& {Allende
  Prieto}}]{reddy03}
{Reddy}, B.~E., {Tomkin}, J., {Lambert}, D.~L., \& {Allende Prieto}, C. 2003,
  \mnras, 340, 304

\bibitem[{{Romano} {et~al.}(2010){Romano}, {Karakas}, {Tosi}, \&
  {Matteucci}}]{romano10}
{Romano}, D., {Karakas}, A.~I., {Tosi}, M., \& {Matteucci}, F. 2010, \aap, 522,
  A32

\bibitem[{{Romano} \& {Matteucci}(2007)}]{romano07}
{Romano}, D. \& {Matteucci}, F. 2007, \mnras, 378, L59

\bibitem[{{Ryabchikova} {et~al.}(2015){Ryabchikova}, {Piskunov}, {Kurucz},
  {Stempels}, {Heiter}, {Pakhomov}, \& {Barklem}}]{vald15}
{Ryabchikova}, T., {Piskunov}, N., {Kurucz}, R.~L., {et~al.} 2015, \physscr,
  90, 054005

\bibitem[{{Saito} {et~al.}(2009){Saito}, {Takada-Hidai}, {Honda}, \&
  {Takeda}}]{saito09}
{Saito}, Y.-J., {Takada-Hidai}, M., {Honda}, S., \& {Takeda}, Y. 2009, \pasj,
  61, 549

\bibitem[{{Santos} {et~al.}(2004){Santos}, {Israelian}, \& {Mayor}}]{santos04}
{Santos}, N.~C., {Israelian}, G., \& {Mayor}, M. 2004, \aap, 415, 1153

\bibitem[{{Santos} {et~al.}(2011){Santos}, {Mayor}, {Bonfils}, {Dumusque},
  {Bouchy}, {Figueira}, {Lovis}, {Melo}, {Pepe}, {Queloz}, {S{\'e}gransan},
  {Sousa}, \& {Udry}}]{santos_harps4}
{Santos}, N.~C., {Mayor}, M., {Bonfils}, X., {et~al.} 2011, \aap, 526, A112

\bibitem[{{Sneden} {et~al.}(2008){Sneden}, {Cowan}, \& {Gallino}}]{sneden08}
{Sneden}, C., {Cowan}, J.~J., \& {Gallino}, R. 2008, \araa, 46, 241

\bibitem[{{Sneden} {et~al.}(1991){Sneden}, {Gratton}, \& {Crocker}}]{sneden91}
{Sneden}, C., {Gratton}, R.~G., \& {Crocker}, D.~A. 1991, \aap, 246, 354

\bibitem[{{Sneden}(1973)}]{sneden}
{Sneden}, C.~A. 1973, PhD thesis, The University of Texas at Austin.

\bibitem[{{Sousa} {et~al.}(2015){Sousa}, {Santos}, {Adibekyan}, {Delgado-Mena},
  \& {Israelian}}]{sousa_ares2}
{Sousa}, S.~G., {Santos}, N.~C., {Adibekyan}, V., {Delgado-Mena}, E., \&
  {Israelian}, G. 2015, \aap, 577, A67

\bibitem[{{Sousa} {et~al.}(2011{\natexlab{a}}){Sousa}, {Santos}, {Israelian},
  {Lovis}, {Mayor}, {Silva}, \& {Udry}}]{sousa_harps4}
{Sousa}, S.~G., {Santos}, N.~C., {Israelian}, G., {et~al.} 2011{\natexlab{a}},
  \aap, 526, A99

\bibitem[{{Sousa} {et~al.}(2007){Sousa}, {Santos}, {Israelian}, {Mayor}, \&
  {Monteiro}}]{sousa_ares}
{Sousa}, S.~G., {Santos}, N.~C., {Israelian}, G., {Mayor}, M., \& {Monteiro},
  M.~J.~P.~F.~G. 2007, \aap, 469, 783

\bibitem[{{Sousa} {et~al.}(2011{\natexlab{b}}){Sousa}, {Santos}, {Israelian},
  {Mayor}, \& {Udry}}]{sousa_harps2}
{Sousa}, S.~G., {Santos}, N.~C., {Israelian}, G., {Mayor}, M., \& {Udry}, S.
  2011{\natexlab{b}}, \aap, 533, A141

\bibitem[{{Sousa} {et~al.}(2008){Sousa}, {Santos}, {Mayor}, {Udry},
  {Casagrande}, {Israelian}, {Pepe}, {Queloz}, \& {Monteiro}}]{sousa08}
{Sousa}, S.~G., {Santos}, N.~C., {Mayor}, M., {et~al.} 2008, \aap, 487, 373

\bibitem[{{Spite} \& {Spite}(1978)}]{spite78}
{Spite}, M. \& {Spite}, F. 1978, \aap, 67, 23

\bibitem[{{Steinmetz}(2003)}]{steinmetz03}
{Steinmetz}, M. 2003, in Astronomical Society of the Pacific Conference Series,
  Vol. 298, GAIA Spectroscopy: Science and Technology, ed. U.~{Munari}, 381

\bibitem[{{Su{\'a}rez-Andr{\'e}s}
  {et~al.}(2016{\natexlab{a}}){Su{\'a}rez-Andr{\'e}s}, {Israelian},
  {Gonz{\'a}lez Hern{\'a}ndez}, {Adibekyan}, {Delgado Mena}, {Santos}, \&
  {Sousa}}]{suarez-andres16}
{Su{\'a}rez-Andr{\'e}s}, L., {Israelian}, G., {Gonz{\'a}lez Hern{\'a}ndez},
  J.~I., {et~al.} 2016{\natexlab{a}}, \aap, 591, A69

\bibitem[{{Su{\'a}rez-Andr{\'e}s}
  {et~al.}(2016{\natexlab{b}}){Su{\'a}rez-Andr{\'e}s}, {Israelian},
  {Gonz{\'a}lez Hern{\'a}ndez}, {Adibekyan}, {Delgado Mena}, {Santos}, \&
  {Sousa}}]{suarez-andres17}
{Su{\'a}rez-Andr{\'e}s}, L., {Israelian}, G., {Gonz{\'a}lez Hern{\'a}ndez},
  J.~I., {et~al.} 2016{\natexlab{b}}, ArXiv e-prints

\bibitem[{{Torres}(2010)}]{torres10}
{Torres}, G. 2010, \aj, 140, 1158

\bibitem[{{Travaglio} {et~al.}(1999){Travaglio}, {Galli}, {Gallino}, {Busso},
  {Ferrini}, \& {Straniero}}]{travaglio99}
{Travaglio}, C., {Galli}, D., {Gallino}, R., {et~al.} 1999, \apj, 521, 691

\bibitem[{{Travaglio} {et~al.}(2004{\natexlab{a}}){Travaglio}, {Gallino},
  {Arnone}, {Cowan}, {Jordan}, \& {Sneden}}]{travaglio04a}
{Travaglio}, C., {Gallino}, R., {Arnone}, E., {et~al.} 2004{\natexlab{a}}, \apj

\bibitem[{{Travaglio} {et~al.}(2004{\natexlab{b}}){Travaglio}, {Hillebrandt},
  {Reinecke}, \& {Thielemann}}]{travaglio04b}
{Travaglio}, C., {Hillebrandt}, W., {Reinecke}, M., \& {Thielemann}, F.-K.
  2004{\natexlab{b}}, \aap

\bibitem[{{Trippella} {et~al.}(2014){Trippella}, {Busso}, {Maiorca},
  {K{\"a}ppeler}, \& {Palmerini}}]{trippella14}
{Trippella}, O., {Busso}, M., {Maiorca}, E., {K{\"a}ppeler}, F., \&
  {Palmerini}, S. 2014, \apj, 787, 41

\bibitem[{{Tsantaki} {et~al.}(2013){Tsantaki}, {Sousa}, {Adibekyan}, {Santos},
  {Mortier}, \& {Israelian}}]{tsantaki13}
{Tsantaki}, M., {Sousa}, S.~G., {Adibekyan}, V.~Z., {et~al.} 2013, \aap, 555,
  A150

\bibitem[{{van Leeuwen}(2007)}]{hip}
{van Leeuwen}, F. 2007, \aap, 474, 653

\bibitem[{{Wilson} {et~al.}(2010){Wilson}, {Hearty}, {Skrutskie}, {Majewski},
  {Schiavon}, {Eisenstein}, {Gunn}, {Blank}, {Henderson}, {Smee}, {Barkhouser},
  {Harding}, {Fitzgerald}, {Stolberg}, {Arns}, {Nelson}, {Brunner}, {Burton},
  {Walker}, {Lam}, {Maseman}, {Barr}, {Leger}, {Carey}, {MacDonald}, {Horne},
  {Young}, {Rieke}, {Rieke}, {O'Brien}, {Hope}, {Krakula}, {Crane}, {Zhao},
  {Carr}, {Harrison}, {Stoll}, {Vernieri}, {Holtzman}, {Shetrone},
  {Allende-Prieto}, {Johnson}, {Frinchaboy}, {Zasowski}, {Bizyaev},
  {Gillespie}, \& {Weinberg}}]{wilson10}
{Wilson}, J.~C., {Hearty}, F., {Skrutskie}, M.~F., {et~al.} 2010, in \procspie,
  Vol. 7735, Ground-based and Airborne Instrumentation for Astronomy III,
  77351C

\bibitem[{{Woosley} \& {Weaver}(1995)}]{woosley95}
{Woosley}, S.~E. \& {Weaver}, T.~A. 1995, \apjs, 101, 181

\bibitem[{{Yan} {et~al.}(2015){Yan}, {Shi}, \& {Zhao}}]{yan15}
{Yan}, H.~L., {Shi}, J.~R., \& {Zhao}, G. 2015, \apj, 802, 36

\bibitem[{{Yanny} {et~al.}(2009){Yanny}, {Rockosi}, {Newberg}, {Knapp},
  {Adelman-McCarthy}, {Alcorn}, {Allam}, {Allende Prieto}, {An}, {Anderson},
  {Anderson}, {Bailer-Jones}, {Bastian}, {Beers}, {Bell}, {Belokurov},
  {Bizyaev}, {Blythe}, {Bochanski}, {Boroski}, {Brinchmann}, {Brinkmann},
  {Brewington}, {Carey}, {Cudworth}, {Evans}, {Evans}, {Gates}, {G{\"a}nsicke},
  {Gillespie}, {Gilmore}, {Nebot Gomez-Moran}, {Grebel}, {Greenwell}, {Gunn},
  {Jordan}, {Jordan}, {Harding}, {Harris}, {Hendry}, {Holder}, {Ivans},
  {Ivezi{\v c}}, {Jester}, {Johnson}, {Kent}, {Kleinman}, {Kniazev},
  {Krzesinski}, {Kron}, {Kuropatkin}, {Lebedeva}, {Lee}, {French Leger},
  {L{\'e}pine}, {Levine}, {Lin}, {Long}, {Loomis}, {Lupton}, {Malanushenko},
  {Malanushenko}, {Margon}, {Martinez-Delgado}, {McGehee}, {Monet}, {Morrison},
  {Munn}, {Neilsen}, {Nitta}, {Norris}, {Oravetz}, {Owen}, {Padmanabhan},
  {Pan}, {Peterson}, {Pier}, {Platson}, {Re Fiorentin}, {Richards}, {Rix},
  {Schlegel}, {Schneider}, {Schreiber}, {Schwope}, {Sibley}, {Simmons},
  {Snedden}, {Allyn Smith}, {Stark}, {Stauffer}, {Steinmetz}, {Stoughton},
  {SubbaRao}, {Szalay}, {Szkody}, {Thakar}, {Sivarani}, {Tucker}, {Uomoto},
  {Vanden Berk}, {Vidrih}, {Wadadekar}, {Watters}, {Wilhelm}, {Wyse}, {Yarger},
  \& {Zucker}}]{yanni09}
{Yanny}, B., {Rockosi}, C., {Newberg}, H.~J., {et~al.} 2009, \aj, 137, 4377

\bibitem[{{Zhao} {et~al.}(2016){Zhao}, {Mashonkina}, {Yan}, {Alexeeva},
  {Kobayashi}, {Pakhomov}, {Shi}, {Sitnova}, {Tan}, {Zhang}, {Zhang}, {Zhou},
  {Bolte}, {Chen}, {Li}, {Liu}, \& {Zhai}}]{zhao16}
{Zhao}, G., {Mashonkina}, L., {Yan}, H.~L., {et~al.} 2016, ArXiv e-prints

\end{thebibliography}

\clearpage

\appendix

\section{}
\begin{figure*}
\centering
\includegraphics[width=19.2cm]{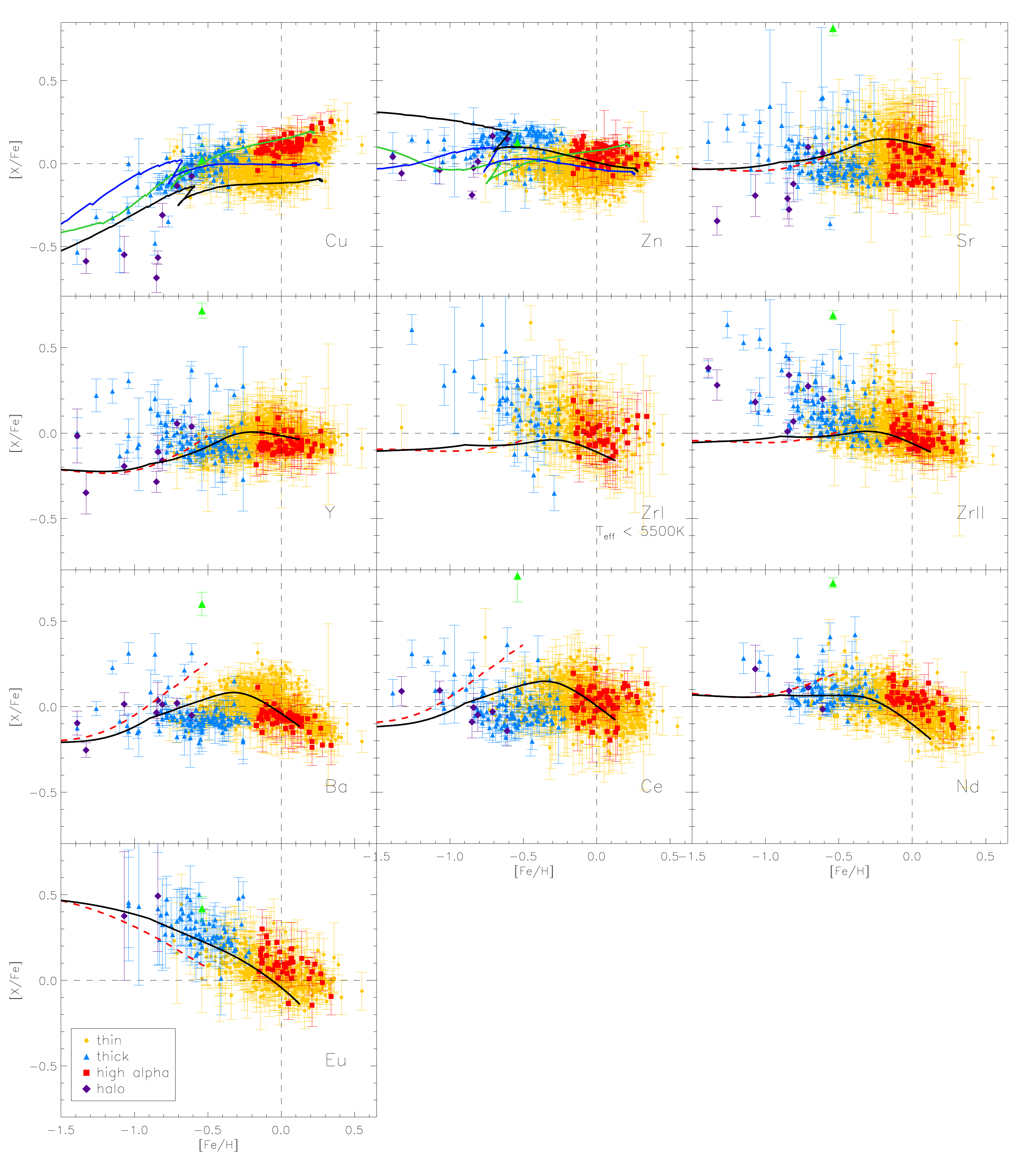}
\caption{Final [X/Fe] ratios as a function of [Fe/H] for the full sample. The different stellar populations are depicted with different colours and symbols as explained in the legend. The green bigger triangle is the s-enriched star HD11397. For Cu and Zn we overplot the GCE models 1, 4 and 5 (green, blue and black lines, respectively) from \citet{romano10}. For the rest of the elements we show the GCE models from \citet{bisterzo17} for the thin disk (black lines) and the thick disk (red dashed lines).} 
\label{all_XFe_Fe}
\end{figure*}

\end{document}